\documentclass[a4paper,fleqn,usenatbib]{mnras}
\usepackage{newtxtext,newtxmath}
\usepackage[T1]{fontenc}
\usepackage{ae,aecompl}

\usepackage{graphicx}	
\usepackage{color}
\usepackage{url}
\usepackage{float}
\usepackage{adjustbox}
\usepackage{subfig}
\usepackage{placeins}
\usepackage{booktabs}
\usepackage{afterpage}

\usepackage{etoolbox}

\definecolor{RED}{RGB}{255,0,0}

\title[The transmission spectrum of WASP-6b]{Detection of Na, K, and H$_2$O in the hazy atmosphere of WASP-6b}

\author[A. L. Carter et al.]{Aarynn L. Carter,$^{1}$\thanks{E-mail: alc227@exeter.ac.uk (ALC)} 
Nikolay Nikolov,$^{2}$
David K. Sing,$^{2}$
Munazza K. Alam,$^{3}$
\newauthor
Jayesh M. Goyal,$^{1,4}$
Thomas Mikal-Evans,$^{5}$
Hannah R. Wakeford,$^{6}$
Gregory W. Henry,$^{7}$
\newauthor
Sam Morrell,$^{1}$
Mercedes L\'opez-Morales,$^{3}$
Barry Smalley,$^{8}$
Panayotis Lavvas,$^{9}$
\newauthor
Joanna K. Barstow,$^{10}$
Antonio Garc\'ia Mu\~noz,$^{11}$
Neale P. Gibson,$^{12}$
\newauthor
and Paul A. Wilson$^{13, 14}$
\\
$^{1}$Astrophysics Group, University of Exeter, Physics Building, Stocker Road, Devon, EX4 4QL, UK\\
$^{2}$Department of Earth and Planetary Sciences, Johns Hopkins University, Baltimore, MD 21218, USA \\
$^{3}$Center for Astrophysics ${\rm \mid}$ Harvard {\rm \&} Smithsonian, 60 Garden St, Cambridge, MA 02138, USA \\
$^{4}$Department of Astronomy and Carl Sagan Institute, Cornell University, New York 14853, USA \\ 
$^{5}$Kavli Institute for Astrophysics and Space Research, Massachusetts Institute of Technology, MA 02139, USA \\
$^{6}$Space Telescope Science Institute, Baltimore, MD 21218, USA \\
$^{7}$Center of Excellence in Information Systems, Tennessee State University, Nashville, TN 37209, USA \\
$^{8}$Astrophysics Group, Keele University, Staffordshire, ST5 5BG, UK \\
$^{9}$Universit\'e de Reims Champagne Ardenne, CNRS, GSMA, UMR 7331, 51097 Reims, France \\
$^{10}$Physics and Astronomy, University College London, London, WC1E 6BT, UK \\
$^{11}$Zentrum f\"ur Astronomie und Astrophysik, Technische Universit\"at Berlin, Berlin, Germany \\
$^{12}$Astrophysics Research Centre, School of Mathematics and Physics, Queens University Belfast, Belfast BT7 1NN, UK \\
$^{13}$Department of Physics, University of Warwick, Coventry, CV4 7AL, UK \\
$^{14}$Centre for Exoplanets and Habitability, University of Warwick, Gibbet Hill Road, Coventry, CV4 7AL, UK \\
}

\date{Accepted XXX. Received YYY; in original form ZZZ}

\pubyear{2019}

\begin{document}
\label{firstpage}
\pagerange{\pageref{firstpage}--\pageref{lastpage}}
\maketitle

\begin{abstract}
We present new observations of the transmission spectrum of the hot Jupiter WASP-6b both from the ground with the \textit{Very Large Telescope} (\textit{VLT}) FOcal Reducer and Spectrograph (FORS2) from 0.45-0.83 $\mu$m, and space with the \textit{Transiting Exoplanet Survey Satellite} (\textit{TESS}) from 0.6-1.0 $\mu$m and the \textit{Hubble Space Telescope} (\textit{HST}) Wide Field Camera 3 from 1.12-1.65 $\mu$m. Archival data from the \textit{HST} Space Telescope Imaging Spectrograph (STIS) and \textit{Spitzer} is also reanalysed on a common Gaussian process framework, of which the STIS data show a good overall agreement with the overlapping FORS2 data. We also explore the effects of stellar heterogeneity on our observations and its resulting implications towards determining the atmospheric characteristics of WASP-6b. Independent of our assumptions for the level of stellar heterogeneity we detect Na {\sc i}, K {\sc i} and H$_2$O absorption features and constrain the elemental oxygen abundance to a value of [O/H] $\simeq -0.9\pm0.3$ relative to solar. In contrast, we find that the stellar heterogeneity correction can have significant effects on the retrieved distributions of the [Na/H] and [K/H] abundances, primarily through its degeneracy with the sloping optical opacity of scattering haze species within the atmosphere. Our results also show that despite this presence of haze, WASP-6b remains a favourable object for future atmospheric characterisation with upcoming missions such as the \textit{James Webb Space Telescope}.
\end{abstract}




\section{Introduction}\label{intro}

Transiting exoplanets currently present one of the best options towards studying the atmospheres of planets outside of the Solar System through observations of wavelength-dependent variations in their apparent radii as they occult their host star. These variations are intrinsically linked to the composition and structure of an exoplanetary atmosphere, as the starlight transmitted through the planetary limb is strongly modulated by the wavelength dependent opacities of its constituent molecular species \citep{Seag00}. Tracing these variations as a function of wavelength, known as transmission spectroscopy, has already been successfully applied across a range of both ground- and space-based observatories, unveiling a host of atomic and molecular species in the atmospheres of exoplanets (e.g. \citealt{Char02, Redf08, Snel08, Sing11, Demi13, Spak18, Evan18}) as well as providing strong insights into their bulk atmospheric properties (e.g. \citealt{Madh11, Evan17, Wake18}). In particular, \citet{Sing16} show a large diversity in the atmospheres of a sample of ten hot Jupiter exoplanets, revealing a continuum in the obscuring effects of haze and clouds on molecular absorption features present in their transmission spectra. Of the ten exoplanets displayed by \citet{Sing16}, WASP-6b and WASP-39b were lacking in near-infrared observations between 1-2 $\mu$m, a region abundant in potential water absorption features. \citet{Wake18} reported such observations for WASP-39b, providing a strong constraint on the water abundance in its atmosphere. In this study we present these observations for WASP-6b, completing the search for water absorption features across this sample of exoplanets. 

Space-based observations, such as those performed with the \textit{Hubble Space Telescope} (\textit{HST}) and \textit{Spitzer}, have thus far proven to be the most prolific method towards the broad spectrophotometric characterisation of exoplanet atmospheres (e.g. \citealt{Char02, Demi13, Sing16}). However, ground-based characterisation through multi-object differential spectrophotometry with the \textit{Very Large Telescope} (\textit{VLT}) FOcal Reducer and Spectrograph (FORS2) \citep{Appe98}, has recently been able to produce \textit{HST}-quality transmission spectra for a variety of exoplanets \citep{Bean11, Niko16, Gibs17, Seda17, Niko18}. As part of a small survey to test the performance of FORS2 and assess the validity of previously observed spectroscopic features with \textit{HST}, the optical spectra of WASP-31b, WASP-39b and WASP-6b have been observed. In the case of WASP-39b and WASP-31b, these results have already been reported in \citet{Niko16} and \citet{Gibs17} respectively. In this study we report the results for WASP-6b, the final target from our ground-based comparative program. 

WASP-6b is an inflated hot Jupiter with a mass of $0.485\, M_{\textrm{Jup}}$, a radius of $1.230\, R_{\textrm{Jup}}$ and an equilibrium temperature of 1184$\,$K \citep{Treg15} discovered by the Wide Angle Search for Planets (\textit{WASP}) ground-based transit survey \citep{Poll06, Gill09}. WASP-6b orbits with a period of $P \simeq 3.36$ d at a separation $a \simeq 0.041$ AU around a mildly metal-poor G8V star \citep{Gill09, Treg15}. \citet{Ibgu10} demonstrate that the planet's inflated radius could be due to tidal-heating brought on by a non-zero eccentricity reported in \citet{Gill09}. Whilst further radial velocity data from \citet{Husn12} demonstrated that this eccentricity is not significantly non-zero, as initially inferred, it does not necessitate a circular orbit and as such the true cause of the inflation has yet to be definitively determined. \citet{Doyl13} refine the bulk properties of the host star WASP-6 through spectroscopy, providing measurements of T$_\textrm{eff}$ = 5375 $\pm$ 65, log($g$) = 4.61 $\pm$ 0.07 and [Fe/H] = -0.15 $\pm$ 0.09. Finally, \citet{Treg15} demonstrated that fluctuations in multiple transit light curves of archival photometry of WASP-6b could be attributed to a single star spot anomaly. This enabled a more precise measurement on the sky projected spin-orbit alignment of $\lambda = 7.2^\circ \pm 3.7^\circ$ in agreement with \citet{Gill09}.

The atmosphere of WASP-6b was initially probed spectrophotmetrically in the optical with the ground-based IMACS instrument on the 6.5-m \textit{Magellan Telescope} by \citet{Jord13} who observed a decrease in transit depth as a function of wavelength, characteristic of a scattering haze, and no evidence of the Na {\sc i} and K {\sc i} absorption lines. Subsequent observations performed in the optical with \textit{HST}'s Space Telescope Imaging Spectrograph (STIS) and \textit{Spitzer}'s InfraRed Array Camera (IRAC) \citep{Niko15} also demonstrated evidence of a scattering haze, however the Na {\sc i} and K {\sc i} lines were resolved in this case with significance levels of 1.2$\sigma$ and 2.7$\sigma$ respectively. WASP-6b's atmosphere has also been observed at secondary eclipse as the planet passes behind its host star from our point of view with \textit{Spitzer} IRAC, providing day side temperature estimates of 1235$\substack{ +70 \\ -77 }\,$K and 1118$\substack{ +68 \\ -74 }\,$K for the 3.6 and 4.5 $\mu$m channels respectively \citep{Kamm15}.

We present new spectrophotometric observations from 1.1 to 1.7 $\mu$m using the \textit{HST} Wide Field Camera 3 (WFC3) instrument with the G141 grism for the exoplanet WASP-6b, the final object in the \citet{Sing16} study without observations in this wavelength range. Additionally, we present new spectrophotometric observations from 0.4 to 0.8 $\mu$m performed from the ground using \textit{VLT} FORS2. Recent photometric observations of WASP-6b performed from space with the Transiting Exoplanet Survey Satellite (\textit{TESS}) \citep{Rick14} are also included in our study. These datasets were analysed in tandem with a reanalysis of the archival STIS and \textit{Spitzer} datasets on a common Gaussian Process (GP) framework \citep{Gibs12}. We also perform light-curve corrections to account for the effects of stellar heterogeneity on the perceived transmission spectrum of WASP-6b, the presence of which can act to mimic the signatures of scattering hazes \citep{Mccu14, Rack18, Pinh18, Alam18, Rack19}.

Descriptions of our observations and the necessary data reduction are shown in Section \ref{obs}. All light curve fitting and analysis is presented in Section \ref{lightcurves}. An accounting of the effects of stellar heterogeneity is shown in Section \ref{stelact}. The resultant transmission spectra and the conclusions drawn from them using both forward and retrieval based models are described in Section \ref{disc}. Finally, we summarise our results in Section \ref{conc}. 

\section{Observations and Data Reduction}\label{obs}
\begin{figure*}
\centering
\includegraphics[width=\textwidth]{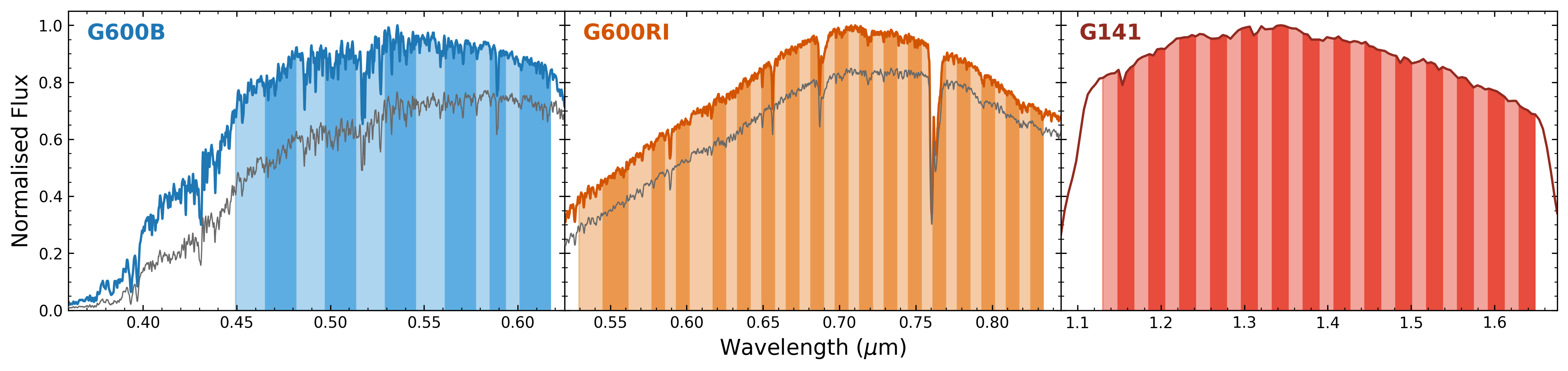}
\vspace*{-5mm}
\caption{Representative observed spectra for the FORS2 G600B, FORS2 G600RI and WFC3 G141 grisms, the thicker coloured lines indicate spectra of WASP-6 whilst thinner grey lines correspond to that of the reference star, both target and reference spectra are normalised to the maximum of the target spectrum for that observation. Shaded bands indicate the selected wavelength binning for each grism. }
\label{grismthroughput}
\end{figure*}

\subsection{\textit{VLT} FORS2}\label{vltobs}
We obtained observations of two primary transits of WASP-6b using the \textit{VLT} FORS2 GRIS600B (G600B) and GRIS600RI (G600RI) grisms in multi-object spectroscopy mode on 2015 October 3 and 2015 November 9 respectively as part of program 096.C-0765 (PI: Nikolov). These observations utilise a mask with broad slits centred on WASP-6 and a nearby reference star (2MASS J23124095-2243232), all slits had a width of 25", the slit lengths used in the G600B and G600RI observations were 31" and 90" respectively. On the night of the G600B observations conditions began clear (less than 10 per cent of the sky covered in clouds, transparency variations under 10 per cent) and moved to photometric (no clouds, transparency variations under 2 per cent) approximately half way through the observations. The exposure time was set at 100 seconds per exposure for a total of 152 exposures. During this night observations were halted for $\sim$30 minutes during transit ingress as the target passed through the zenith and was outside the observable region of the telescope. On the night of the G600RI observations, conditions began clear but moved to photometric for the bulk of the observation and the exposure time was set to 60 seconds per exposure for a total of 184 exposures. Towards the end of the transit an earthquake caused a guide star loss and as such observations were halted for $\sim$15 minutes. 

We begin the data reduction by performing bias- and flat-field corrections on the raw data frames, followed by cosmic ray correction using two iterations of the \texttt{L.A.Cosmic} algorithm \citep{Dokk01}. Background flux subtraction for each spectrum was conducted using the median of a box of pixels outside of each spectral trace. Spectra were then extracted using the \texttt{APALL} procedure within the \texttt{IRAF} package \citep{Tody93}. Aperture widths for the spectral extraction were varied and values of 14 and 15 pixels were selected as they minimised the dispersion in the out-of-transit flux for the G600B and G600RI white light curves respectively. We produce a wavelength solution for both observations using the spectra of an emission lamp taken with the calibration mask following each observation. In particular, a low-order Chebyshev polynomial was fit to a multitude of emission lines, the centres of which were determined through individual Gaussian fits. This wavelength solution was then applied to a single data frame to produce a reference spectrum for each observation. Finally, each extracted spectrum was then cross-correlated against its respective reference in order to account for sub-pixel shifts in the dispersion direction, the maximum resultant shifts were $\sim$1.2 pixels and $\sim$0.3 pixels for the G600B and G600RI datasets respectively. Representative spectra of both WASP-6 and the reference star are shown in Figure \ref{grismthroughput} for both the G600B and G600RI observations. 

\subsection{\textit{HST} WFC3}\label{hstobs}
A primary transit of WASP-6b was also observed using the \textit{HST} WFC3 G141 grism on 2017 May 6 as part of General Observer (GO) program 14767 (PI: Sing and L\'opez-Morales). All exposures were taken in sequence across five \textit{HST} orbits, with 13 exposures per orbit, except for the first orbit which only consisted of 10 exposures. Each exposure was performed in forward spatial scanning mode \citep{mccu12}, where the telescope slews in the cross dispersion axis during the exposure, allowing for longer exposure times whilst avoiding saturation on the detector. For the first orbit the exposure times were set to $\sim$184 seconds, whilst the remaining orbits had exposure times of $\sim$138 seconds. All exposures employed the SPARS25 readout mode and used a scan rate of $\sim$0.46 pixels per second.

Reduction of the spectra began with the \texttt{.ima} files output from the \texttt{CALWF3} pipeline. Each \texttt{.ima} file contains multiple reads for each individual spatial scan, up to the final full scan image. We do not however perform spectral extraction on the final frame of each scan but rather the sum of differenced frames, following \citet{Demi13}. This has the advantage of reducing the impact of cosmic rays and hot pixels, whilst also reducing the overall sky background. For each differenced read, pixels beyond a mask of 35 pixels above and below the centre of the spectral trace were zeroed before extraction of the differenced frame following \citep{Evan16}. Finally, we then sum all of the differenced frames for each spatial scan to produce a final differenced frame scan. 

To perform cosmic ray correction these frames were stacked into a single cube so that the variation of each pixel could be tracked as a function of time. Each pixel was smoothed temporally with a Gaussian filter and pixel deviations between this and the initial datacube larger than 8$\sigma$ were flagged as cosmic rays. Static bad pixels were also flagged by searching for deviations greater than 10$\sigma$ between each individual unsmoothed pixel and the median of a span of 5 pixels in the cross-dispersion direction, centred on the initial pixel. These cosmic rays and static pixels were then replaced by a linear interpolation of the pixel to the PSF of the same median span. Using a second mask of 50 pixels above and below the centre of the final scans, the 2D spectra were summed along the cross-dispersion axis to produce a 1D spectrum for each scan. This mask width was selected as it provided the minimal white light curve out-of-transit scatter across a range of 30 to 80 pixels in steps of 5 pixels. The background was subtracted from each spectrum using the median of a box of pixels in a region of the detector unpolluted by the diffuse light from the edges of the spatial scan. 

Wavelength solutions were obtained by cross-correlating each individual spectrum with an \texttt{ATLAS}\footnote{\url{http://kurucz.harvard.edu/}} \citep{Kuru93} stellar spectrum, with parameters similar to WASP-6 (T$_\textrm{eff}$=5500K, log($g$)=4.5, [M/H]=-0.2), convolved with the throughput of the G141 grism. Before cross-correlation, both spectra were smoothed with a Gaussian filter to inhibit the effects of spectral lines and focus the correlation on the steep edges of the G141 throughput. This process revealed shifts in the dispersion direction across the course of observation within $\sim$0.12 pixels. An example 1D spectrum from the G141 observations is shown in Figure \ref{grismthroughput}.

\subsection{\textit{TESS}}\label{tessobs}
The \textit{Transiting Exoplanet Survey-Satellite} (\textit{TESS}) is currently performing an all sky search for transiting exoplanets in a single broadband filter from 0.6 to 1.0 $\mu$m \citep{Rick14}. Due to the broad 24$^\circ$ $\times$ 96$^\circ$ field of view, \textit{TESS} holds enormous potential not only for discovering new exoplanets, but also observing transits of already known transiting systems. With the public release of the \textit{TESS} Sector 2 data, 7 clear transits of WASP-6b can be readily identified from 2018 Aug 23 to 2018 Sep 19.

To obtain the \textit{TESS} light curve spanning this time period we initially used the pre-calibrated and extracted light curve held in the \texttt{lc.fits} file. However, on closer inspection we found indications of a non-optimal pipeline correction and as such choose to perform our own correction on the uncorrected light curve in the same file. We follow a Pixel Level Decorrelation (PLD) systematics removal method on the raw data as implemented by the \texttt{lightkurve} python package \citep{lkurv}. PLD has already been used successfully as a systematics correction technique on both \textit{Spitzer} \citep{Demi15} and K2 data \citep{Luge16, Luge18} and we refer the reader to these references for further information on the PLD technique itself. Finally, to prepare for the transit light curve analysis, we extract seven separate portions from the complete light curve, each centred on one of the observed transits. Each individual extracted light curve spans from roughly 5 hours pre-transit to 5 hours post transit, in order to facilitate an effective out-of-transit baseline determination. 

\subsection{Archival Data}
In order to fully exploit the data that are available to us we opt to perform a reanalysis of the previously reported \textit{HST} STIS and \textit{Spitzer} IRAC data \citep{Niko15}. Specifically, there were two spectroscopic transit observations with the STIS G430L grism from 0.33-0.57 $\mu$m, one spectroscopic transit using the STIS G750L grism from 0.55-1.03 $\mu$m, and one photometric transit for each of the \textit{Spitzer} IRAC 3.6 and 4.5 $\mu$m bandpasses. Performing such a reanalysis can account for transit depth baseline offsets between these datasets and those in this study by fitting all light curves under a common set of prior system parameters. Furthermore, the implementation of a stellar heterogeneity correction, and its changes to the system parameters (Section \ref{stelact}) necessitates further light curve fitting. A complete reanalysis ensures that any comparisons between the spot corrected and uncorrected datasets are not influenced by the differing light curve fitting methodologies of this study and that of \citet{Niko15}. 

With respect to the data reduction of the observations themselves, all light curves were extracted following the same methodology outlined in \citet{Niko15}. For the STIS data this involves spectral extraction following the \texttt{APALL} procedure in \texttt{IRAF} \citep{Tody93}, and photometry is performed for the \textit{Spitzer} data through time-variable aperture extraction. For the \textit{Spitzer} IRAC light curves there are thousands of independent photometric measurements throughout each observation and to reduce the computational intensity of the light curve fitting procedure described in Section \ref{lightcurves} we bin each light curve into 1000 bins, corresponding to a cadence of $\sim$15 and $\sim$16 seconds for the 3.6 and 4.5 $\mu$m bands respectively.

\section{Light Curve Analysis}\label{lightcurves}
White light curves for the G600B, G600RI and G141 datasets were produced by summing the flux for each individual spectrum along the dispersion axis from 0.449 to 0.617 $\mu$m, 0.529 to 0.833 $\mu$m, and from 1.0 to 1.8 $\mu$m respectively. Spectrophotometric light curves were produced for the G600B, G600RI, and G141 datasets by summing the flux within 12, 34, and 28 respective bins across the wavelength ranges displayed in Figure \ref{grismthroughput}. 

Below $\sim$ 0.45 $\mu$m the G600B flux levels are the lowest of both of the FORS2 datasets and inherently contain a limited amount of information due to the higher photon error. Whilst using a larger bin size could alleviate this, the contribution of differential extinction due to a spectral type mismatch between the target and reference must also be considered. In the case of our observations such a mismatch is evident in the different spectral profiles of the target and reference star in Figure \ref{grismthroughput}. The flux of the reference star from 0.40 to 0.45 $\mu$m is ~50\% that of the target, whereas at 0.6 $\mu$m this value is ~80\%. Therefore, the data below 0.45$\mu$m not only contain the lowest flux levels of our FORS2 observations, but their accuracy is impacted the most by the differential extinction. Furthermore, including such a wavelength range would also impart further differential extinction effects on every other spectrophotometric bin in the G600B dataset due to the nature of the common-mode, white-light correction performed during the light curve fitting. In an effort to mitigate the impact of differential extinction on our final transmission spectra we therefore choose to exclude the G600B data below $\sim$0.45 $\mu$m.

In the case of the G600B and G600RI observations, all light curves were also produced for the reference star. Before fitting any of the G600B or G600RI light curves, we first correct for dominant atmospheric effects by dividing the raw flux of the target by that of the corresponding wavelength range reference. The spectrophotometric bins for all observations are displayed in Figure \ref{grismthroughput}. As the \textit{TESS} observations are photometric they hold no spectral information and were treated as white light curves in terms of fitting. Finally, we obtain the archival STIS and \textit{Spitzer} light curves across identical wavelength ranges as described in \citet{Niko15}.

During both the G600B and G600RI observations, the target needed to be reacquired and as such all light curves suffer from incomplete phase coverage, this also results in the separate pieces of each light curve exhibiting differing systematics effects. Throughout our analyses we were unable to accurately and effectively account for these systematic offsets due to the significant, or complete, absence of in transit observations for one piece of each light curve. As such, in the analysis presented here we exclude the pre-ingress data for the G600B observation and the post-egress data for the G600RI observation. The first orbit, and first spectrum of all other orbits, of the G141 observation exhibit much stronger systematics than the other obtained spectra due to charge trapping in the detector \citep{Zhou17}. We therefore opt to remove these data from our analysis in line with many other studies (e.g. \citealt{Knut14, Sing16, Wake18, Evan19}) that have been performed since the first spatial scanning WFC3 transit observations were made \citep{Demi13}. 

\begin{table}
\centering
\begin{tabular}{c c c c}
\hline
\hline
 & Uncorrected & \textit{TESS} Corrected & \textit{AIT} Corrected \\ 
\hline
$i$ ($^{\circ}$) & 88.78$\substack{ +0.13 \\ -0.13 }$ & 88.73$\substack{ +0.13 \\ -0.12 }$ & 88.72$\substack{ +0.013 \\ -0.012 }$ \\[1ex]
$a/R_*$ & 11.154$\substack{ +0.049 \\ -0.072 }$ & 11.135$\substack{ +0.050 \\ -0.072 }$ & 11.123$\substack{ +0.050 \\ -0.072 }$ \\
\hline
\end{tabular}
\caption{Weighted average values of the orbital inclination and normalised semi-major axis for the uncorrected and spot corrected light curve analyses.}
\label{bestfitpars}
\end{table}

\subsection{White Light Curves}\label{wlcs}

\begin{figure*}
\centering
\includegraphics[width=\textwidth]{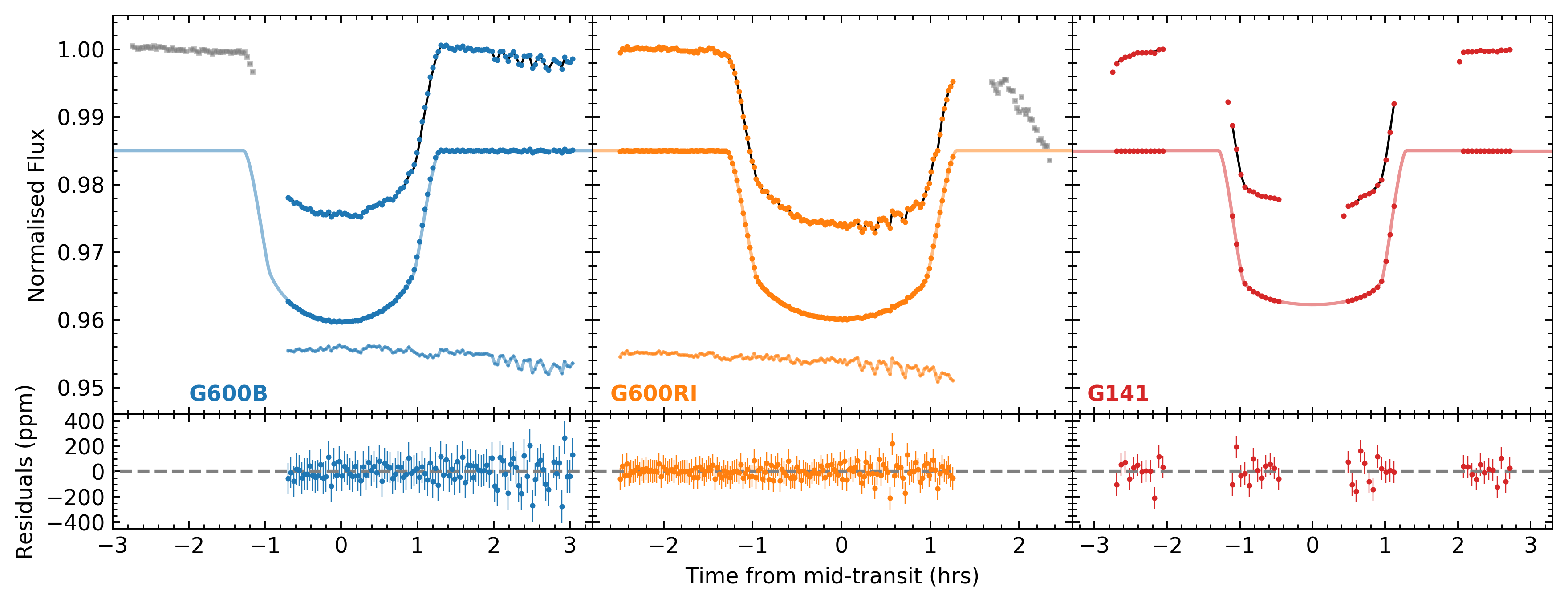}
\vspace*{-5mm}
\caption{Normalised white light curves and residuals of WASP-6b for the G600B, G600RI and G141 grism observations as labelled. \textit{Left:} Data shown from top to bottom are: the raw light curve following reference star correction (grey squares indicating the excluded sections of the light curve) with the black line indicating the GP transit plus systematic model fit, the light curve after removal of the GP systematic component overplotted with the best fitting transit model from \citet{Mand02}, and the computed common-mode correction following division of the raw data by the best fitting transit model. \textit{Centre:} As in the left panel. \textit{Right:} The upper light curve is the raw flux with the black line indicating the GP transit plus systematic model fit, whilst the lower is the light curve after removal of the GP systematic component overplotted with the best fitting transit model from \citet{Mand02}. All lower panels display residuals following subtraction of the corresponding corrected light curves by their respective best fitting models.}
\label{whitelightcurve}
\end{figure*}

\begin{figure*}
\centering
\includegraphics[width=\textwidth]{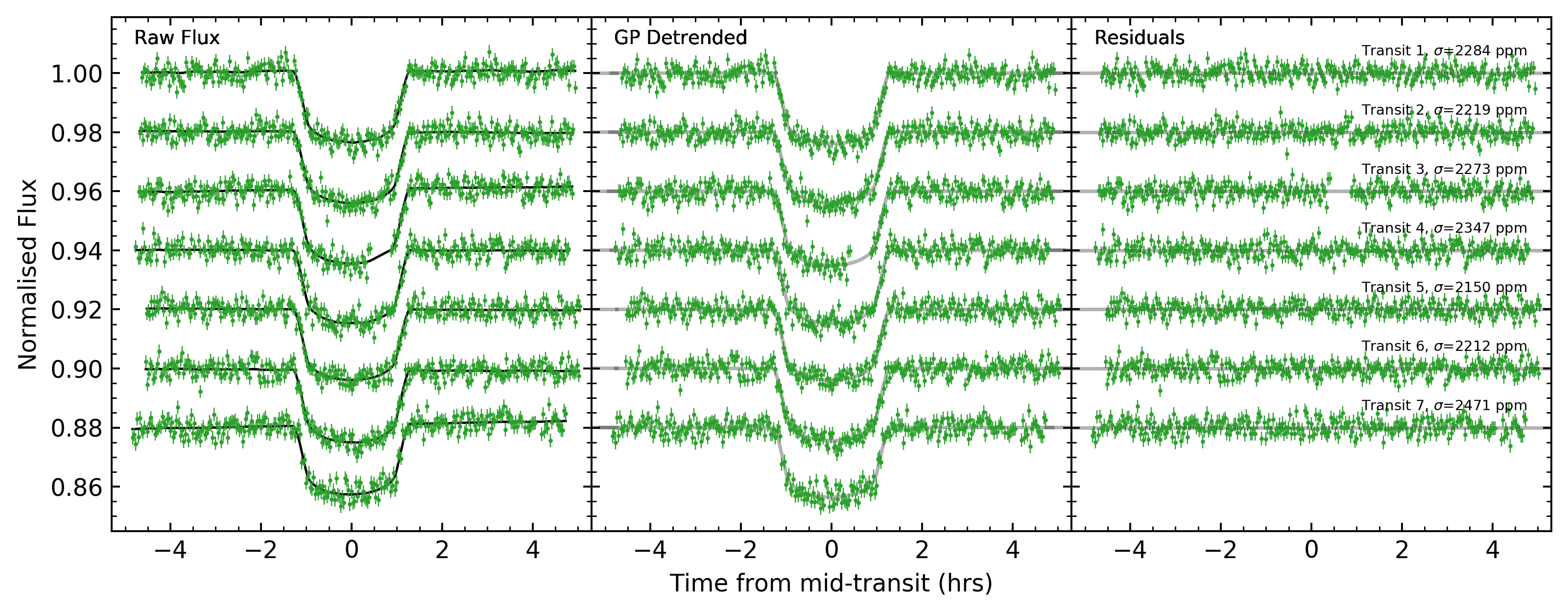}
\vspace*{-5mm}
\caption{Normalised \textit{TESS} photometric light curves multiplied by an arbitrary constant. \textit{Left:} Raw extracted light curves with black lines indicating the GP transit plus systematic model fits. \textit{Centre:} Light curves after removal of GP systematic component. The best fitting transit models from \citet{Mand02} are displayed in grey. \textit{Right:} Residuals after subtraction of best fitting models from the GP systematic corrected light curves. }
\label{tesslightcurves}
\end{figure*}

To perform all lightcurve fitting we follow \citet{Gibs12}, accounting for the transit and instrumental signals simultaneously by treating the data for each light curve as a Gaussian process (GP) using the Python library \texttt{George} \citep{Ambi14}. GP fitting methodologies have been successfully applied to a range of transit observations \citep{Gibs12, Gibs12b, Gibs17, Evan13, Evan15, Evan16, Evan17, Evan18, Evan19, Cart17, Kirk17, Kirk18, Kirk19, Loud17, Seda17} from both the ground and space thus far and enable the measurement of the systematic signal without assuming any prior knowledge on its functional form. We obtain the best fit model to each light curve by marginalising over the constructed GP likelihoods using Markov chain Monte Carlo (MCMC) as implemented by the Python library \texttt{emcee} \citep{Fore13}. When executing each MCMC, we first initialised a group of 150 walkers near the maximum likelihood solution, identified using a Nelder-Mead simplex algorithm as implemented by the \texttt{fmin} function within the \texttt{scipy} library. We run a group for 500 samples and then use the best run to initialise a second group of 150 walkers in a narrow space of this solution. This second group was then run for 3000 samples, with the first 500 samples being discarded as burn-in. 

We list the individual subtleties for each dataset throughout the GP fitting procedure below, however there are some aspects which remained unchanged regardless of the dataset. For the GP covariance amplitudes of all datasets we utilise gamma priors of the form $p(A) \propto e^{-100A}$ as in \citet{Evan18} in order to favour smaller correlation amplitudes and reduce the effects of outliers. Additionally, we follow previous studies and fit for the natural logarithm of the inverse length scale hyperparameters (e.g. \citealt{Evan17, Gibs17, Evan18}), but limit these quantities with a uniform prior ranging between the cadence of the observation and twice the length of the observation. This prescription encourages the GP to fit the broader systematic variations that occur during the transit, with shorter variations described by white noise and longer variations accounted for by the linear baseline trend. Finally, in all cases the orbital period was held fixed to the value of $P$ = 3.36100239 d from \citet{Niko15} and the eccentricity was held fixed to the value of $e$ = 0.041 from \citet{Husn12}.

\subsubsection{G600B \& G600RI}
To describe the mean function of the GP we use the model transit light curves of \citet{Mand02} generated using the \texttt{batman} Python library \citep{Krei15} multiplied by a linear airmass baseline trend. We initially tested a time baseline trend however found that this restricted the final GP fitting of shorter frequency variations within the light curves, by utilising a linear airmass baseline trend the non-linear sloping of the light curves was better matched and the GP had more freedom to fit these shorter frequency variations. Whilst the observed airmass trend can be included in the GP directly as a decorrelation parameter we found this necessitated stricter priors on the length scale hyperparameters and did not measurably improve the fitting. As such, we opt to include this term through the baseline trend. To construct the covariance matrix of the GP we use the Mat\' ern $\nu$ = 3/2 kernel, with time as the decorrelation parameter. Other decorrelation parameters were also tested both individually and in combination such as: spectral dispersion drift, cross-dispersion drift, full-width half maximum, ambient temperature, ambient pressure, and telescope rotation angle. Despite this, no clear correlations were observed and therefore we excluded these parameters from the final analysis.

Unlike the other datasets, for the FORS2 analysis we account for limb-darkening following the two-parameter quadratic law. The treatment is different as these observations were performed from the ground where the Earth's atmosphere acts as an filter for the incoming light. Crucially, the response of the atmosphere is a function of wavelength and varies as a function of the zenith distance, which varies throughout the observations. Instead of making explicit assumptions about this atmospheric transmission and including it directly in our determination of the precomputed limb darkening coefficients we choose to fit for the coefficients themselves. We select the quadratic limb darkening law in order to improve computational efficiency by reducing the number of fit parameters whilst still providing an accurate description for the true limb darkening of WASP-6 given its temperature \citep{Espi16}.

For the G600RI observation we allow the transit depth $R_\textrm{p} / R_*$, inclination $i$, normalised semi-major axis $a/R_*$, transit central time $T_0$, linear trend parameters and quadratic limb darkening parameters $c_1$ and $c_2$ to vary throughout the fit. However, in the case of the G600B observation, we found that the paucity of transit coverage provided imprecise determinations of $i$ and $a/R_*$ and as such perform a simpler fit after retrieving the weighted average best fit parameters, see Section \ref{bfm_para}. 

The presence of high frequency variations from $\sim$2-3 hours and $\sim$0-1 hours after mid-transit for the G600B and G600RI light curves, respectively, strongly constrain the hyperparameters of the GP fit which leads to over fitting of other variations within the light curve. In order to assess the impact on the fit transit parameters we restricted the priors on these hyperparameters such that the high frequency variations could no longer bias the GP fitting. Whilst this significantly reduced the perceived overfitting, we find that all fit transit parameters are unaffected by this change and lie within 1$\sigma$ of the original fit. Therefore, and in addition to the lack of prior knowledge on these hyperparameters, we opt not to perform such a restriction for any of the final white light curve fits.

\subsubsection{STIS \& G141}
The mean function of the GP is described identically to the G600B and G600RI mean functions, except using a linear time baseline trend. To construct the covariance matrix of the GP we use the Mat\' ern $\nu$ = 3/2 kernel, with \textit{HST} orbital phase, dispersion shift and cross dispersion shift identified as the optimal decorrelation parameters. Limb darkening was accounted for through the four-parameter non-linear law. During the fitting we allow the transit depth $R_\textrm{p} / R_*$, inclination $i$, normalised semi-major axis $a/R_*$, transit central time $T_0$ and linear trend parameters to vary throughout the fit and we fixed all four non-linear limb darkening values to values calculated from the ATLAS model described in Section \ref{obs}, following \citet{Sing10}. Finally, as there are two independent light curves in the STIS G430L observations we performed a joint fit between them, only allowing the transit central time for each light curve to vary independently.

\subsubsection{Spitzer}
The mean function of the GP is described identically to the G600B and G600RI mean functions, except using a linear time baseline trend. We construct the covariance matrix following \citet{Evan15}. Specifically, we construct a kernel $k = k_{xy} + k_t$ where $k_{xy}$ is a squared exponential kernel, with the photometric centroid $x$ and $y$ coordinates as the decorrelation parameters, and $k_t$ is a Mat\' ern $\nu$ = 3/2 kernel, with time as the decorrelation parameter. Constructing such a kernel allows us to account for the smooth variations in pixel sensitivities as well as residual correlated noise in the light curve. Limb darkening was accounted for through the four-parameter non-linear law. During the fitting we allow the transit depth $R_\textrm{p} / R_*$, inclination $i$, normalised semi-major axis $a/R_*$, transit central time $T_0$ and linear trend parameters to vary throughout the fit and we fixed all four non-linear limb darkening values similarly to the STIS and G141 observations.

\subsubsection{TESS}
The mean function of the GP is described identically to the G600B and G600RI mean functions, except using a linear time baseline trend. To construct the covariance matrix of the GP we use the Mat\' ern $\nu$ = 3/2 kernel, with time as the decorrelation parameter. Limb darkening was accounted for through the four-parameter non-linear law. During the fitting we allow the transit depth $R_\textrm{p} / R_*$, inclination $i$, normalised semi-major axis $a/R_*$, transit central time $T_0$ and linear trend parameters to vary throughout the fit and we fixed all four non-linear limb darkening values similarly to the STIS and G141 observations.

\subsubsection{Best Fit Models}\label{bfm_para}
In order to obtain the best fit model to each dataset we determine the weighted average values of the orbital inclination and the normalised semi-major axis (Table \ref{bestfitpars}). Using these values we performed the fit to the G600B dataset, where we allowed the transit depth $R_\textrm{p} / R_*$, transit central time $T_0$, linear trend parameters, and the quadratic limb darkening parameters $u_1$ and $u_2$ to vary. In addition, we repeat the fit for each other light curve, with the orbital inclination and normalised semi-major axis fixed to the weighted average values and the transit central time to that of its respective original fit. The G600B, G600RI and G141 light curves, alongside the systematics corrected light curves are displayed in Figure \ref{whitelightcurve}, all \textit{TESS} light curves are displayed in Figure \ref{tesslightcurves}, all STIS light curves are displayed in Figure \ref{stis_wlcs}, all \textit{Spitzer} light curves are displayed in Figure \ref{spitzer_wlcs}, and all relevant MCMC results are displayed in Table \ref{wlcparams}.

\begin{table*} 
\centering 
\fontsize{8}{10}\selectfont 
\begin{tabular}{l c c c c c c c c} 
\hline 
\hline 
\vspace{-8pt} \\ 
Instrument & $R_\textrm{p}/R_*$ & $T_0$ (MJD) & $i$ ($^\circ$) & $a/R_*$ & $c_1$ & $c_2$ & $c_3$ & $c_4$  \\ 
\hline 
\vspace{-9pt} \\ 
\textbf{Uncorrected} & & & & & & & & \\ 
G600B & $0.14425\substack{+0.00161 \\ -0.00176}$ & $57298.172234\substack{+0.000549 \\ -0.000516}$ & - & - & $0.510\substack{+0.117 \\ -0.122}$ & $0.126\substack{+0.272 \\ -0.252}$ & - & - \\[1ex] 
G600RI & $0.14602\substack{+0.00057 \\ -0.00058}$ & $57335.146823\substack{+0.000173 \\ -0.000172}$ & $88.67\substack{+0.53 \\ -0.38}$ & $10.98\substack{+0.19 \\ -0.20}$ & $0.469\substack{+0.054 \\ -0.054}$ & $0.004\substack{+0.088 \\ -0.086}$ & - & - \\[1ex] 
STIS430 V1 & $0.14618\substack{+0.00065 \\ -0.00061}$ & $56088.216349\substack{+0.000183 \\ -0.000155}$ & $89.38\substack{+0.41 \\ -0.54}$ & $11.33\substack{+0.10 \\ -0.21}$ & 0.4593 & -0.0641 & 0.8327 & -0.3729 \\[1ex] 
STIS430 V2 & $0.14618\substack{+0.00065 \\ -0.00061}$ & $56094.937530\substack{+0.000263 \\ -0.000250}$ & $89.38\substack{+0.41 \\ -0.54}$ & $11.33\substack{+0.10 \\ -0.21}$ & 0.4593 & -0.0641 & 0.8327 & -0.3729 \\[1ex] 
STIS750 & $0.14505\substack{+0.00058 \\ -0.00060}$ & $56131.906130\substack{+0.000306 \\ -0.000292}$ & $89.17\substack{+0.54 \\ -0.74}$ & $11.33\substack{+0.17 \\ -0.38}$ & 0.6068 & -0.1441 & 0.4857 & -0.2312 \\[1ex] 
\textit{TESS}$_1$ & $0.14315\substack{+0.00127 \\ -0.00126}$ & $58357.394027\substack{+0.000429 \\ -0.000446}$ & $88.35\substack{+1.07 \\ -0.91}$ & $10.91\substack{+0.45 \\ -0.67}$ & 0.6990 & -0.4538 & 0.9531 & -0.4668 \\[1ex] 
\textit{TESS}$_2$ & $0.14336\substack{+0.00115 \\ -0.00116}$ & $58360.755834\substack{+0.000405 \\ -0.000381}$ & $88.68\substack{+0.83 \\ -0.88}$ & $10.99\substack{+0.31 \\ -0.56}$ & 0.6990 & -0.4538 & 0.9531 & -0.4668 \\[1ex] 
\textit{TESS}$_3$ & $0.14653\substack{+0.00124 \\ -0.00124}$ & $58364.116237\substack{+0.000384 \\ -0.000391}$ & $88.62\substack{+0.90 \\ -0.91}$ & $10.91\substack{+0.34 \\ -0.59}$ & 0.6990 & -0.4538 & 0.9531 & -0.4668 \\[1ex] 
\textit{TESS}$_4$ & $0.14538\substack{+0.00112 \\ -0.00114}$ & $58370.838845\substack{+0.000401 \\ -0.000390}$ & $88.37\substack{+0.99 \\ -0.83}$ & $10.93\substack{+0.45 \\ -0.59}$ & 0.6990 & -0.4538 & 0.9531 & -0.4668 \\[1ex] 
\textit{TESS}$_5$ & $0.14303\substack{+0.00106 \\ -0.00108}$ & $58374.199389\substack{+0.000365 \\ -0.000353}$ & $88.66\substack{+0.86 \\ -0.87}$ & $11.03\substack{+0.32 \\ -0.57}$ & 0.6990 & -0.4538 & 0.9531 & -0.4668 \\[1ex] 
\textit{TESS}$_6$ & $0.14467\substack{+0.00112 \\ -0.00112}$ & $58377.559940\substack{+0.000384 \\ -0.000389}$ & $88.45\substack{+0.91 \\ -0.85}$ & $10.91\substack{+0.40 \\ -0.59}$ & 0.6990 & -0.4538 & 0.9531 & -0.4668 \\[1ex] 
\textit{TESS}$_7$ & $0.14193\substack{+0.00124 \\ -0.00122}$ & $58380.922042\substack{+0.000420 \\ -0.000426}$ & $89.24\substack{+0.50 \\ -0.77}$ & $11.34\substack{+0.18 \\ -0.36}$ & 0.6990 & -0.4538 & 0.9531 & -0.4668 \\[1ex] 
G141 & $0.14374\substack{+0.00048 \\ -0.00041}$ & $57880.135381\substack{+0.000057 \\ -0.000061}$ & $88.64\substack{+0.20 \\ -0.17}$ & $11.09\substack{+0.12 \\ -0.10}$ & 0.5692 & 0.1519 & -0.2305 & 0.0672 \\[1ex] 
\textit{Spitzer} CH1 & $0.14124\substack{+0.00142 \\ -0.00132}$ & $56313.405391\substack{+0.000244 \\ -0.000246}$ & $89.37\substack{+0.44 \\ -0.59}$ & $11.26\substack{+0.11 \\ -0.22}$ & 0.4839 & -0.3558 & 0.3447 & -0.1402 \\[1ex] 
\textit{Spitzer} CH2 & $0.14148\substack{+0.00191 \\ -0.00187}$ & $56306.683284\substack{+0.000335 \\ -0.000358}$ & $88.44\substack{+0.85 \\ -0.65}$ & $10.95\substack{+0.41 \\ -0.48}$ & 0.5652 & -0.7296 & 0.7488 & -0.2845 \\[1ex] 
\textbf{Corrected \textit{TESS}} & & & & & & & & \\ 
G600B & $0.14330\substack{+0.00156 \\ -0.00177}$ & $57298.172253\substack{+0.000552 \\ -0.000527}$ & - & - & $0.513\substack{+0.118 \\ -0.123}$ & $0.123\substack{+0.270 \\ -0.254}$ & - & - \\[0.90ex] 
G600RI & $0.14513\substack{+0.00055 \\ -0.00057}$ & $57335.146826\substack{+0.000178 \\ -0.000178}$ & $88.57\substack{+0.53 \\ -0.37}$ & $10.93\substack{+0.22 \\ -0.20}$ & $0.474\substack{+0.053 \\ -0.054}$ & $-0.006\substack{+0.087 \\ -0.086}$ & - & - \\[0.90ex] 
STIS430 V1 & $0.14518\substack{+0.00062 \\ -0.00063}$ & $56088.216348\substack{+0.000171 \\ -0.000168}$ & $89.35\substack{+0.40 \\ -0.51}$ & $11.32\substack{+0.09 \\ -0.20}$ & 0.4593 & -0.0641 & 0.8327 & -0.3729 \\[0.90ex] 
STIS430 V2 & $0.14518\substack{+0.00062 \\ -0.00063}$ & $56094.937581\substack{+0.000267 \\ -0.000254}$ & $89.35\substack{+0.40 \\ -0.51}$ & $11.32\substack{+0.09 \\ -0.20}$ & 0.4593 & -0.0641 & 0.8327 & -0.3729 \\[0.90ex] 
STIS750 & $0.14424\substack{+0.00056 \\ -0.00059}$ & $56131.906148\substack{+0.000305 \\ -0.000289}$ & $89.19\substack{+0.55 \\ -0.69}$ & $11.33\substack{+0.16 \\ -0.35}$ & 0.6068 & -0.1441 & 0.4857 & -0.2312 \\[0.90ex] 
\textit{TESS}$_1$ & $0.14211\substack{+0.00121 \\ -0.00119}$ & $58357.393949\substack{+0.000410 \\ -0.000412}$ & $88.13\substack{+1.09 \\ -0.80}$ & $10.77\substack{+0.54 \\ -0.61}$ & 0.6990 & -0.4538 & 0.9531 & -0.4668 \\[0.90ex] 
\textit{TESS}$_2$ & $0.14214\substack{+0.00110 \\ -0.00111}$ & $58360.755686\substack{+0.000364 \\ -0.000375}$ & $88.63\substack{+0.90 \\ -0.86}$ & $10.98\substack{+0.34 \\ -0.56}$ & 0.6990 & -0.4538 & 0.9531 & -0.4668 \\[0.90ex] 
\textit{TESS}$_3$ & $0.14564\substack{+0.00119 \\ -0.00118}$ & $58364.116453\substack{+0.000370 \\ -0.000363}$ & $88.78\substack{+0.79 \\ -0.91}$ & $11.06\substack{+0.29 \\ -0.56}$ & 0.6990 & -0.4538 & 0.9531 & -0.4668 \\[0.90ex] 
\textit{TESS}$_4$ & $0.14399\substack{+0.00108 \\ -0.00108}$ & $58370.839042\substack{+0.000375 \\ -0.000386}$ & $88.19\substack{+0.93 \\ -0.73}$ & $10.79\substack{+0.50 \\ -0.54}$ & 0.6990 & -0.4538 & 0.9531 & -0.4668 \\[0.90ex] 
\textit{TESS}$_5$ & $0.14249\substack{+0.00107 \\ -0.00106}$ & $58374.199361\substack{+0.000358 \\ -0.000368}$ & $88.47\substack{+0.94 \\ -0.83}$ & $10.94\substack{+0.40 \\ -0.58}$ & 0.6990 & -0.4538 & 0.9531 & -0.4668 \\[0.90ex] 
\textit{TESS}$_6$ & $0.14371\substack{+0.00110 \\ -0.00108}$ & $58377.560027\substack{+0.000382 \\ -0.000383}$ & $88.61\substack{+0.91 \\ -0.84}$ & $10.98\substack{+0.34 \\ -0.55}$ & 0.6990 & -0.4538 & 0.9531 & -0.4668 \\[0.90ex] 
\textit{TESS}$_7$ & $0.14139\substack{+0.00123 \\ -0.00122}$ & $58380.922106\substack{+0.000422 \\ -0.000431}$ & $89.27\substack{+0.49 \\ -0.73}$ & $11.28\substack{+0.17 \\ -0.32}$ & 0.6990 & -0.4538 & 0.9531 & -0.4668 \\[0.90ex] 
G141 & $0.14305\substack{+0.00046 \\ -0.00041}$ & $57880.135372\substack{+0.000058 \\ -0.000061}$ & $88.59\substack{+0.20 \\ -0.17}$ & $11.06\substack{+0.12 \\ -0.11}$ & 0.5692 & 0.1519 & -0.2305 & 0.0672 \\[0.90ex] 
\textit{Spitzer} CH1 & $0.14078\substack{+0.00138 \\ -0.00132}$ & $56313.405403\substack{+0.000243 \\ -0.000244}$ & $89.31\substack{+0.47 \\ -0.63}$ & $11.25\substack{+0.12 \\ -0.27}$ & 0.4839 & -0.3558 & 0.3447 & -0.1402 \\[0.90ex] 
\textit{Spitzer} CH2 & $0.14114\substack{+0.00194 \\ -0.00186}$ & $56306.683297\substack{+0.000339 \\ -0.000359}$ & $88.41\substack{+0.82 \\ -0.62}$ & $10.93\substack{+0.41 \\ -0.47}$ & 0.5652 & -0.7296 & 0.7488 & -0.2845 \\[0.90ex] 
\textbf{Corrected \textit{AIT}} & & & & & & & & \\ 
G600B & $0.14280\substack{+0.00161 \\ -0.00175}$ & $57298.172246\substack{+0.000556 \\ -0.000532}$ & - & - & $0.512\substack{+0.118 \\ -0.125}$ & $0.125\substack{+0.275 \\ -0.256}$ & - & - \\[0.90ex] 
G600RI & $0.14464\substack{+0.00055 \\ -0.00055}$ & $57335.146821\substack{+0.000176 \\ -0.000174}$ & $88.54\substack{+0.47 \\ -0.35}$ & $10.91\substack{+0.19 \\ -0.20}$ & $0.472\substack{+0.054 \\ -0.054}$ & $-0.000\substack{+0.087 \\ -0.085}$ & - & - \\[0.90ex] 
STIS430 V1 & $0.14469\substack{+0.00069 \\ -0.00065}$ & $56088.216337\substack{+0.000175 \\ -0.000161}$ & $89.32\substack{+0.42 \\ -0.58}$ & $11.32\substack{+0.10 \\ -0.24}$ & 0.4593 & -0.0641 & 0.8327 & -0.3729 \\[0.90ex] 
STIS430 V2 & $0.14469\substack{+0.00069 \\ -0.00065}$ & $56094.937540\substack{+0.000279 \\ -0.000251}$ & $89.32\substack{+0.42 \\ -0.58}$ & $11.32\substack{+0.10 \\ -0.24}$ & 0.4593 & -0.0641 & 0.8327 & -0.3729 \\[0.90ex] 
STIS750 & $0.14379\substack{+0.00058 \\ -0.00060}$ & $56131.906103\substack{+0.000310 \\ -0.000298}$ & $89.24\substack{+0.51 \\ -0.71}$ & $11.35\substack{+0.15 \\ -0.34}$ & 0.6068 & -0.1441 & 0.4857 & -0.2312 \\[0.90ex] 
\textit{TESS}$_1$ & $0.14171\substack{+0.00122 \\ -0.00122}$ & $58357.393948\substack{+0.000398 \\ -0.000399}$ & $88.14\substack{+1.05 \\ -0.80}$ & $10.78\substack{+0.52 \\ -0.61}$ & 0.6990 & -0.4538 & 0.9531 & -0.4668 \\[0.90ex] 
\textit{TESS}$_2$ & $0.14170\substack{+0.00112 \\ -0.00110}$ & $58360.755698\substack{+0.000377 \\ -0.000372}$ & $88.61\substack{+0.85 \\ -0.85}$ & $10.96\substack{+0.34 \\ -0.56}$ & 0.6990 & -0.4538 & 0.9531 & -0.4668 \\[0.90ex] 
\textit{TESS}$_3$ & $0.14528\substack{+0.00118 \\ -0.00118}$ & $58364.116468\substack{+0.000370 \\ -0.000364}$ & $88.81\substack{+0.79 \\ -0.91}$ & $11.07\substack{+0.28 \\ -0.56}$ & 0.6990 & -0.4538 & 0.9531 & -0.4668 \\[0.90ex] 
\textit{TESS}$_4$ & $0.14353\substack{+0.00109 \\ -0.00108}$ & $58370.839017\substack{+0.000381 \\ -0.000367}$ & $88.20\substack{+0.96 \\ -0.72}$ & $10.80\substack{+0.50 \\ -0.54}$ & 0.6990 & -0.4538 & 0.9531 & -0.4668 \\[0.90ex] 
\textit{TESS}$_5$ & $0.14211\substack{+0.00107 \\ -0.00109}$ & $58374.199350\substack{+0.000362 \\ -0.000371}$ & $88.54\substack{+0.91 \\ -0.85}$ & $10.98\substack{+0.38 \\ -0.58}$ & 0.6990 & -0.4538 & 0.9531 & -0.4668 \\[0.90ex] 
\textit{TESS}$_6$ & $0.14332\substack{+0.00108 \\ -0.00111}$ & $58377.559989\substack{+0.000382 \\ -0.000377}$ & $88.49\substack{+0.97 \\ -0.83}$ & $10.92\substack{+0.38 \\ -0.57}$ & 0.6990 & -0.4538 & 0.9531 & -0.4668 \\[0.90ex] 
\textit{TESS}$_7$ & $0.14096\substack{+0.00124 \\ -0.00121}$ & $58380.922103\substack{+0.000422 \\ -0.000428}$ & $89.29\substack{+0.48 \\ -0.72}$ & $11.29\substack{+0.16 \\ -0.32}$ & 0.6990 & -0.4538 & 0.9531 & -0.4668 \\[0.90ex] 
G141 & $0.14280\substack{+0.00048 \\ -0.00042}$ & $57880.135374\substack{+0.000059 \\ -0.000059}$ & $88.59\substack{+0.19 \\ -0.16}$ & $11.05\substack{+0.11 \\ -0.10}$ & 0.5692 & 0.1519 & -0.2305 & 0.0672 \\[0.90ex] 
\textit{Spitzer} CH1 & $0.14060\substack{+0.00145 \\ -0.00134}$ & $56313.405406\substack{+0.000240 \\ -0.000239}$ & $89.30\substack{+0.48 \\ -0.61}$ & $11.25\substack{+0.12 \\ -0.26}$ & 0.4839 & -0.3558 & 0.3447 & -0.1402 \\[0.90ex] 
\textit{Spitzer} CH2 & $0.14098\substack{+0.00185 \\ -0.00188}$ & $56306.683310\substack{+0.000341 \\ -0.000353}$ & $88.50\substack{+0.91 \\ -0.65}$ & $10.99\substack{+0.40 \\ -0.47}$ & 0.5652 & -0.7296 & 0.7488 & -0.2845 \\[0.90ex] 
\hline 
\vspace{-5mm} \\ 
\end{tabular} 
\caption{Measured parameters for WASP-6b from fits to the photometric \textit{TESS} and \textit{Spitzer} light curves, and the white light curves of the G600B, G600RI, STIS430, STIS750 and G141 datasets. Transit depths are those calculated following the fixing of the system parameters to the weighted average values. } 
\label{wlcparams} 
\end{table*}

\subsection{Spectrophotometric Light Curves}\label{slcs}
Prior to the full spectrophotometric fits, we correct all of the spectrophotometric light curves for wavelength independent (common-mode) systematics. In the case of the G600B and G600RI datasets we follow \citet{Niko16} and determine a common-mode correction by dividing each uncorrected transit white light curve by its final best fit transit model. To apply the correction we divide all spectrophotometric light curves by the common-mode calculated from their parent white light curve. For the G141 dataset we correct for common-mode systematics following the shift-and-fit method of \citet{Demi13}. In this case a reference spectrum was first produced by averaging all of the out-of-transit spectra. Each individual spectrum was then matched against this reference through stretching vertically in flux and shifting horizontally in wavelength following a linear least-squares optimisation. We then separate the spectral residuals of the previous fit into 28 wavelength bins spanning 1.13 to 1.65 $\mu$m. Each spectrophotometric residual was then added to a transit model constructed using the best fit parameters from the white light curve fit and limb-darkening calculated for the relative wavelength bin to produce the spectrophotometric light curves. All corrections can be seen under each systematics corrected light curve in Figure \ref{whitelightcurve}. 

\begin{figure*}
\centering
\includegraphics[width=\textwidth]{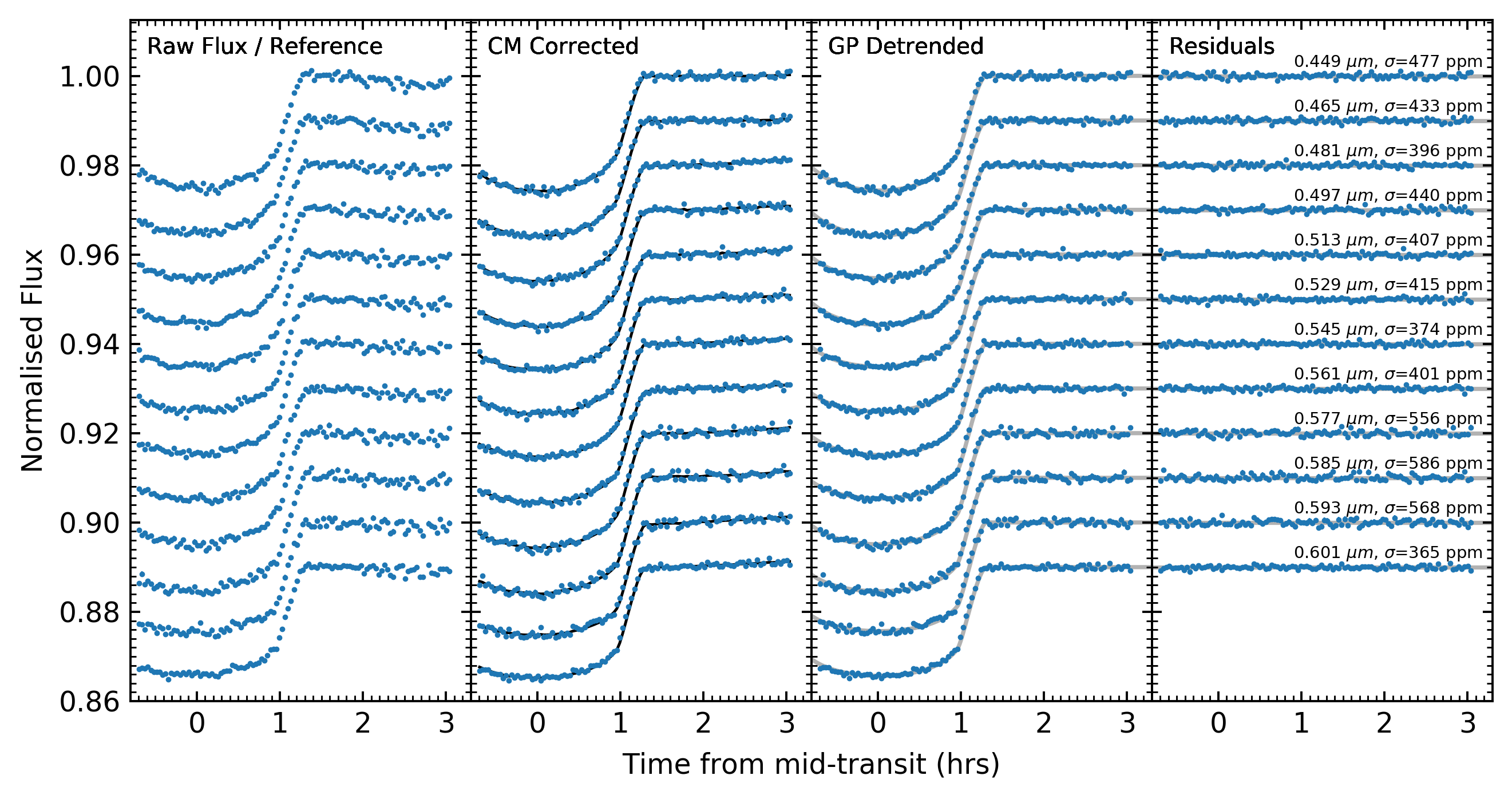}
\vspace*{-5mm}
\caption{Normalised spectrophotometric light curves for the G600B dataset of WASP-6b, light curves are offset from one another by an arbitrary constant. \textit{Left:} Raw light curves following reference star correction. \textit{Centre-Left:} Light curves after common-mode correction with black lines indicating the best GP transit plus systematic model fit. \textit{Centre-Right:} Light curves after common-mode correction and removal of GP systematic component. The best fitting transit models from \citet{Mand02} are displayed in grey. \textit{Right:} Residuals following subtraction of best fitting model.}
\label{spectro_b}
\end{figure*}

\begin{figure*}
\centering
\includegraphics[width=\textwidth]{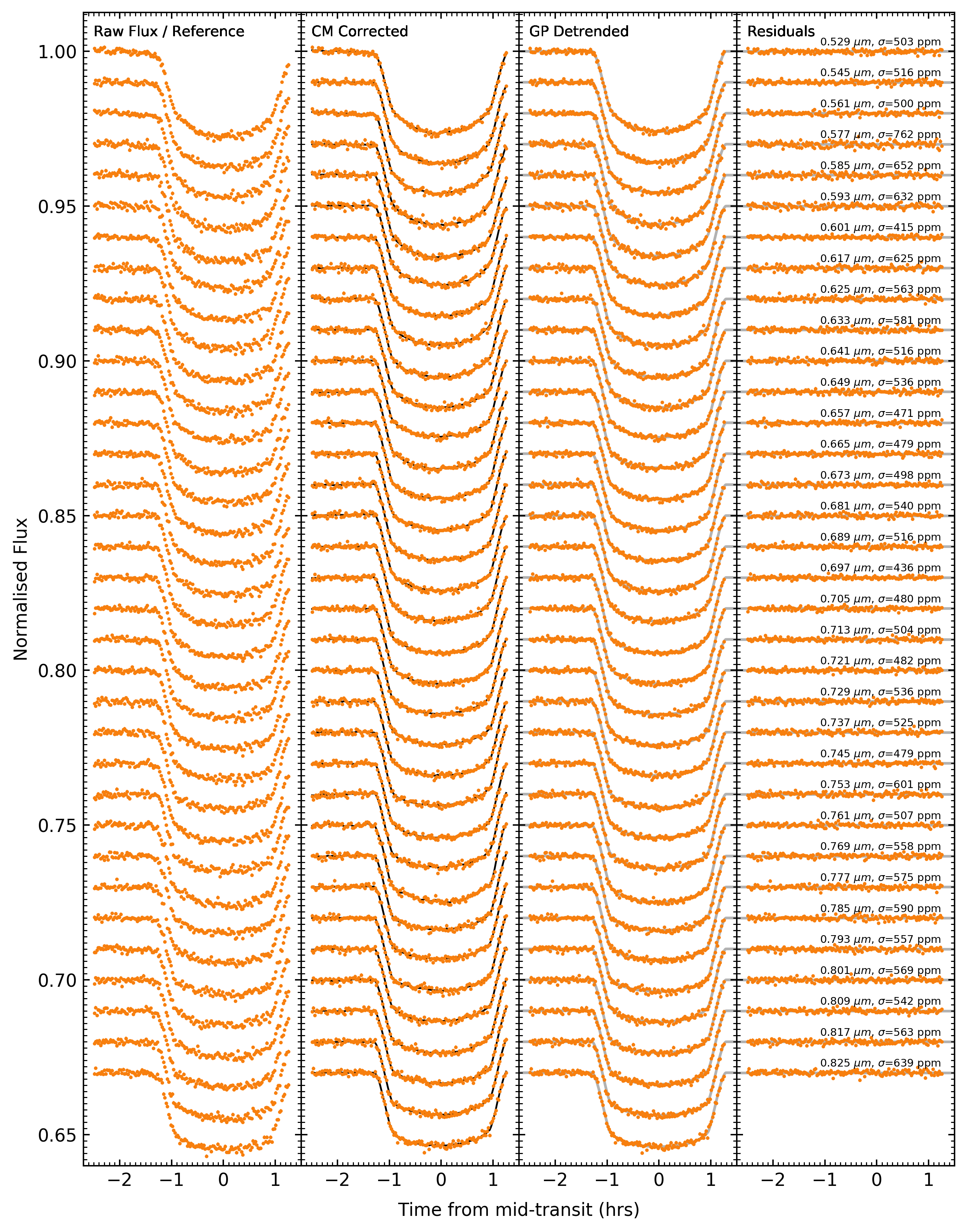}
\vspace*{-5mm}
\caption{As in Figure \ref{spectro_b}, but for the G600RI dataset.}
\label{spectro_ri}
\end{figure*}

\begin{figure*}
\centering
\includegraphics[width=\textwidth]{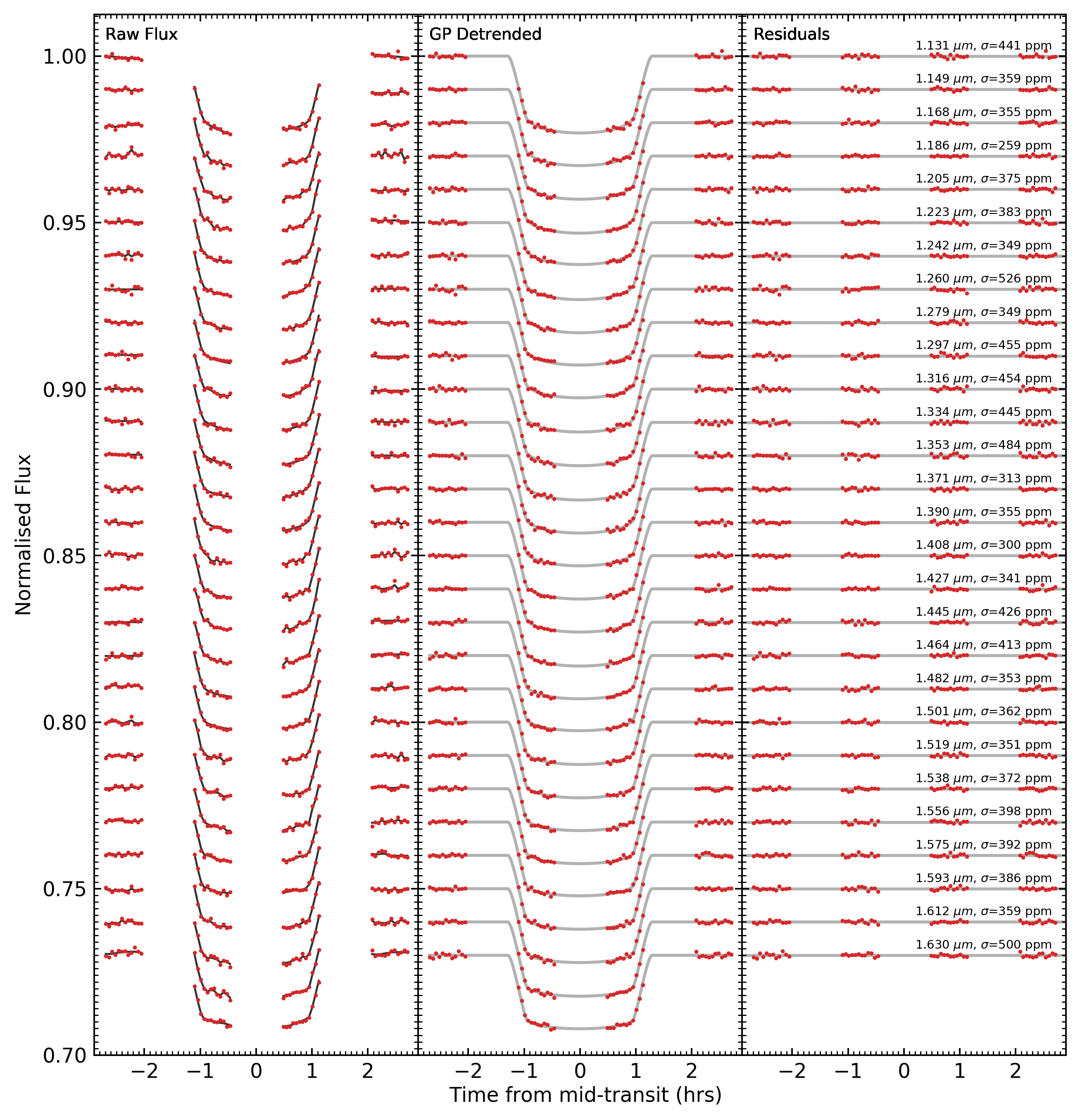}
\vspace*{-5mm}
\caption{Normalised spectrophotometric light curves for the G141 dataset of WASP-6b, light curves are offset from one another by an arbitrary constant. \textit{Left:} Raw extracted light curves with black lines indicating the GP transit plus systematic model fit. \textit{Centre:} Light curves after removal of GP systematic component. The best fitting transit models from \citet{Mand02} are displayed in grey. \textit{Right:} Residuals following subtraction of best fitting model.}
\label{spectro_w}
\end{figure*}

All spectrophotometric light curves were then fit following the same process as their corresponding white light curves. In each case however, the inclination and normalised semi-major axis were fixed to the weighted average values calculated from the white light curve fits and the transit central time was fixed to that of each respective white light curve fit. Additionally, for the G600B and G600RI light curves the quadratic limb darkening parameter $u_2$ was fixed to a value calculated from the ATLAS model described in Section \ref{obs} for each individual wavelength bin. The results for all best fit transit depths are displayed in Tables \ref{slcparams} and \ref{slcparams2} and all spectrophotometric light curves for the G600B, G600RI, G141 and STIS datasets are displayed in Figures \ref{spectro_b}, \ref{spectro_ri}, \ref{spectro_w} and \ref{stis_slcs} respectively.

\begin{figure}
\centering
\includegraphics[width=\columnwidth]{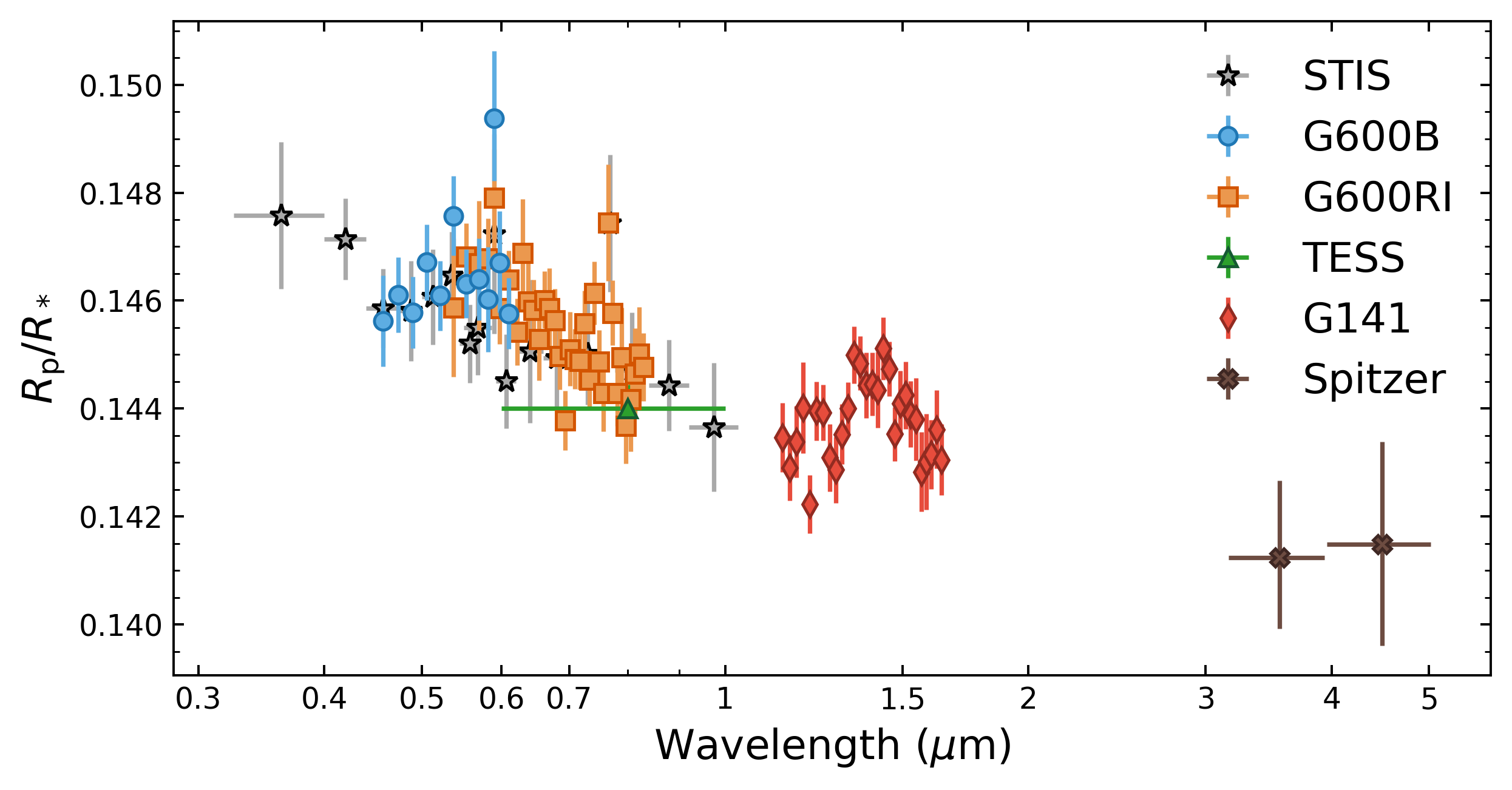}
\vspace*{-5mm}
\caption{The measured transmission spectrum of WASP-6b obtained from the G600B, G600RI, \textit{TESS}, STIS, G141 and \textit{Spitzer} datasets.}
\label{reduced_transpec}
\end{figure}

The initial transmission spectrum of these spectrophotometric light curves revealed an offset in transit depth between the G600B and G600RI datasets. Whilst activity of the host star can lead to such offsets, the stellar variability monitoring performed in \citet{Niko15} shows that potential offsets are of a magnitude $\Delta R_\textrm{p}/R_* \simeq 0.00022$, much too small to account for the observed offset of $\Delta R_\textrm{p}/R_* \sim 0.002$. Furthermore, the very good agreement of the G600RI dataset with the STIS measurements (Section \ref{archivalcomparison}) of \citet{Niko15} demonstrates that the cause of this offset most likely lies with the G600B dataset. Due to the poor phase coverage of the G600B dataset there are almost no observations during ingress, this produces a large uncertainty in the transit central time and subsequently the absolute transit depth, which may be responsible for the offset we see. Therefore, to account for this offset we apply a vertical shift to the G600B dataset by performing a weighted least-squares minimisation on the difference between the spectrophotometric bins in the overlapping region between the G600B and G600RI datasets, leaving the relative vertical shift of the G600B dataset as a free parameter in the minimisation. This results in a shift of $\Delta R_\textrm{p}/R_* = 0.00248$, equivalent to $\sim1.5\sigma$ of the error on the transit depth of the G600B white light curve. A full transmission spectrum with this offset included is shown in Figure \ref{reduced_transpec}.

\section{Correcting for Stellar Heterogeneity}\label{stelact}
Stellar activity leads to the presence of heterogeneities on stellar surfaces through the magnetically driven formation of cooler regions known as star spots and hotter regions known as faculae. The presence of spots (or faculae) on the surface of a star results in a wavelength dependent variation in the stellar baseline flux due to the respective differences in the emission profiles of the relatively cool spot (or relatively hot faculae) and the stellar surface itself. As the stellar baseline flux is crucial in determining transit depth, the presence of an unocculted star spot during a transit observation will necessarily produce a wavelength dependent variation in the measured transit depth \citep{Rack18, Rack19}. If significant enough, this variation can produce an artificial slope in the optical region of the final measured transmission spectrum, potentially mimicking the effects of haze in the atmosphere \citep{Pont08, Sing11b, Mccu14, Alam18, Pinh18}. These wavelength dependent variations can also impact individual spectral features due to the differential emission of specific stellar lines. Previous studies have displayed small decreases in the amplitude of Na {\sc{i}} absorption following a stellar heterogeneity correction (e.g. \citealt{Sing11b, Alam18}), however this effect is typically secondary to the artificially induced optical slope.

To estimate the impact surface stellar heterogeneities may have on our observations we obtained a proxy of the magnetic activity level of WASP-6 using a measurement of $\log(R^\prime_{HK})$. This value has been previously quoted without uncertainties as -4.741 in \citet{Sing16}, however analysis of the emission cores of the Ca {\sc ii} H and K lines in the HARPS spectra of \citet{Gill09} results in a direct measurement of -4.511 $\pm$ 0.037, indicating that WASP-6 is a moderately active star compared to the broader population of cool stars \citep{Saik18}. We therefore endeavour to account for the effects of unocculted star spots following the methodology of \citet{Alam18}.

\subsection{Photometric Monitoring of WASP-6}\label{photom}
We estimate the long baseline variability of WASP-6 by considering all 18,317 images from the \textit{TESS} observations previously described in Section \ref{tessobs} in addition to 435 $R$-band images from the Tennessee State University 14-inch Celestron \textit{Automated Imaging Telescope} (\textit{AIT}) taken from September 2011 to January 2019 (Figure \ref{spotgps}). Initially, we also incorporated 738 $V$-band images taken from November 2013 to July 2018 as part of The Ohio State University \textit{All-Sky Automated Survey for Supernovae} (\textit{ASAS-SN}) \citep{Shap14, Jaya18} into our photometric monitoring dataset as in \citet{Alam18}. However, on comparing the contemporaneous \textit{ASAS-SN} and \textit{TESS} data we find a $\sim$4 times larger photometric scatter in the \textit{ASAS-SN} dataset compared to the more precise \textit{TESS} sample and, as such, exclude it from our analysis in order to avoid influencing the variability amplitude estimation with such a noise-dominated dataset.

\subsection{The Stellar Rotation Period}\label{period}
In order to perform an accurate fit to the photometric monitoring data, it is necessary to have a measurement of the stellar rotation period. However, a range of rotation periods have been reported for WASP-6. In particular, \citet{Jord13} find a period of 16 $\pm$ 3 d based on the $v$sin$I$ = 2.4 $\pm$ 0.5 km $\textrm{s}^{-1}$ measurement from \citet{Doyl13}, \citet{Niko15} determine a period of 23.6 $\pm$ 0.5 d from a portion of their \textit{AIT} photometric monitoring, and by tracking transit star spot crossings \citet{Treg15} find a period of 23.80 $\pm$ 0.15 d, assuming the star had rotated only once between successive observed crossings.

We also perform a measurement of this rotation period through virtue of the very high cadence \textit{TESS} observations. Even from an initial inspection of the light curve shown in Figure \ref{spotgps} a clear sinusoidal variation can be seen. In order to determine that this variation is not due to an instrumental effect we inspect the light curves and background flux of the four closest neighbouring stars to WASP-6 with \textit{TESS} light curve observations. We find that none of the stars exhibit the same sinusoidal variation as WASP-6, and they all exhibit similar variations in their background flux. To determine the rotation period itself, we perform a least-squares minimisation using a simplistic sinusoidal model on the data with all transit events removed. This resulted in an inferred period of 12.18 $\pm$ 0.05 d. 

Even though this method of model fitting is quite rudimentary, the determined period is clearly in contradiction to current estimates of the stellar rotation period. This contradiction suggests that the variability observed is likely not that of a single spot feature rotating with a period equal to that of the stellar rotation period. Alternatively, the perceived \textit{TESS} period can be explained by the spot coverage during the \textit{TESS} epoch being concentrated on opposite hemispheres of the star, rather than one single hemisphere. During a period of \textit{AIT} photometry performed shortly after the \textit{TESS} observations from September 2018 to January 2019 we find a standard deviation of 3.8 mmag, in contrast to previous seasons where this reached up to 8.1 mmag. This reduced variability is further justification of the measured \textit{TESS} period being a result of hemispherically varying star spot coverage and not intrinsic to the \textit{TESS} instrument itself. Further high-quality photometric monitoring will likely be necessary to fully resolve the discrepancy between these observations. For subsequent analysis however, we adopt the stellar rotation period of 23.6 $\pm$ 0.5 d from \citet{Niko15} as this estimate was made over much longer timescales compared to the estimates of \citet{Jord13} and \citet{Treg15}.

\subsection{Modelling and Correction of Unocculted Star Spots}\label{stelcorr}
The variability of WASP-6 was modelled following the methodology of \citet{Alam18}. We perform a Gaussian process (GP) regression model fit to the photometric monitoring data constructed with a three component kernel which models: the quasi-periodicity of the data, irregularities in the amplitude, and stellar noise. A gradient based optimisation routine was used to locate the best-fit hyperparameters and a uniform prior was placed on the stellar rotation period, centred on the value of 23.6 $\pm$ 0.5 d from \citet{Niko15} with a width three times that of the standard deviation. The \textit{TESS} bandpass ranges from 0.6-1.0 $\mu$m and is less susceptible to active photometric variations compared to the \textit{AIT} $R$-band observations. This should not affect the wavelength dependence of our determined spot correction however, as the estimated variability amplitude is ultimately used as a reference to normalise the true model wavelength-dependent correction factor (Equation \ref{wdcf}). Despite this, the discrepancy of the measured \textit{TESS} period from the measured period in other studies \citep{Jord13, Niko15, Treg15}, and the reduced variation in a subset of \textit{AIT} data described in Section \ref{period}, does indicate that the variability of the star as a whole was also lower during this epoch. Because the variability amplitude is crucial in determining the spot correction, we opt to perform separate fits to the \textit{TESS} and \textit{AIT} datasets. To avoid influencing the GP fitting with the lower variance \textit{AIT} data, we exclude 41 measurements obtained shortly after the \textit{TESS} epoch which correspond to the subset described in Section \ref{period}. Due to the large size of the \textit{TESS} dataset ($\sim$18,000 data points) we bin the data down by a factor of 10 in order to make the GP fitting computationally tractable. 

\begin{figure*}
\centering
\includegraphics[width=\textwidth]{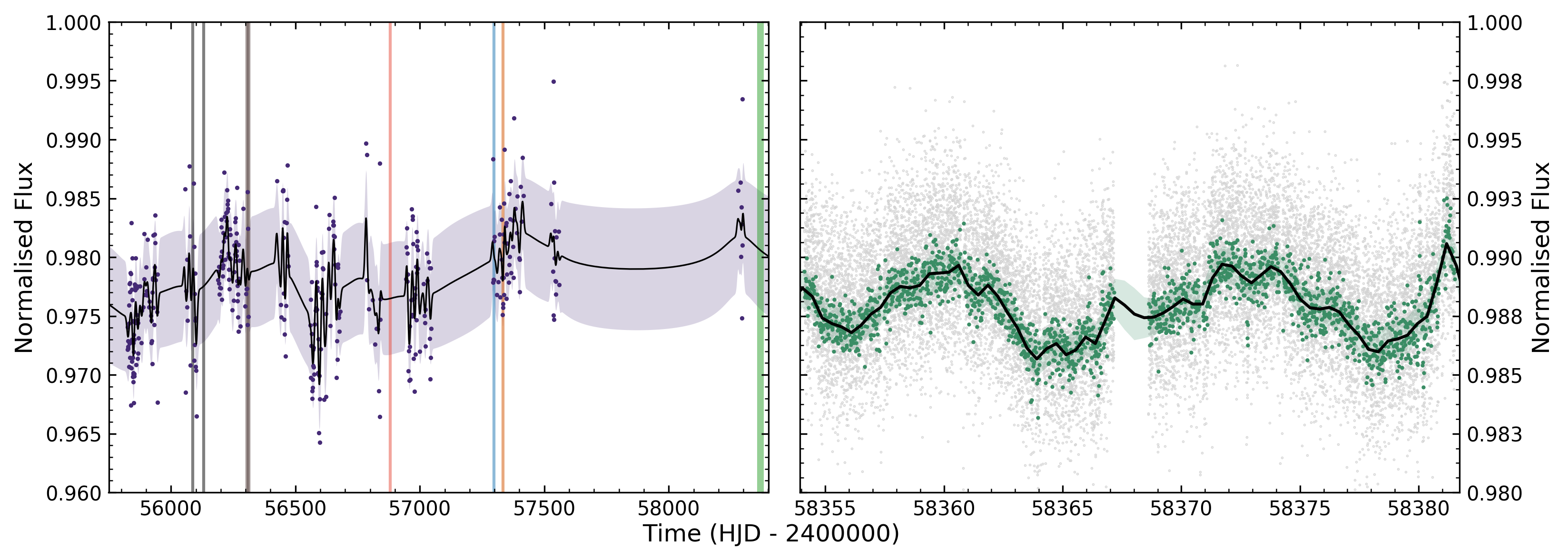}
\vspace*{-5mm}
\caption{Photometric monitoring and modelling of the stellar variability of WASP-6. \textit{Left}: \textit{AIT} monitoring data prior to the \textit{TESS} epoch(purple dots) with best fit GP model represented by the solid black line, the shaded area represents the 1$\sigma$ confidence region. Additional vertical lines are plotted corresponding to the best fit transit central times of each observation as shown in Table \ref{wlcparams}, the broader green region nearest the latest observations corresponds to the full \textit{TESS} epoch. \textit{Right:} Unbinned (grey) and binned (cyan) \textit{TESS} monitoring data with best fit GP model represented by the solid black line, the shaded area represents the 1$\sigma$ confidence region. For both the \textit{AIT} and \textit{TESS} datasets the flux has been normalised with the maximum stellar flux obtained from their respective GP model fits corresponding to unity.}
\label{spotgps}
\end{figure*}

Whilst the \textit{TESS} data is well sampled and more precise than the \textit{AIT} data, we may be perceiving a lower level of variability due to the \textit{TESS} bandpass or the lower intrinsic variability of WASP-6 during the \textit{TESS} epoch (Section \ref{period}). Comparatively, the \textit{AIT} data has a much broader temporal coverage and could therefore be more indicative of the longer-term variability of WASP-6, though as there are no contemporaneous measurements with the \textit{TESS} dataset their accuracy is not guaranteed. The \textit{TESS} and \textit{AIT} model fits therefore provide respectively more conservative or realistic estimates of the true stellar variability. All such fits to the photometric monitoring data are displayed in Figure \ref{spotgps}.

We are then able to correct for the unocculted spots in the transit light curves following \citet{Huit13}. Under the assumption that there is always some level of spot coverage on the stellar surface, the maximum observed stellar flux does not correspond to the flux emitted by an entirely unspotted surface. Using the amplitude of the GP fit to both the \textit{TESS} and \textit{AIT} photometric monitoring data we determine different estimates for the unspotted stellar flux $F'$ = max($F$)+$k\sigma$, where $F$ is the observed photometric monitoring data, $\sigma$ is the dispersion of these photometric measurements, and $k$ is a value fixed to unity. Whilst an accurate value of $k$ can be difficult to determine a $k$ = 1 has been shown to be suitable for active stars \citep{Aigr12}. Furthermore, varying the chosen value of $k$ does not significantly influence the wavelength dependence of the correction and mainly influences the offset of the transmission spectrum baseline \citep{Alam18}. For each estimate, the fractional dimming due to stellar spots was then calculated as $f_\textrm{norm} = \overline{F/F'}$, giving the amplitude of the spot correction at the variability monitoring wavelength as $\Delta f_0$ = 1 - $f_\textrm{norm}$.

\begin{figure}
\centering
\includegraphics[width=\columnwidth]{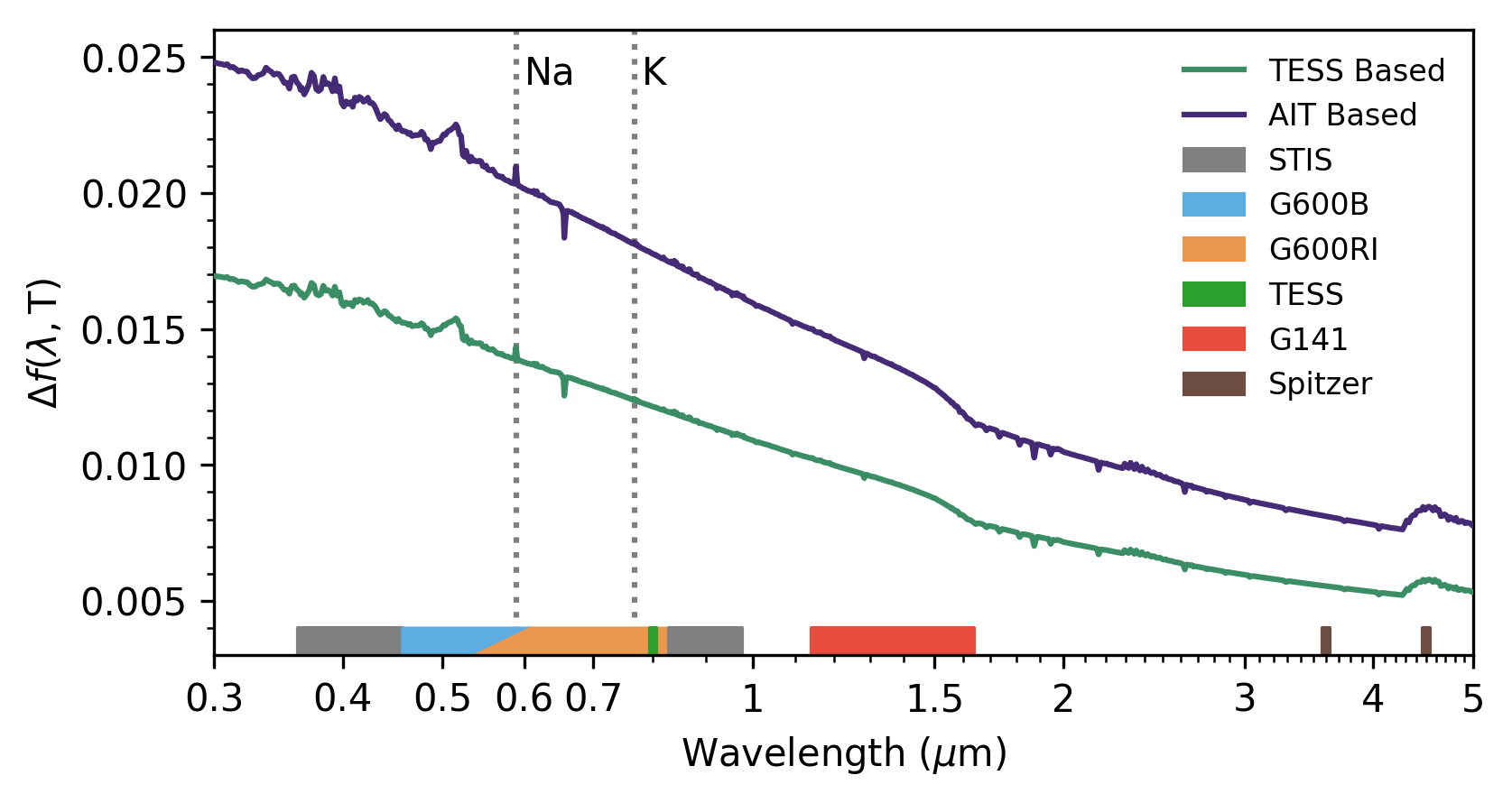}
\vspace*{-5mm}
\caption{Calculated spot corrections based on the \textit{TESS} (teal, bottom) and \textit{AIT} (purple, top) photometric data. Regions of wavelength coverage for all observations performed in this study are also shown, the photometric \textit{TESS} and \textit{Spitzer} data points are represented as lines at the centre of their respective bandpasses. }
\label{spotcorrfig}
\end{figure}

In order to determine each wavelength dependent spot correction we must compute the wavelength-dependent correction factor shown in \citet{Sing11}:
\begin{equation}\label{wdcf}
f(\lambda ,T)  =  \bigg{(} 1-  \frac{F_{\lambda ,T_{\textrm{spot}}}}{F_{\lambda ,T_{\textrm{star}}}} \bigg{)} \Bigg{/} \bigg{(} 1-  \frac{F_{\lambda_{0} ,T_{\textrm{spot}}}}{F_{\lambda_{0} ,T_{\textrm{star}}}} \bigg{)}
\end{equation}
where $F_{\lambda ,T_{\textrm{spot}}}$ is the wavelength dependent stellar flux at temperature $T_\textrm{spot}$, $F_{\lambda ,T_{\textrm{star}}}$ is the wavelength dependent stellar flux at temperature $T_\textrm{star}$, $F_{\lambda_{0} ,T_{\textrm{spot}}}$ is the stellar flux at the wavelength of the photometric monitoring data at temperature $T_\textrm{spot}$, and $F_{\lambda_{0} ,T_{\textrm{star}}}$ is the stellar flux at the wavelength of the photometric monitoring data at temperature $T_\textrm{star}$. In order to determine the stellar and spot fluxes described we use the \texttt{ATLAS} stellar model described in Section \ref{obs}. The only difference between the stellar flux and spot models is that they differ by a temperature of 1500K, assumed from an empirically determined relation \citep{Berd05}. Finally, we compute wavelength dependent spot corrections based on both the \textit{AIT} and \textit{TESS} photometry following $\Delta f = \Delta f_0 \times f(\lambda ,T)$ (Figure \ref{spotcorrfig}).

Each spot correction was then independently applied to both the white and spectrophotometric light curves using:
\begin{equation}
y_\textrm{corr} = y + \frac{\Delta f}{(1-\Delta f)}\overline{y_{\textrm{oot}}}
\end{equation}
where $y_\textrm{corr}$ is the corrected light curve flux, $y$ is the uncorrected flux, and $\overline{y_{\textrm{oot}}}$ is the out-of-transit mean flux. These corrected light curves, informed by either the \textit{TESS} or \textit{AIT} photometry, were then refit following the same method as demonstrated in Section \ref{lightcurves} and are hereafter defined as the \textit{TESS} corrected or \textit{AIT} corrected datatsets. Both \textit{TESS} and \textit{AIT} corrected G600B spectrophotometric light curves exhibited comparable offsets to the uncorrected dataset (Section \ref{slcs}) of $\Delta R_\textrm{p}/R_* $ = 0.00244 and 0.00242 respectively and thus similar vertical shifts are performed. All best fit parameters from the white light curve fits are displayed in Tables \ref{bestfitpars} and \ref{wlcparams}, and all best fit spectrophotometric transit depths are displayed in Table \ref{slcparams} and \ref{slcparams2}.

\section{Discussion}\label{disc}

\begin{figure*}
\captionsetup[subfigure]{labelformat=empty}
\centering
	\subfloat[\label{Full}]{%
	\includegraphics[clip, width=\textwidth]{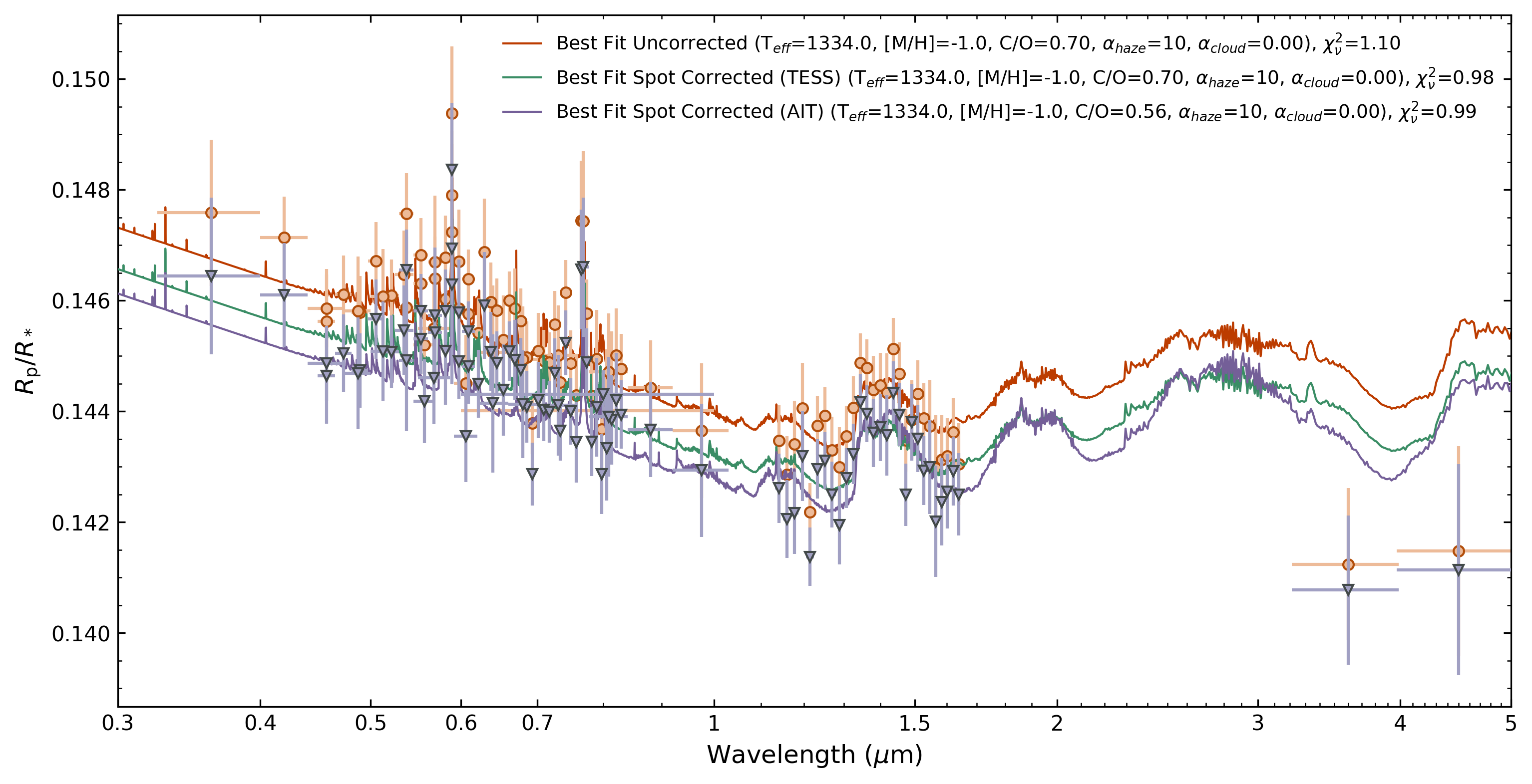}}
	\vspace{-41pt}
	\subfloat[\label{Full}]{%
	\includegraphics[clip, width=\textwidth]{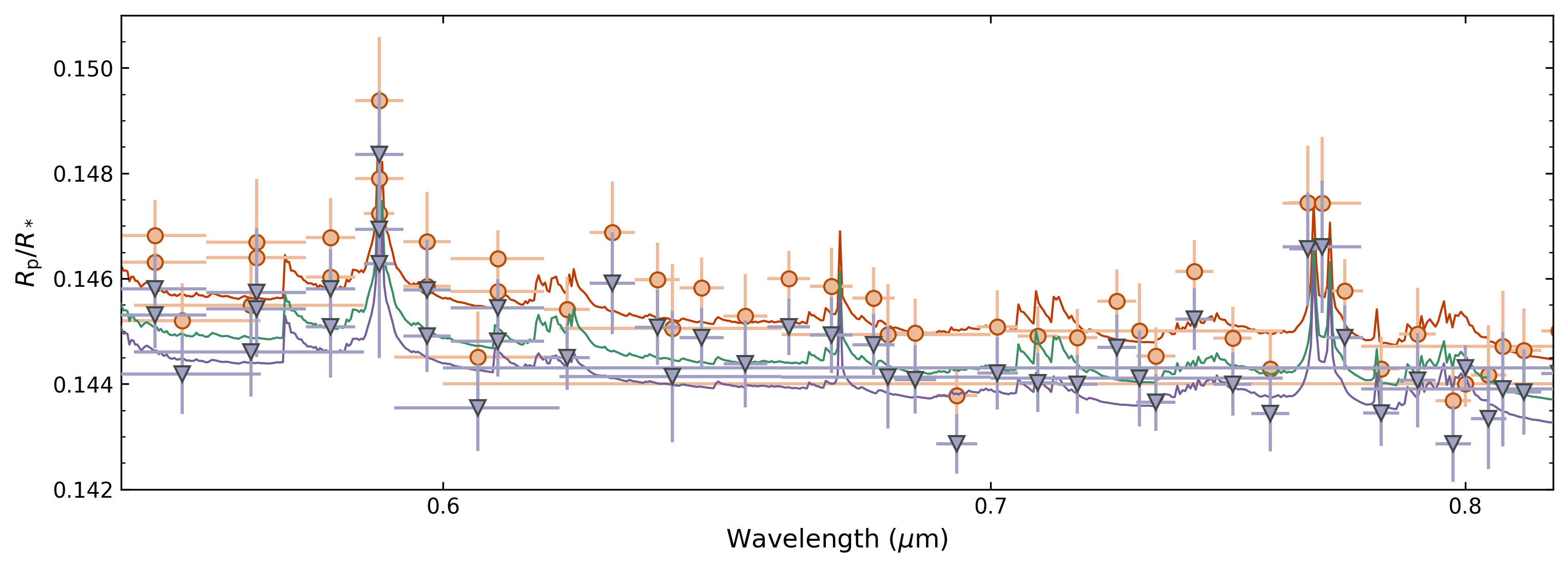}}
	\vspace*{-5mm}
\caption{\textit{Top}: The uncorrected (orange circles) and \textit{AIT} spot corrected (purple triangles) transmission spectra of WASP-6b as determined from the performed G600B, G600RI, G141, \textit{TESS}, and archival STIS and \textit{Spitzer} observations with the best fit models from the \citet{Goya18} forward grid. For reasons of clarity the \textit{TESS} spot corrected dataset is not shown, however the best fit model is displayed in order to demonstrate the differences in transit depth. \textit{Bottom}: As in the top panel, except zoomed in to the wavelength region spanning the Na {\sc i} and K {\sc i} lines}
\label{grid_transpec}
\end{figure*}

The observed transmission spectrum of WASP-6b reveals a variety of spectroscopic features present both in the uncorrected and spot corrected analyses (Figure \ref{grid_transpec}). In particular, the broad absorption feature at 1.4 $\mu$m indicates the presence of H$_2$O in the atmosphere. Additionally, narrow band absorption features at 0.589 and 0.767 $\mu$m due to Na {\sc i} and K {\sc i} are also evident in the optical. Finally, a distinct increase in transit depth across optical wavelengths is seen, indicative of a scattering haze and in agreement with \citet{Niko15}. The primary difference between the uncorrected and spot corrected datasets is the presence of a vertical offset across the full wavelength range. This offset is not wavelength independent however and the spot correction has acted to slightly reduce the gradient across the optical slope. This wavelength dependence is clearly identified by the difference in transit depth between the uncorrected and \textit{AIT} corrected datasets at the shortest wavelength bin compared to that of the longest wavelength (Figure \ref{grid_transpec}). 

\subsection{Archival Data Comparisons}\label{archivalcomparison}
\begin{figure}
\centering
\includegraphics[width=\columnwidth]{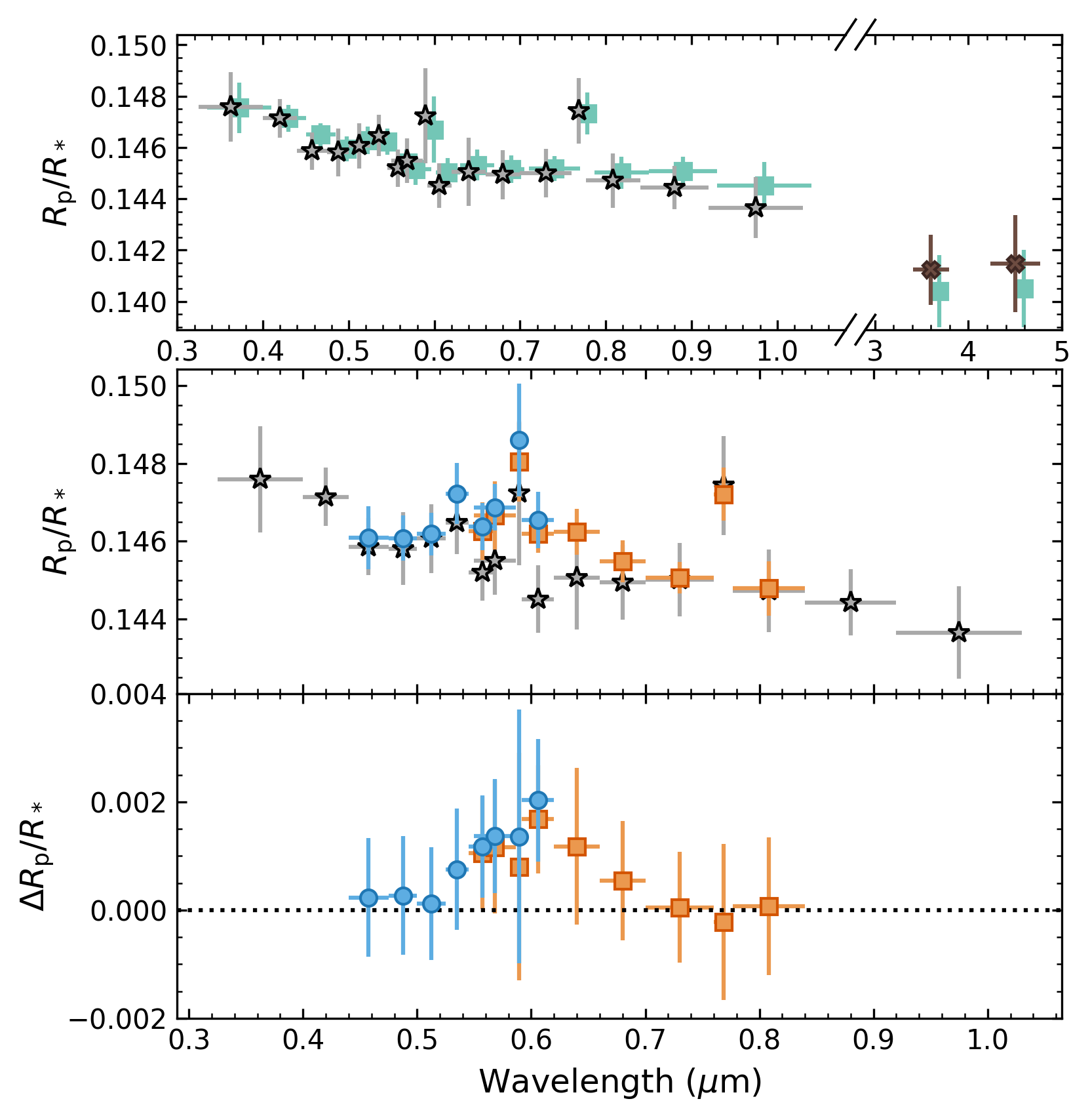}
\vspace*{-5mm}
\caption{\textit{Top}: A comparison of the measured STIS and \textit{Spitzer} transit depths from this study (grey stars/brown crosses) and those published in \citet{Niko15} (teal squares). A small wavelength offset has been added to the literature datasets for clarity. \textit{Middle}: The measured uncorrected transit depths of the STIS (grey stars) dataset in comparison to the G600B (blue circles) and G600RI (orange squares) datasets, binned down to an identical resolution where possible. \textit{Bottom}: Differenced transit depths following subtraction of the STIS dataset from the G600B and G600RI datasets, a slight disparity is seen within the Na {\sc i} line.}
\label{stiscomp_transpec}
\end{figure}

The transmission spectrum of WASP-6b had already been measured using the available \textit{HST} STIS and \textit{Spitzer} IRAC datasets \citep{Niko15}. In order to compare our independent reduction against these results we overplot both the uncorrected transit depths from this study, with those from this prior published study (Figure \ref{stiscomp_transpec}). The different reductions agree quite well, with all measurements within 1$\sigma$ of one another. A minor discrepancy in transit depth is seen for the longest wavelength STIS bins and the \textit{Spitzer} photometry. These discrepancies are likely due to the slightly different measured system parameters which were held fixed during the independent fittings in addition to slight differences in the adopted stellar limb darkening parameters. The error bars for the reduction performed in this study are larger than those of those from the original reduction, primarily due to the difference between the model marginalisation and Gaussian process approaches towards light curve fitting.

As the STIS and \textit{VLT} FORS2 datasets have a broad overlapping wavelength range we reproduce the \textit{VLT} FORS2 transmission spectrum using an identical wavelength binning as the \textit{HST} STIS measurements to facilitate a comparison between the results (Figure \ref{stiscomp_transpec}). It is evident from this comparison that whilst our results agree very well at the shortest and longest wavelengths, there is a small disparity in the measurements centred around the Na {\sc i} absorption line. We calculate a weighted average transit depth across 5 wavelengths bins centred on the Na {\sc i} absorption line for the G600RI dataset and the STIS dataset, resulting in $R_\textrm{p}/R_*$'s of 0.14628$\pm$0.00031 and 0.14520$\pm$0.00043 respectively. We exclude the G600B dataset from the calculation to avoid any bias due to the applied vertical shift as described in Section \ref{lightcurves}. 

As the offset reduces proportionally with separation from the Na {\sc i} line center, this signal could be indicative of an observation of the pressure-broadened wings from the full Na {\sc i} feature in the FORS2 datasets. Such wings have recently been definitively observed in the atmosphere of the hot Jupiter WASP-96b \citep{Niko18}. Given these wings are not present in the STIS dataset, this could suggest we are observing variability in the atmosphere of WASP-6b. However, this offset being of an instrumental or systematic origin cannot be excluded, particularly as the FORS2 observations are taken from the ground where systematic variations are not as well understood and harder to model. The possibility that this discrepancy has been caused by the STIS observations in particular also cannot be excluded as there exists robust evidence that systematics in STIS observations resulted in a spurious detection of K in WASP-31b \citep{Gibs17, Gibs19}. The true cause of the discrepancy, be it physical or systematic, can not be determined with these data and additional observations at higher signal to noise and over long timescales will be required to investigate this further. 

\subsection{Goyal Forward Models}\label{goyalmodels}
In order to explore the bulk properties of WASP-6b we fit the observed transmission spectrum to a grid of forward models \citep{Goya18, Goya19a}. These models are generated using the 1D radiative-convective equilibrium code \texttt{ATMO}. Initially we opted to use the more recent generic model grid \citep{Goya19a} in our analysis as it allowed for a broader coverage of the parameter space than the WASP-6b specific grid from \citet{Goya18}. However, as sub-solar metallicity forward models have yet to be implemented into the generic grid our ability to accurately fit the observed data was ultimately restricted. As such, we used the WASP-6b specific grid \citep{Goya18} in order cover the sub-solar metallicity range of parameter space.

With the arrival of the \textit{Gaia} Data Release 2 \citep{Gaia18} the distance to WASP-6 has been more accurately determined as $d = 197.1\substack{ +0.4 \\ -1.6 }$ pc \citep{Bail18}, significantly different to the prior measurement of 307 pc. This re-estimation has significant effects on the inferred stellar radius of WASP-6 which in turn affects the estimation of planetary radius from the observed transit depths. A mismeasurement of the planetary radius naturally leads to a mismeasurement of the planetary gravity, a currently fixed parameter for the planet specific forward model grid of \citet{Goya18}. Following the methodology of \citet{Morr19}, we performed spectral energy distribution (SED) fitting on WASP-6 using NUV, optical and NIR broadband photometry. The fitted integrated flux allows us to measure its luminosity, and the shape of the SED determines its so-called $T_{\rm SED}$ \citep[see][for details]{Morr19}. By combining this with the revised distance measurement, we obtained an updated estimate of the radius of WASP-6, and subsequently the radius of WASP-6b. This radius results in a new value for the planetary gravity of $g = 10.55\substack{ +0.19 \\ -0.39 }$ ms$^{-2}$, notably different from the previous estimate of $g = 8.71 \pm 0.55$ ms$^{-2}$ \citep{Gill09}. Changes in gravity can have significant effects on the computed forward models \citep{Goya18, Goya19a} and therefore to fit our observed data we use a more updated forward model grid for WASP-6, identical to the original shown in \citet{Goya18} except recomputed for a value of $g=10.5$. 

The model grid used consists of 3920 different transmission spectra varying in temperature, metallicity, C/O ratio, scattering haze and uniform cloud. The scattering haze is implemented through the use of a haze enhancement factor $\alpha_{\textrm{haze}}$ which simulates an increase in the total scattering of small aerosol particles in the atmosphere. Similarly, the uniform cloud is implemented through a variable cloudiness factor $\alpha_{\textrm{cloud}}$, which produces the effects of a cloud deck through a modification to the wavelength dependent scattering using the strength of grey scattering due to H$_2$ at 350 nm. Irrespective of the true cloud composition, implementing a grey cloud is appropriate for our observations as at the observed wavelengths Mie scattering predicts essentially grey scattering profiles \citep{Wake15}. Further details on the grid parameters, including their ranges and implementations, can be found in \citet{Goya18}. 

Each model spectrum was fit in turn by producing a binned version of the spectrum which matches the selected spectrophotometric bands from the data reduction and then averaged to produce a single value of transit depth in each bin. A $\chi^2$ measurement between the observed and model data was then computed following a least-squares minimisation scheme with a varying wavelength-independent vertical offset. These fits were performed for both the uncorrected and both spot corrected transmission spectra and the best fitting models for each are presented in Figure \ref{grid_transpec}.

\begin{figure}
\centering\captionsetup[subfloat]{labelfont=bf}
   \subfloat[\label{Uncorrected}]{%
      \includegraphics[clip, width=\columnwidth]{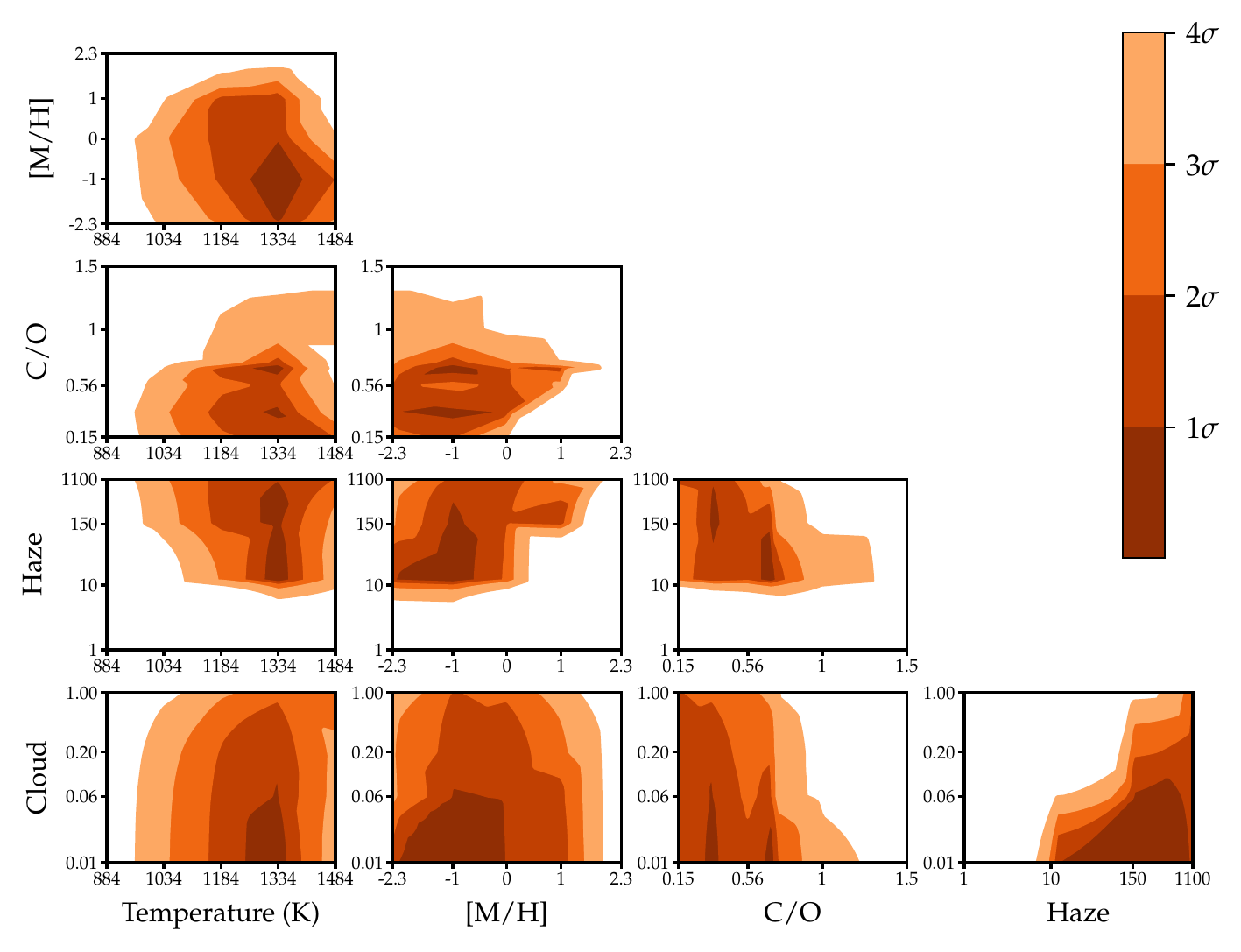}}
    \vspace{-10pt}
   \subfloat[\label{TESS Corrected} ]{%
      \includegraphics[clip, width=\columnwidth]{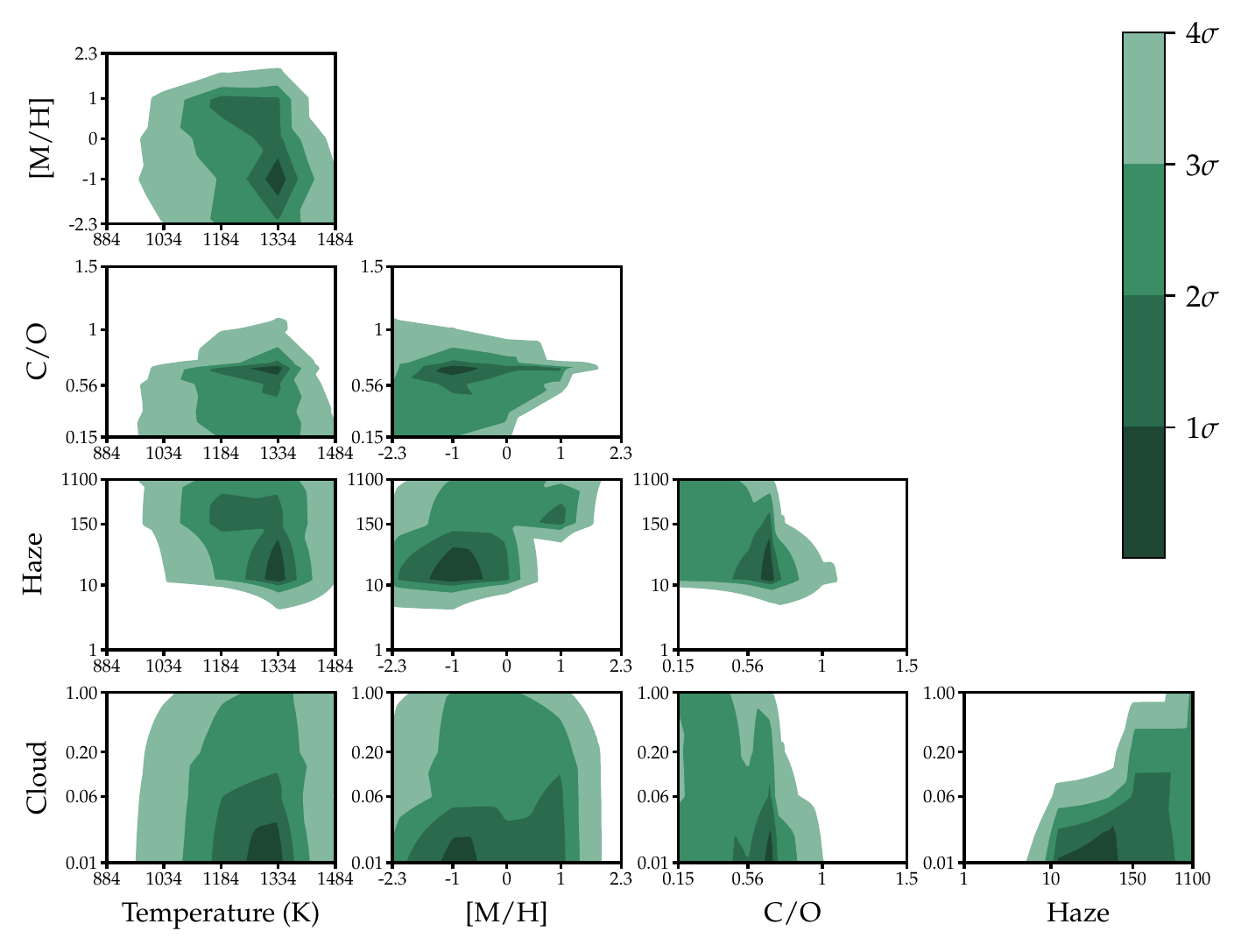}}
       \vspace{-10pt}   
   \subfloat[\label{AIT Corrected}]{%
      \includegraphics[clip, width=\columnwidth]{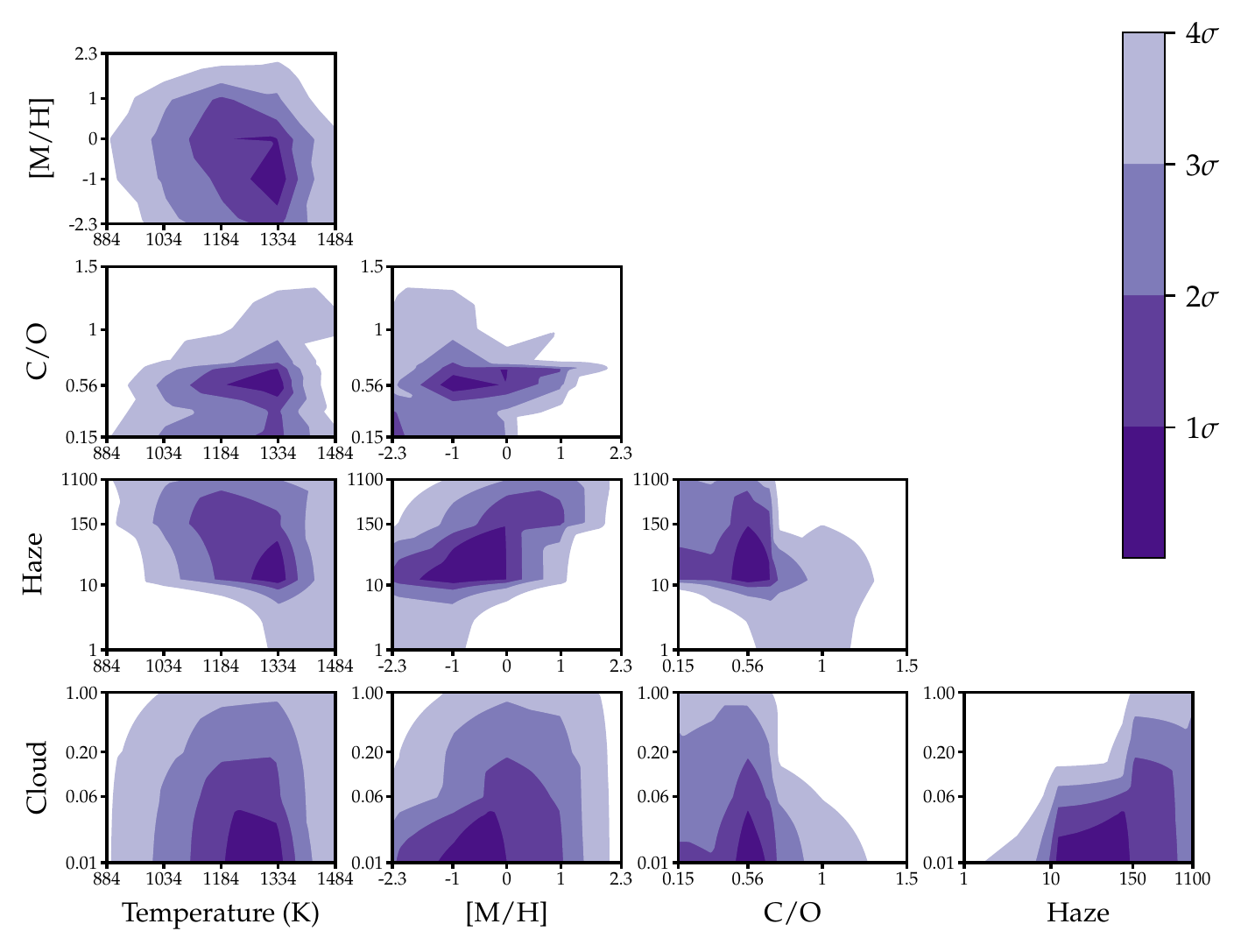}}\\
\caption{\label{all_gridplots} $\chi^2$ contour maps produced when fitting the complete transmission spectrum of WASP-6b to forward model grids of \citet{Goya18} considering \textbf{(a)} no correction for stellar heterogeneity, \textbf{(b)} correction using \textit{TESS} photometry, and \textbf{(c)} correction using \textit{AIT} photometry. Shaded regions indicate models in the parameter space which are at least $N-\sigma$ from the best fit model. Preferences towards the lowest metallicity, highest haze enhancement factors, and lower C/O ratios are present for the uncorrected dataset, whereas this is not the case for the \textit{TESS} or \textit{AIT} spot corrected datasets.}
\end{figure}

For the uncorrected and the \textit{TESS} corrected transmission spectra we find a best fitting model of $T = 1334\,$K, sub-solar metallcity [M/H] = $-1.0$, slightly super-solar C/O ratio of [C/O] = $0.70$, moderate hazes $\alpha_\textrm{haze} = 10$ and no evidence of clouds $\alpha_\textrm{cloud} = 0$ corresponding to a $\chi^2_\nu$ = 1.10 and 0.98 respectively. For the \textit{AIT} corrected transmission spectrum however, we find a best fitting model of $T = 1334\,$K, sub-solar metallcity [M/H] = $-1.0$, solar C/O ratio of [C/O] = $0.56$, moderate hazes $\alpha_\textrm{haze} = 10$ and no evidence of clouds $\alpha_\textrm{cloud} = 0$ corresponding to a $\chi^2_\nu$ = 0.99. To explore the discrepancies and commonalties between the grid fits to the uncorrected and corrected datasets we produce $\chi^2$ contour maps \citep{Madh09} as shown in Figure \ref{all_gridplots}. We begin by constructing 2D grids of every possible pair of model parameters. In each separate grid, and at every individual grid point dictated by the resolution of the model parameters, we vary all the remaining model parameters in turn and determine the model with the smallest $\chi^2$. Across these new $\chi^2$ spaces we determine contours which correspond to models in the parameter space which are $N$-$\sigma$ from the overall best fit model following \citet{Goya18}.

The primary differences between the datasets are the existence of subsets of model fits more favoured by the lowest metallicities and the highest haze enhancement factors for only the uncorrected dataset. These subsets are present because the wavelength dependence of stellar heterogeneity acts to increase the gradient of the optical slope in the observed data, an effect that is somewhat degenerate with lower metallicity and hazy atmospheres \citep{Goya18}. Whilst both the lowest metallicities and highest haze enhancements factors are not as favoured in tandem, they both correspond to model fits favouring a lower level of C/O ratio. This is because both low metallicity and high haze enhancement factor act to suppress the H$_2$O absorption features beyond the constraints set by the G141 dataset and as such the C/O ratio must be reduced in order to re-inflate the H$_2$O features to match the observations. In summary, the $\chi^2$ contour map for even the conservative \textit{TESS} corrected dataset indicates that these highest haze enhancement factors, lowest metallicities, and lowest C/O ratios are likely effects of stellar heterogeneity on the transmission spectrum of WASP-6b and not truly symptomatic of its atmosphere. However, a moderate haze enhancement of at least $\alpha_\textrm{haze} = 10$ is strongly constrained, and a preference towards sub-solar metallicities is still evident, independent of the addition of a spot correction.

Whether or not a spot correction is used, temperatures of 1334$\,$K are primarily preferred for each grid fit. Comparatively, the measured dayside temperatures for WASP-6b are 1235$\substack{ +70 \\ -77 }$ and 1118$\substack{ +68 \\ -74 }$ from the 3.6 and 4.5 $\mu$m \textit{Spitzer} IRAC channels respectively \citep{Kamm15}. As these values are within $\sim1\sigma$ they do not suggest a disagreement, however, it is worthwhile assessing the source of the slight preference of the grid model fits towards limb temperatures higher than that measured from the dayside. As the model grid varies in temperature steps of 150$\,$K the model cannot settle on a precise temperature estimate and is therefore likely to be somewhat discrepant from the true value. However, there are models at a temperature $T = 1184\,$K which should in theory match the true temperature of WASP-6b's limb more accurately. Looking to Figure \ref{all_gridplots}, the preferred temperature is strongly constrained below the 1484$\,$K grid models, as at approximately this temperature absorption features due to TiO and VO start to become significant in the optical \citep{Fort08} and are strongly disfavoured by the observed FORS2 and STIS datasets. As temperature acts to increase the gradient of the optical slope \citep{Goya18} it is also degenerate with the effects of stellar heterogeneity. Therefore the models at 1334$\,$K are the most favoured as it is the highest temperature, and thus steepest slope, that the model grid can produce without generating conflicting TiO and VO features. Figure \ref{all_gridplots} demonstrates this as the model preferences for the highest temperatures are slightly reduced upon application of the spot corrections, with the most significant difference being for the \textit{AIT} corrected dataset. As the best fit temperature for the \textit{AIT} correction is still beyond what we would expect given the day side temperatures already reported it could even suggest that the spot correction used has been underestimated. However, a subset of $1184\,$K models are comfortably within the 2$\sigma$ region for every dataset and therefore conclusively determining the true effect of stellar heterogeneity on the best fit model temperature will require further investigation with observations at a higher signal to noise.

To determine the significance of the perceived detections of the Na {\sc i} and K {\sc i} features we begin by performing a quadratic interpolation of the baseline of the best fit model to each dataset from 0.4-0.9 $\mu$m using regions of the optical slope with no clear absorption features as anchors for the interpolation. The interpolation then served as a comparison against the weighted mean value of the G600B, G600RI, STIS 430 and STIS 750 data contained with the Na {\sc i} and K {\sc i} lines. Detection significances are summarised in Table \ref{nak_sigma}, these values indicate at least a 3$\sigma$ detection of the Na {\sc i} and K {\sc i} narrow line signatures in the atmosphere WASP-6b, irrespective of an applied spot correction.

\afterpage{
\begin{figure*}
\centering
\includegraphics[width=\textwidth]{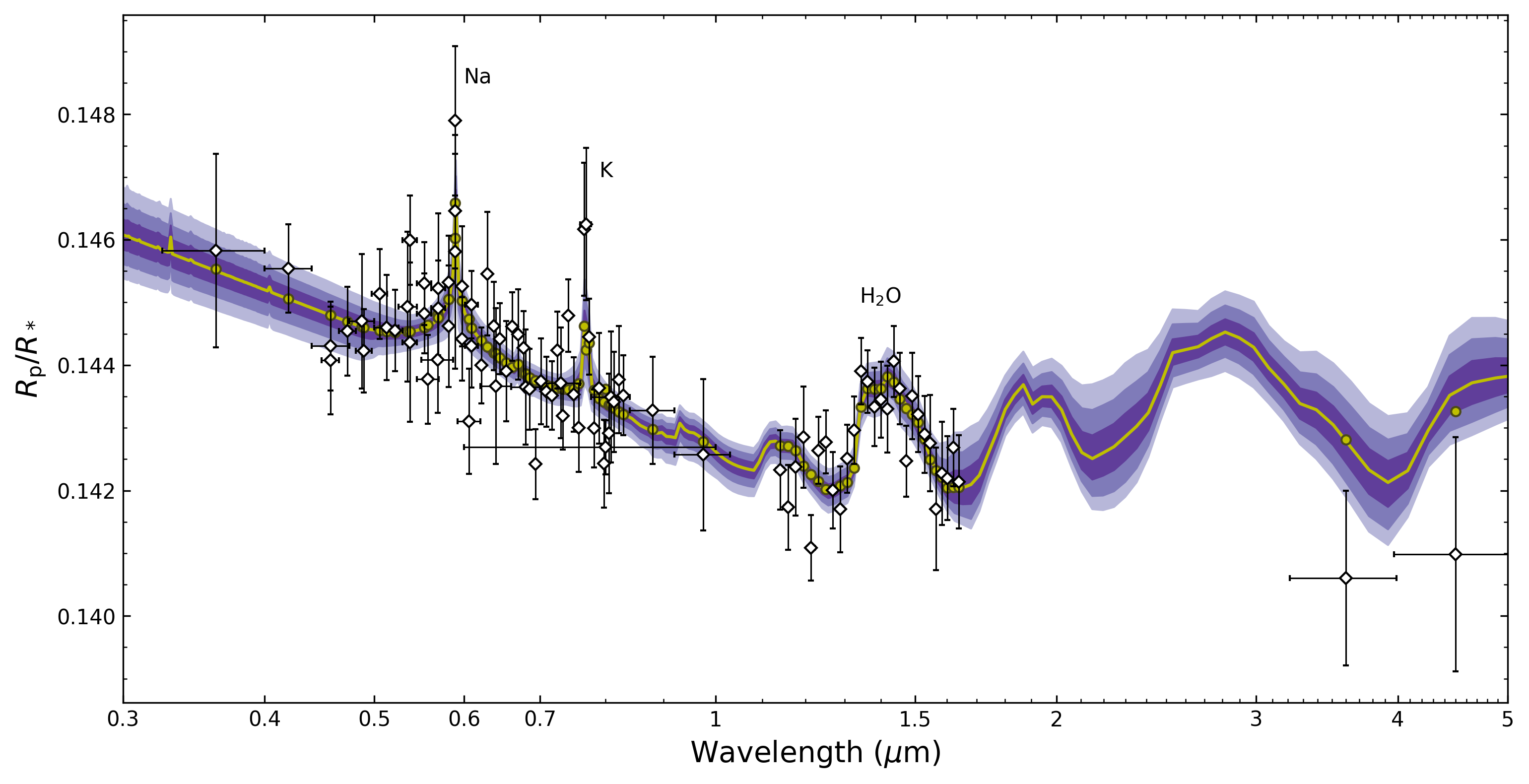}
\vspace*{-5mm}
\caption{The measured \textit{AIT} spot corrected transmission spectrum of WASP-6b (white diamonds) in addition to the best fit \texttt{ARC} retrieval model (yellow line) and its corresponding 1, 2 and 3$\sigma$ bounds (purple shaded regions).}
\label{retrieval_spectrum}
\end{figure*}

\begin{table*} 
\centering 
\begin{tabular}{l c c c c c c c c} 
\hline 
\hline 
Dataset & T$_{\textrm{eq}}$ (K) & log(M$_{\textrm{trace}}$/H) & Radius (R$_\textrm{J}$) & Haze Opacity ln($\frac{\sigma}{\sigma_0}$-1) & log(C/H) & log(O/H) & log(Na/H) & log(K/H)  \\ 
\hline 
Uncorrected & $1312^{\scalebox{0.8}{$+$91}}_{\scalebox{0.8}{$-$89}}$ & $-1.30^{\scalebox{0.8}{$+$0.59}}_{\scalebox{0.8}{$-$0.45}}$ &  $1.140^{\scalebox{0.8}{$+$0.005}}_{\scalebox{0.8}{$-$0.003}}$ &  $3.85^{\scalebox{0.8}{$+$0.59}}_{\scalebox{0.8}{$-$0.83}}$ &  <$$0.26 & $-0.99^{\scalebox{0.8}{$+$0.31}}_{\scalebox{0.8}{$-$0.31}}$ & $1.33^{\scalebox{0.8}{$+$0.42}}_{\scalebox{0.8}{$-$0.67}}$ & $0.22^{\scalebox{0.8}{$+$0.65}}_{\scalebox{0.8}{$-$0.74}}$ \\[1ex] 
$\textit{TESS}$ Corrected & $1202^{\scalebox{0.8}{$+$80}}_{\scalebox{0.8}{$-$74}}$ & $-1.04^{\scalebox{0.8}{$+$0.71}}_{\scalebox{0.8}{$-$0.61}}$ &  $1.133^{\scalebox{0.8}{$+$0.003}}_{\scalebox{0.8}{$-$0.003}}$ &  $3.72^{\scalebox{0.8}{$+$0.69}}_{\scalebox{0.8}{$-$0.62}}$ &  <$$0.26 & $-0.83^{\scalebox{0.8}{$+$0.31}}_{\scalebox{0.8}{$-$0.29}}$ & $1.37^{\scalebox{0.8}{$+$0.38}}_{\scalebox{0.8}{$-$0.48}}$ & $0.44^{\scalebox{0.8}{$+$0.57}}_{\scalebox{0.8}{$-$0.65}}$ \\[1ex] 
$\textit{AIT}$ Corrected & $1199^{\scalebox{0.8}{$+$94}}_{\scalebox{0.8}{$-$80}}$ & $-1.10^{\scalebox{0.8}{$+$0.80}}_{\scalebox{0.8}{$-$0.56}}$ &  $1.132^{\scalebox{0.8}{$+$0.006}}_{\scalebox{0.8}{$-$0.005}}$ &  $3.08^{\scalebox{0.8}{$+$0.89}}_{\scalebox{0.8}{$-$0.92}}$ &  <$$0.64 & $-0.84^{\scalebox{0.8}{$+$0.40}}_{\scalebox{0.8}{$-$0.39}}$ & $0.83^{\scalebox{0.8}{$+$0.67}}_{\scalebox{0.8}{$-$0.80}}$ & $-0.12^{\scalebox{0.8}{$+$0.71}}_{\scalebox{0.8}{$-$0.74}}$ \\[1ex] 
\hline 
\end{tabular} 
\caption{Mean retrieved parameters for the uncorrected and corrected datasets using \texttt{ARC}. All abundances are quoted relative to the solar abundances of \citep{Aspl09} and as the log(C/H) abundances are largely unconstrained, we quote 3$\sigma$ upper limits.} 
\label{retrieval_results} 
\end{table*}

}

\subsection{\texttt{ATMO} Retrieval Modelling}\label{retrieval}
The previously available transmission spectra of WASP-6b has been the subject of multiple retrieval based model analyses thus far. Firstly by \citet{Bars17} who utilize the \texttt{NEMESIS} retrieval code to demonstrate that the atmosphere of WASP-6b is best described by Rayleigh scattering clouds at high altitudes. In addition, \citet{Pinh18} perform a retrieval using the \texttt{AURA} code, demonstrating that the atmosphere of WASP-6b is best described as a combination of the effects of stellar heterogeneity and atmospheric hazes. However, in an effort to fit the widely disparate STIS and \textit{Spitzer} points this retrieval predicts a very low H$_2$O abundance, a claim that has not been possible to verify or refute until the recent acquisition of \textit{HST} WFC3 data from this study. 

\begin{table}
\centering
\begin{tabular}{l c c}
\hline
\hline
Dataset & Na {\sc i} Significance & K {\sc i} significance \\
\hline
Uncorrected & 4.2 $\sigma$ & 3.5 $\sigma$ \\
\textit{TESS} Corrected & 3.9 $\sigma$ & 3.2 $\sigma$ \\
\textit{AIT} Corrected & 3.9 $\sigma$ & 3.4 $\sigma$ \\
\hline
\end{tabular}
\caption{Sigma confidence levels of the Na {\sc i} and K {\sc i} line detections with respect to the model baseline level. }
\label{nak_sigma}
\end{table}

Due to the wealth of new data available with the addition of the FORS2, WFC3 and \textit{TESS} observations we perform our own atmospheric retrieval on the uncorrected and spot corrected datasets using the \texttt{ATMO} Retrieval Code (\texttt{ARC}) which has already been used for a variety of transmission spectra to date \citep{Wake17b, Wake18, Niko18, Spak18, Evan18}. For the retrieval model, the relative elemental abundances for each model were calculated in equilibrium. For each model, equilibrium chemistry was calculated on the fly, using input elemental abundances fit by assuming solar values and we allowed for non-solar scaled elemental compositions by fitting the carbon, oxygen, sodium, and potassium elemental abundances ([C/H], [O/H], [Na/H], [K/H]), which can potentially all be constrained by the transmission spectrum. We fit all remaining species by varying a single quantity for the trace metallicity, [M$_\textrm{trace}$/H]. Throughout this study, all abundances are quoted as [X/H] which is logarithmic relative to the Sun, with all solar abundances taken from \citet{Aspl09}. The resulting chemical network consisted of 175 neutral gas phase species, 93 condensates, and the ionized species e$^{-}$, H$^+$, H$^{-}$, He$^+$, Na$^+$, K$^+$, C$^+$, Ca$^+$, and Si$^+$. By varying both C and O separately, we mitigate several important modelling deficiencies and assumptions compared to varying the C/O ratio as a single parameter \citep{Drum19}. For the spectral synthesis, we included the spectroscopically active molecules of H$_2$, He, H$_2$O, CO$_2$, CO, CH$_4$, NH$_3$, Na, K, Li, TiO, VO, FeH, and Fe. The temperature was assumed to be isothermal, fit with one parameter, and we also included a uniform haze fit with the enhancement factor. A differential-evolution MCMC was used to infer the posterior probability distribution which was then marginalised \citep{East13}, we ran twelve chains each for 30,000 steps, discarding the burn-in before combining them into a single chain. Uniform priors were adopted, with the log$_{10}$ abundances allowed to vary between -12 and -1.3. 

\begin{figure*}
\centering
\includegraphics[width=\textwidth]{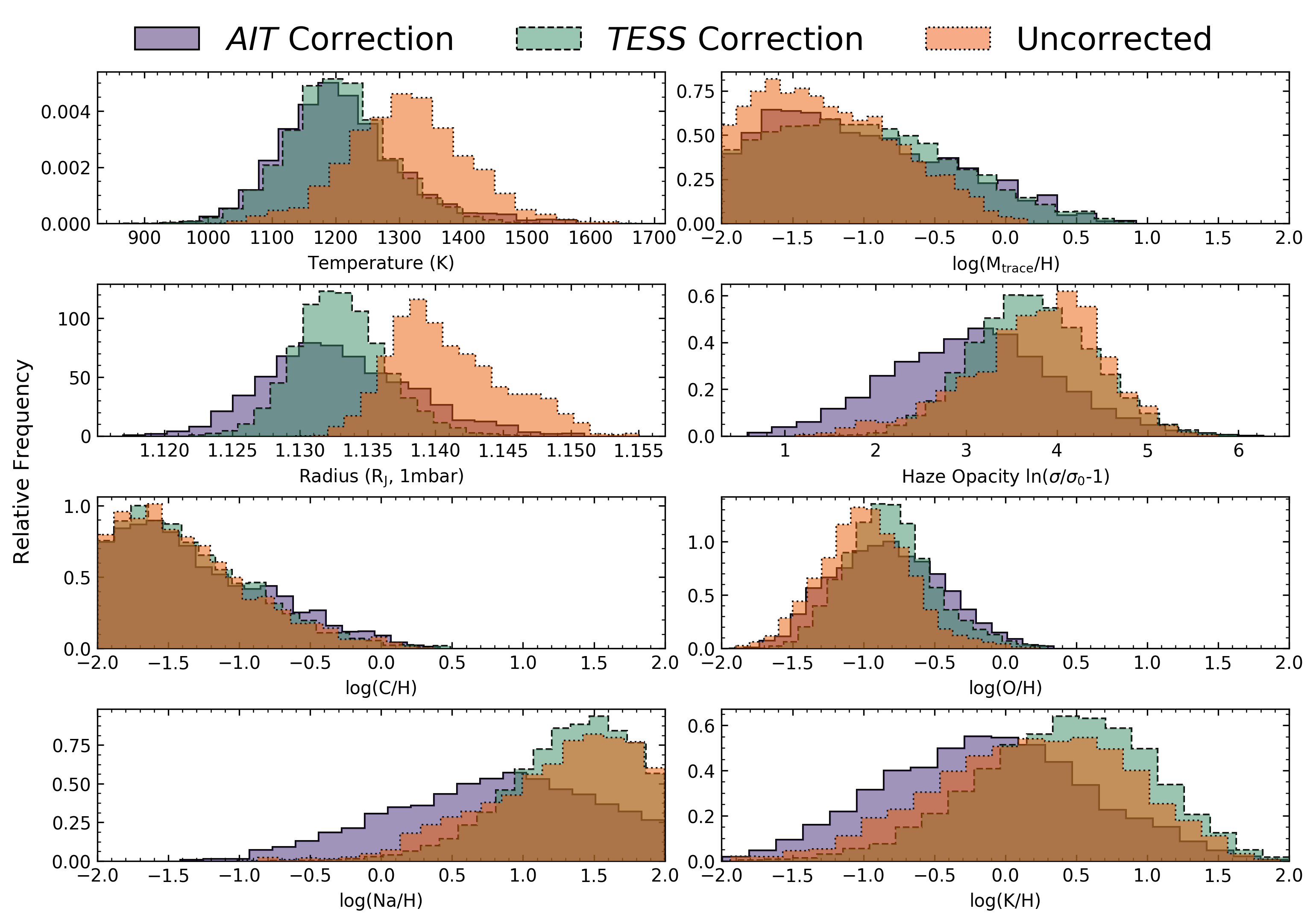}
\vspace*{-5mm}
\caption{Retrieval posteriors from the \texttt{ARC} analysis of the uncorrected (orange, dotted line), \textit{TESS} spot corrected (teal, dashed line) and \textit{AIT} spot corrected (purple, solid line) datasets for WASP-6b. The metallicity and abundances of Na, K, C, and O are given with reference to solar values as taken from \citet{Aspl09}. All distributions have been normalised so that their integral is equal to unity.}
\label{retrieval_posteriors}
\end{figure*}

The resulting best fit retrieval models for the uncorrected, \textit{TESS} corrected, and \textit{AIT} corrected datasets all provide good fits to the data, with $\chi^2$ = 75, 71, and 73 respectively for 86 degrees of freedom. We show a visual representation of the retrieval for the \textit{AIT} corrected dataset in Figure \ref{retrieval_spectrum} and the mean values for each individual retrieval are shown in Table \ref{retrieval_results}. To facilitate comparisons between the uncorrected and corrected datasets we plot the retrieval posteriors for each together in Figure \ref{retrieval_posteriors}. As with the forward model grid fits shown in Section \ref{goyalmodels} there are clear differences between the uncorrected and spot corrected datasets, particularly for the temperature, radius, and haze opacity. The difference in radius is a natural result of performing the spot correction, as this results is a wavelength dependent shift in the transmission baseline to lower transit depths. Given the square root of the transit depth $\delta$ = $R_\textrm{p}/R_*$, and that the stellar radius is fixed during the retrieval, any decrease in the transit depth will subsequently produce a decrease in the estimated planetary radius. In a similar fashion to the forward model grid fits, the highest temperatures and highest levels of haze opacity are favoured by the uncorrected dataset, the cause of which being the degeneracy between these properties and the effects of stellar heterogeneity on the uncorrected transmission spectrum. Upon performing a spot correction, the best fit temperature and haze opacity falls as the gradient of the optical slope has been reduced. However at least a moderate amount of haze is still required irrespective of spot correction. 

Due to the freedom of the retrieval analyses we were also able to investigate the specific elemental abundances inferred from the measured transmission spectra. Firstly, as the C, O, Na, and K abundances were fit independently throughout the retrieval analysis the measured metallicity only encompasses the other elemental constituents of the atmosphere. The sub-solar metallicity measured across all retrieval analyses therefore show that no other substantial absorber is required to fit the measured transmission spectra. The \textit{Spitzer} data points are the only observations sensitive to carbon bearing species in the atmosphere such as CH$_4$, CO and CO$_2$, however, given their non-negligible uncertainties and minimal relative offset the retrieved carbon abundance is largely unconstrained and merely represents an upper limit. This is true across all datasets as the addition of a stellar heterogeneity correction has a marginal effect towards the infrared. We constrain the carbon abundance to sub-solar at 3$\sigma$ for the uncorrected and \textit{AIT} datasets, and at 2$\sigma$ for the \textit{TESS} dataset. Our limit on the carbon abundance suggests that H$_2$O is the primary oxygen-bearing species, and from the observed feature we constrain the oxygen abundance to a sub-solar value, irrespective of a spot correction. For the best fit retrieval model to the \textit{AIT} corrected dataset our oxygen abundance corresponds to a water abundance of log(H$_2$O) = -4.87. Given the lack of WFC3 data available to previous studies of WASP-6b this water abundance is the first to be informed by an observed water absorption feature in transmission. Furthermore, given the extensive optical data from FORS2 and STIS, this result is robust to previously observed degeneracies of water abundance and reference pressure \citep{Grif14, Pinh18}. Contrasting to oxygen, the Na and K abundances are relaxed to lower values following the application of a spot correction as the lone Na {\sc i} and K {\sc i} absorption features lie in the optical region where stellar heterogeneity has a significant effect on the observed slope. Upon a reduction in the slope opacity, these abundances must necessarily drop to fit the observed data. Specifically for the \textit{AIT} correction, we see variations in sodium of super-solar, [Na/H] = $1.33^{+0.42}_{-0.67}$, to solar/super-solar, [Na/H] = $0.83^{+0.67}_{-0.80}$, and potassium of solar/super-solar, [K/H] = $ 0.22^{+0.65}_{-0.74}$, to sub-solar/solar, [K/H] = $ -0.12^{+0.71}_{-0.74}$.  Given the measurement precision we cannot explicitly quantify the impact of the correction as both the [Na/H] and [K/H] abundances lie within 1$\sigma$ of their inferred uncorrected abundances. Despite this, the broader shifts of their full retrieved distributions (Figure \ref{retrieval_posteriors}) indicate that neglecting to account for the affects of stellar heterogeneity in future, higher precision, observations may lead to strictly incorrect determinations of their abundances.  

As the metallicity we retrieve excludes C, O, Na, and K, we cannot perform a comparison to the [M/H] distributions obtained as part of the forward model analysis in Section \ref{goyalmodels}. However, comparing the retrieved [O/H] to the forward model [M/H] we see similar distributions indicating a sub-solar metallicity. Additionally, whilst the slightly super-solar abundances of [Na/H] and [K/H] do not completely agree with the sub-solar [M/H] the large uncertainties of these distributions indicate that an overall sub-solar metallicity cannot be ruled out.

\subsection{WASP-6b In Context}\label{context}
Our determined, spot corrected, oxygen abundance of [O/H] = $-0.84^{+0.40}_{-0.39}$ and sodium abundance of [Na/H] = $0.83^{+0.67}_{-0.80}$ are slightly disparate to the determined sub-solar metallicity of the host star of [Fe/H] = $-0.15 \pm 0.09$ \citep{Doyl13}, whilst the potassium abundance is in good agreement at [K/H] = $ -0.12^{+0.71}_{-0.74}$. Variations in these elemental abundances relative to the host star could be indicative of formation history (e.g. \citealt{Ober11}), however in the case of WASP-6b the current uncertainties are not sufficiently constrained to make such determinations, with all values lying within 2$\sigma$ of the host star metallicity. Further observations of the atmosphere of WASP-6b will be necessary to provide more detailed constraints on these elemental abundances. In particular, due to the presence of carbon-bearing molecular features beyond 2 $\mu$m such as CO, CO$_2$, and CH$_4$, spectroscopic observations with the upcoming \textit{James Webb Space Telescope} (\textit{JWST}) will provide stronger constraints on the carbon abundance, of which this study could only provide an upper limit. This in turn will enable robust constraints on the C/O ratio and progress our understanding of the formation history of WASP-6b. 

Irrespective of the application of a stellar heterogeneity correction, both the forward and retrieval models require some level of haze opacity enhancement in order to describe the steep optical slope of the transmission spectrum. In the context of hot Jupiter atmospheres, this haze is often thought of as either photochemically produced, or condensate dust, scattering species within the atmosphere \citep{Marl13}. In the case of the condensate species it is thought that the lofting of particles from deeper atmospheric cloud decks can serve to populate the upper atmosphere and lead to the observed scattering we see (e.g. \citealt{ Parm13}). Despite this, the most recent simulations of condensate particle formation in the atmosphere of the hot Jupiter HD 189733b \citep{Bouc05} fail to fully reproduce its observed scattering slope \citep{Lee17, Powe18}. At the temperature of WASP-6b, generation of hydrocarbons through photochemistry was initially thought to be inhibited \citep{Lian04} and whilst sulphur photochemistry may play a role \citep{Zahn09}, it primarily induces a scattering slope below 0.45 $\mu$m, whereas the observed slope of WASP-6b extends further into the optical. However, recent laboratory experiments have shown that hydrocarbons may form not just in cool exoplanet atmospheres \citep{Hors18, He18}, but also in hot atmospheres beyond 1000 K with a sufficiently high [C/O] = 1 \citep{Fleu19}, a possibility our observations cannot definitively rule out. Additionally, the effects of wind-driven chemistry act to homogenise the atmospheres of tidally locked hot Jupiters such as WASP-6b and can lead to significant increases in the abundance of CH$_4$ compared to standard equilibrium models \citep{Drum18a, Drum18b}. Given photolysis of CH$_4$ can drive the formation of haze precursors \citep{Lavv08}, this increase in abundance may naturally lead to their more efficient production. Furthermore, of the well characterised hot Jupiter atmospheres, WASP-6b and HD 189733b present an interesting comparison as they have similar temperatures, both orbit active stars ($\log(R^\prime_{HK})$ = -4.511 and -4.501 respectively), and both exhibit strong haze scattering slopes across the optical \citep{Sing16}. Recent simulations of HD 189733b by \citet{Lavv17} have shown that the formation of photochemical haze "soots" higher in the atmosphere are not excluded and can match its observed transmission spectrum. Moreover, the increased UV flux that these two planets are subject to due to their large host star activity levels is likely acting to enhance the rate of photochemical haze production in their atmospheres \citep{Kawa19}. Possible evidence to this conclusion is seen in the potential trend towards stronger scattering haze signatures with reducing $\log(R^\prime_{HK})$ (increasing activity) observed in the hot Jupiter population study of \citet{Sing16}. An exact determination of whether the haze produced in the atmosphere of WASP-6b is of photochemical origin, condensate dust origin, or a combination of the two, was not possible as part of this study due to their similar opacities at the wavelengths of these observations (e.g. \citealt{Niko15}). In future analyses however, the relative contributions of both photochemical and condensate haze components should be considered in order to describe this observed scattering. 

Amongst the population of spectroscopically studied exoplanets, the atmosphere of WASP-6b is one of the haziest. Previous studies of its atmosphere predicted a small \citep{Niko15, Sing16} amplitude H$_2$O feature at 1.4 $\mu$m, however the feature observed as part of this study is slightly larger than anticipated. This increase is likely due to the seemingly small \textit{Spitzer} transit depths biasing the model estimates prior to the acquisition of the FORS2 and WFC3 datasets. To quantify the size of the H$_2$O feature relative to an assumed clear atmosphere for WASP-6b we determine the scaled amplitude of the water feature following \citet{Wake19}. Specifically, we begin by taking a clear atmosphere forward model from the grid used throughout this paper \citep{Goya18} with: the equilibrium temperature of WASP-6b, solar metallicity, solar C/O ratio, and no haze or cloud opacity components. We then scale this model to fit the data using a model defined as $S1 = (S0\times p_0) + p_1$, where $S0$ is the clear atmosphere model, $p_0$ is the model amplitude scale factor and $p_1$ is a baseline offset. For the \textit{AIT} corrected dataset we determine $p_0$ = 64 $\pm$ 12 per cent, in contrast to the median amplitude across the observed population of $p_0$ = 33 $\pm$ 24 per cent \citep{Wake19}. These new observations indicate that despite the presence of haze, WASP-6b remains a favourable target for atmospheric characterisation, particularly with \textit{JWST}. This potential for \textit{JWST} to characterise hazy hot Jupiters such as WASP-6b is in contrast to those who exhibit flat, cloudy spectra such as WASP-31b \citep{Gibs17} and WASP-101b \citep{Wake17}.

\section{Conclusions}\label{conc}
We present the most complete optical to infrared transmission spectrum of the hot Jupiter WASP-6b to date utilising new observations performed with \textit{HST} WFC3, \textit{VLT} FORS2 and \textit{TESS} in addition to reanalysed existing \textit{HST} STIS and \textit{Spitzer} IRAC data. The impact of host star heterogeneity on the transmission spectrum was investigated and we correct the observed light curves to account for these effects under different assumptions for the level of stellar activity. All reduced transmission spectra then undergo a retrieval analysis fitting in addition to being fit to a grid of forward atmospheric models.

Across all datasets we find clear evidence for Na {\sc i}, K {\sc i} and H$_2$O within the atmosphere of WASP-6b in addition to a steep increase in transit depth towards the optical. After applying both forward model and retrieval analyses we find that at least a moderate haze enhancement is required to describe the optical slope, however when neglecting even a conservative stellar heterogeneity correction, higher and potentially erroneous haze enhancement factors are more preferred. An analogous effect is also seen in the estimated temperature, where higher and potentially unphysical temperatures are preferred when there is no stellar heterogeneity correction. Both of these effects likely stem from the degeneracy of these properties and the impact of stellar heterogeneity towards increasing the optical slope of the transmission spectrum. 

Whilst the precision of current observations is not sufficient to definitively estimate the impact of stellar heterogeneity on the transmission spectrum of WASP-6b, the parameter differences observed upon the application of a stellar heterogeneity correction indicate that its effect should not be neglected for future observations of exoplanetary atmospheres around moderately active stars. Despite the presence of haze in its atmosphere, WASP-6b remains a favourable target for further characterisation. Contemporaneous and broader wavelength measurements of its transmission spectrum with missions such as \textit{JWST} will enable a more detailed characterisation of its atmosphere in addition to the precisely determining the effects stellar heterogeneity has on its appearance.

\section*{Acknowledgements}
This work is based on observations collected at the European Organization for Astronomical Research in the Southern Hemisphere under European Southern Observatory programme 196.C-0765 in addition to observations associated with program GO-14767 made with the NASA/ESA \textit{Hubble Space Telescope} that were obtained at the Space Telescope Science Institute, which is operated by the Association of Universities for Research in Astronomy, Inc, under NASA contract NAS 5-2655. This work has made use of data from the European Space Agency (ESA) mission {\it Gaia} (\url{https://www.cosmos.esa.int/gaia}), processed by the {\it Gaia} Data Processing and Analysis Consortium (DPAC, \url{https://www.cosmos.esa.int/web/gaia/dpac/consortium}). Funding for the DPAC has been provided by national institutions, in particular the institutions participating in the {\it Gaia} Multilateral Agreement. This publication makes use of data products from the Two Micron All Sky Survey, which is a joint project of the University of Massachusetts and the Infrared Processing and Analysis Center / California Institute of Technology, funded by the National Aeronautics and Space Administration and the National Science Foundation. Based on observations made with the NASA Galaxy Evolution Explorer. \textit{GALEX} is operated for NASA by the California Institute of Technology under NASA contract NAS5-98034. ALC and SM are funded by a UK Science and Technology Facilities Council (STFC) studentship. MKA acknowledges support by the National Science Foundation through a Graduate Research Fellowship. HRW acknowledges support from the Giacconi Prize Fellowship at STScI, operated by AURA. This research has made use of NASAs Astrophysics Data System and the Python modules \texttt{NumPy}, \texttt{Matplotlib}, and \texttt{SciPy}.



\FloatBarrier
\bibliographystyle{mnras}
\bibliography{main_wasp6} 

\bsp	
\label{lastpage}


\appendix
\section{Archival Light Curve Fits}
\begin{figure*}
\centering
\includegraphics[width=\textwidth]{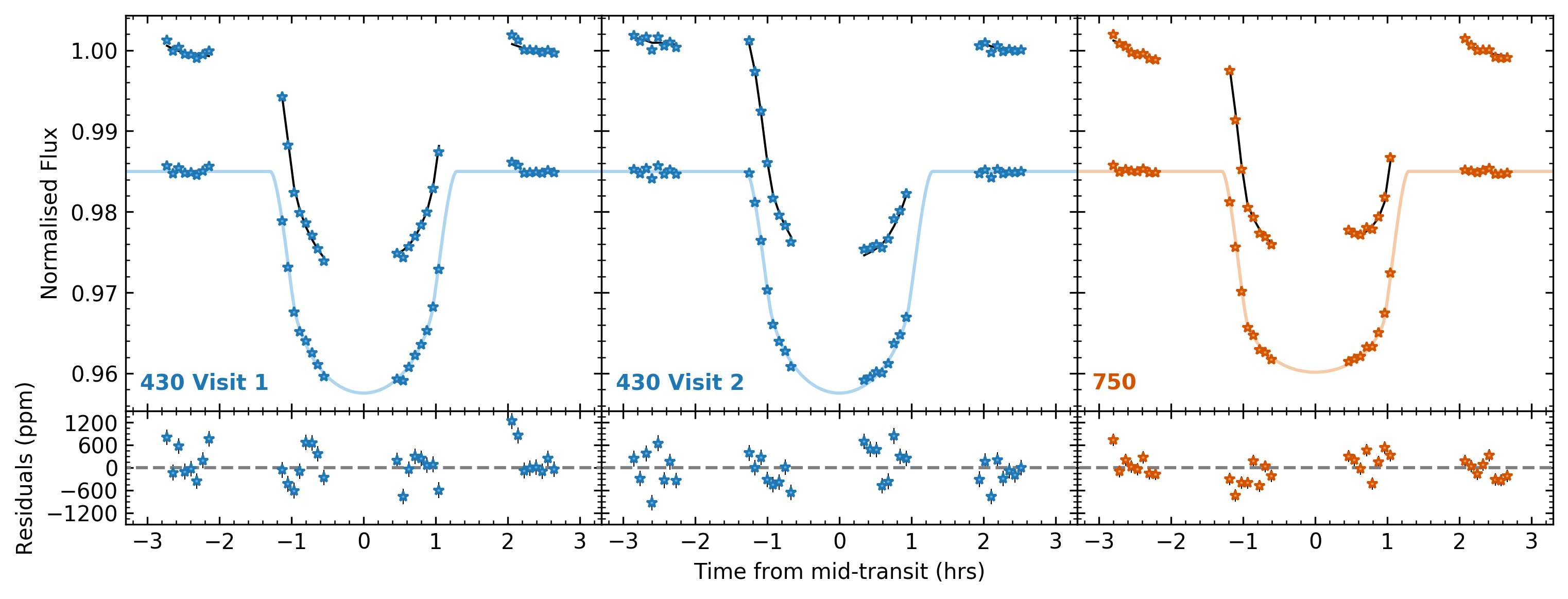}
\vspace*{-5mm}
\caption{Normalised white light curves and residuals of WASP-6b for the STIS 430 and STIS 750 grism observations as labelled. In each panel the upper light curve is the raw flux with black line indicating the GP transit plus systematic model fit, whilst the lower is the light curve after removal of the GP systematic component overplotted with the best fitting transit model from \citet{Mand02}. All lower panels display residuals following subtraction of the corresponding corrected light curves by their respective best fitting models.}
\label{stis_wlcs}
\end{figure*}

\begin{figure*}
\centering
\includegraphics[width=\textwidth]{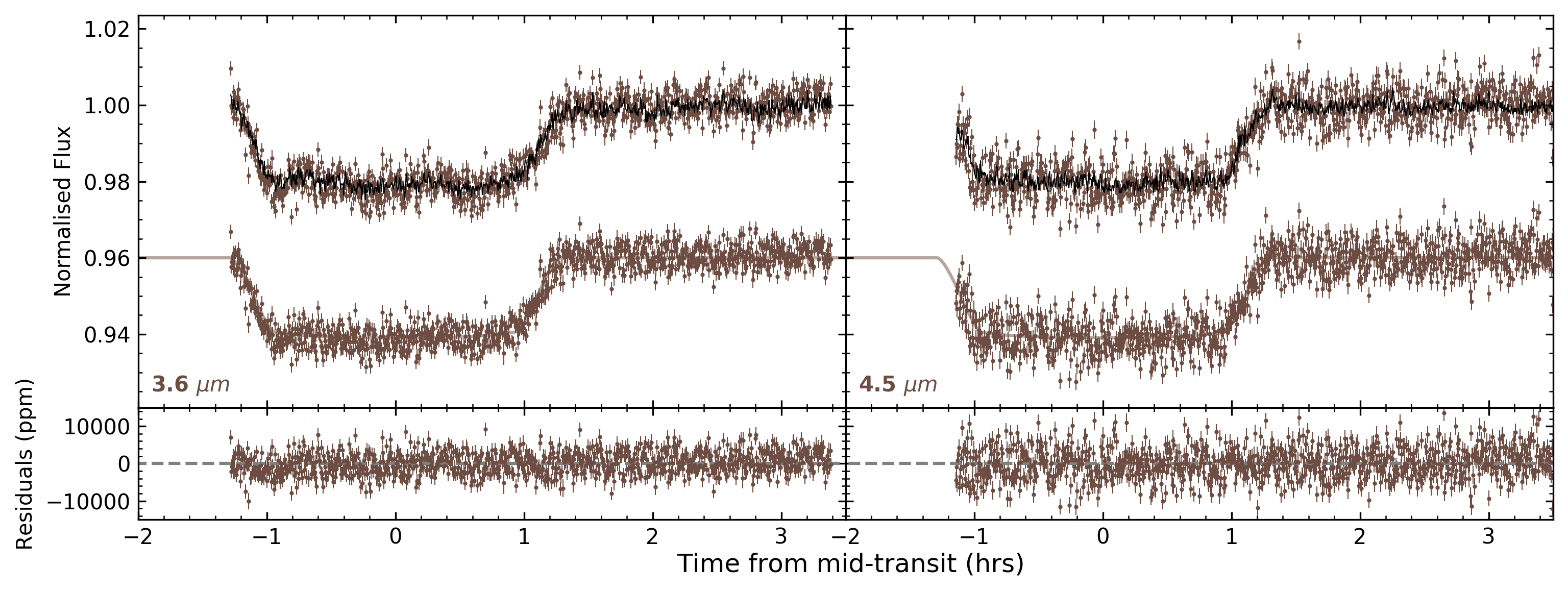}
\vspace*{-5mm}
\caption{As in Figure \ref{stis_wlcs} but for the \textit{Spitzer} IRAC observations as labelled.}
\label{spitzer_wlcs}
\end{figure*}

\begin{figure*}
\centering
\includegraphics[width=\textwidth]{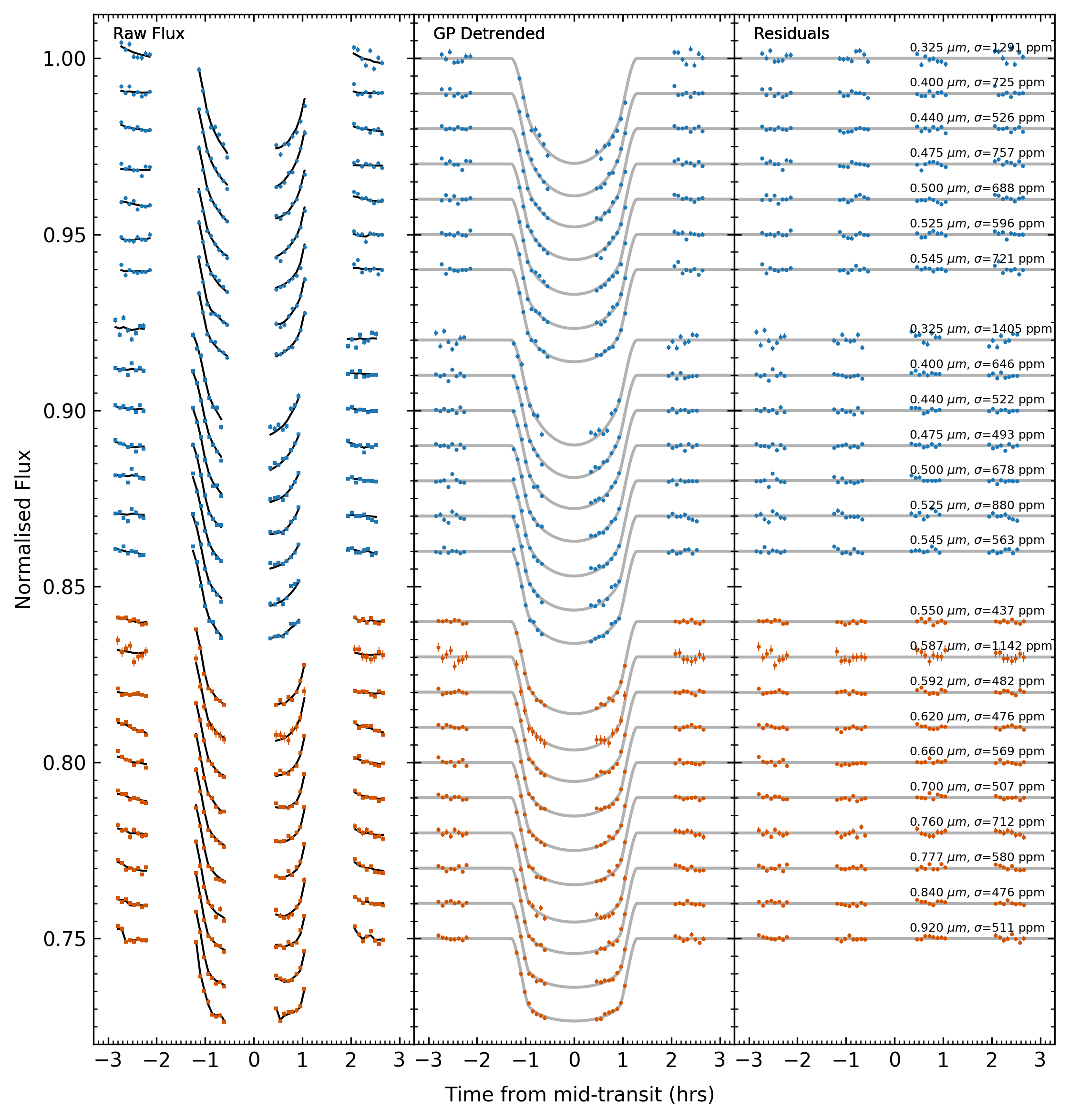}
\vspace*{-5mm}
\caption{Normalised spectrophotometric light curves for both STIS 430 datasets (top, middle groups) and the STIS 750 dataset (bottom group) of WASP-6b, light curves are offset from one another by an arbitrary constant. \textit{Left:} Raw extracted light curves with black lines indicating the GP transit plus systematic model fit. \textit{Centre:} Light curves after removal of GP systematic component. The best fitting transit models from \citet{Mand02} are displayed in grey. \textit{Right:} Residuals following subtraction of best fitting model.}
\label{stis_slcs}
\end{figure*}

\section{Spectrophotometric Light Curve Fits}
\begin{table*} 
\centering 
\fontsize{7}{9}\selectfont 
\begin{tabular}{l c c c c c c c} 
\hline 
\hline 
\vspace{-8pt} \\ 
Wavelength ($\mu$m) & $R_\textrm{p}/R_*$ & $R_\textrm{p}/R_{*,\textrm{\textit{TESS}}}$ & $R_\textrm{p}/R_{*,\textrm{\textit{AIT}}}$ & $c_1$ & $c_2$ & $c_3$ & $c_4$  \\ 
\hline 
\vspace{-5pt} \\ 
\textbf{FORS2 G600B} & & & & & & & \\[1ex] 
$0.4493$--$0.4653$ & $0.14563^{\scalebox{0.8}{$+$0.00084}}_{\scalebox{0.8}{$-$0.00085}}$ & $0.14464^{\scalebox{0.8}{$+$0.00085}}_{\scalebox{0.8}{$-$0.00089}}$ & $0.14408^{\scalebox{0.8}{$+$0.00084}}_{\scalebox{0.8}{$-$0.00087}}$ & $0.563^{\scalebox{0.8}{$+$0.020}}_{\scalebox{0.8}{$-$0.020}}$ & $0.1783$ & - & - \\[1ex] 
$0.4653$--$0.4813$ & $0.14611^{\scalebox{0.8}{$+$0.00069}}_{\scalebox{0.8}{$-$0.00070}}$ & $0.14505^{\scalebox{0.8}{$+$0.00070}}_{\scalebox{0.8}{$-$0.00071}}$ & $0.14454^{\scalebox{0.8}{$+$0.00070}}_{\scalebox{0.8}{$-$0.00071}}$ & $0.547^{\scalebox{0.8}{$+$0.017}}_{\scalebox{0.8}{$-$0.017}}$ & $0.1882$ & - & - \\[1ex] 
$0.4813$--$0.4973$ & $0.14578^{\scalebox{0.8}{$+$0.00067}}_{\scalebox{0.8}{$-$0.00066}}$ & $0.14474^{\scalebox{0.8}{$+$0.00067}}_{\scalebox{0.8}{$-$0.00067}}$ & $0.14422^{\scalebox{0.8}{$+$0.00067}}_{\scalebox{0.8}{$-$0.00066}}$ & $0.507^{\scalebox{0.8}{$+$0.016}}_{\scalebox{0.8}{$-$0.016}}$ & $0.2015$ & - & - \\[1ex] 
$0.4973$--$0.5133$ & $0.14671^{\scalebox{0.8}{$+$0.00070}}_{\scalebox{0.8}{$-$0.00071}}$ & $0.14567^{\scalebox{0.8}{$+$0.00071}}_{\scalebox{0.8}{$-$0.00072}}$ & $0.14514^{\scalebox{0.8}{$+$0.00071}}_{\scalebox{0.8}{$-$0.00072}}$ & $0.520^{\scalebox{0.8}{$+$0.017}}_{\scalebox{0.8}{$-$0.017}}$ & $0.1951$ & - & - \\[1ex] 
$0.5133$--$0.5293$ & $0.14609^{\scalebox{0.8}{$+$0.00065}}_{\scalebox{0.8}{$-$0.00065}}$ & $0.14508^{\scalebox{0.8}{$+$0.00065}}_{\scalebox{0.8}{$-$0.00066}}$ & $0.14456^{\scalebox{0.8}{$+$0.00064}}_{\scalebox{0.8}{$-$0.00066}}$ & $0.464^{\scalebox{0.8}{$+$0.016}}_{\scalebox{0.8}{$-$0.016}}$ & $0.2298$ & - & - \\[1ex] 
$0.5293$--$0.5453$ & $0.14757^{\scalebox{0.8}{$+$0.00074}}_{\scalebox{0.8}{$-$0.00073}}$ & $0.14655^{\scalebox{0.8}{$+$0.00074}}_{\scalebox{0.8}{$-$0.00073}}$ & $0.14599^{\scalebox{0.8}{$+$0.00073}}_{\scalebox{0.8}{$-$0.00070}}$ & $0.426^{\scalebox{0.8}{$+$0.018}}_{\scalebox{0.8}{$-$0.018}}$ & $0.2111$ & - & - \\[1ex] 
$0.5453$--$0.5613$ & $0.14631^{\scalebox{0.8}{$+$0.00064}}_{\scalebox{0.8}{$-$0.00063}}$ & $0.14531^{\scalebox{0.8}{$+$0.00064}}_{\scalebox{0.8}{$-$0.00063}}$ & $0.14483^{\scalebox{0.8}{$+$0.00065}}_{\scalebox{0.8}{$-$0.00062}}$ & $0.427^{\scalebox{0.8}{$+$0.015}}_{\scalebox{0.8}{$-$0.016}}$ & $0.2332$ & - & - \\[1ex] 
$0.5613$--$0.5773$ & $0.14639^{\scalebox{0.8}{$+$0.00076}}_{\scalebox{0.8}{$-$0.00076}}$ & $0.14542^{\scalebox{0.8}{$+$0.00075}}_{\scalebox{0.8}{$-$0.00077}}$ & $0.14491^{\scalebox{0.8}{$+$0.00074}}_{\scalebox{0.8}{$-$0.00077}}$ & $0.395^{\scalebox{0.8}{$+$0.019}}_{\scalebox{0.8}{$-$0.019}}$ & $0.2347$ & - & - \\[1ex] 
$0.5773$--$0.5853$ & $0.14603^{\scalebox{0.8}{$+$0.00096}}_{\scalebox{0.8}{$-$0.00098}}$ & $0.14509^{\scalebox{0.8}{$+$0.00096}}_{\scalebox{0.8}{$-$0.00098}}$ & $0.14462^{\scalebox{0.8}{$+$0.00095}}_{\scalebox{0.8}{$-$0.00099}}$ & $0.437^{\scalebox{0.8}{$+$0.024}}_{\scalebox{0.8}{$-$0.025}}$ & $0.2348$ & - & - \\[1ex] 
$0.5853$--$0.5933$ & $0.14938^{\scalebox{0.8}{$+$0.00125}}_{\scalebox{0.8}{$-$0.00116}}$ & $0.14836^{\scalebox{0.8}{$+$0.00125}}_{\scalebox{0.8}{$-$0.00116}}$ & $0.14790^{\scalebox{0.8}{$+$0.00123}}_{\scalebox{0.8}{$-$0.00115}}$ & $0.355^{\scalebox{0.8}{$+$0.030}}_{\scalebox{0.8}{$-$0.036}}$ & $0.2579$ & - & - \\[1ex] 
$0.5933$--$0.6013$ & $0.14671^{\scalebox{0.8}{$+$0.00095}}_{\scalebox{0.8}{$-$0.00093}}$ & $0.14578^{\scalebox{0.8}{$+$0.00097}}_{\scalebox{0.8}{$-$0.00094}}$ & $0.14526^{\scalebox{0.8}{$+$0.00096}}_{\scalebox{0.8}{$-$0.00095}}$ & $0.342^{\scalebox{0.8}{$+$0.024}}_{\scalebox{0.8}{$-$0.025}}$ & $0.2477$ & - & - \\[1ex] 
$0.6013$--$0.6173$ & $0.14576^{\scalebox{0.8}{$+$0.00066}}_{\scalebox{0.8}{$-$0.00066}}$ & $0.14481^{\scalebox{0.8}{$+$0.00068}}_{\scalebox{0.8}{$-$0.00066}}$ & $0.14431^{\scalebox{0.8}{$+$0.00067}}_{\scalebox{0.8}{$-$0.00066}}$ & $0.384^{\scalebox{0.8}{$+$0.017}}_{\scalebox{0.8}{$-$0.017}}$ & $0.2367$ & - & - \\[1ex] 
\textbf{FORS2 G600RI} & & & & & & & \\[1ex] 
$0.5293$--$0.5453$ & $0.14587^{\scalebox{0.8}{$+$0.00128}}_{\scalebox{0.8}{$-$0.00128}}$ & $0.14492^{\scalebox{0.8}{$+$0.00128}}_{\scalebox{0.8}{$-$0.00127}}$ & $0.14436^{\scalebox{0.8}{$+$0.00127}}_{\scalebox{0.8}{$-$0.00127}}$ & $0.469^{\scalebox{0.8}{$+$0.031}}_{\scalebox{0.8}{$-$0.032}}$ & $0.2113$ & - & - \\[1ex] 
$0.5453$--$0.5613$ & $0.14682^{\scalebox{0.8}{$+$0.00061}}_{\scalebox{0.8}{$-$0.00073}}$ & $0.14581^{\scalebox{0.8}{$+$0.00062}}_{\scalebox{0.8}{$-$0.00072}}$ & $0.14530^{\scalebox{0.8}{$+$0.00062}}_{\scalebox{0.8}{$-$0.00070}}$ & $0.422^{\scalebox{0.8}{$+$0.017}}_{\scalebox{0.8}{$-$0.016}}$ & $0.2332$ & - & - \\[1ex] 
$0.5613$--$0.5773$ & $0.14670^{\scalebox{0.8}{$+$0.00115}}_{\scalebox{0.8}{$-$0.00125}}$ & $0.14573^{\scalebox{0.8}{$+$0.00115}}_{\scalebox{0.8}{$-$0.00130}}$ & $0.14523^{\scalebox{0.8}{$+$0.00114}}_{\scalebox{0.8}{$-$0.00124}}$ & $0.397^{\scalebox{0.8}{$+$0.029}}_{\scalebox{0.8}{$-$0.032}}$ & $0.2346$ & - & - \\[1ex] 
$0.5773$--$0.5853$ & $0.14678^{\scalebox{0.8}{$+$0.00074}}_{\scalebox{0.8}{$-$0.00076}}$ & $0.14581^{\scalebox{0.8}{$+$0.00074}}_{\scalebox{0.8}{$-$0.00076}}$ & $0.14532^{\scalebox{0.8}{$+$0.00074}}_{\scalebox{0.8}{$-$0.00075}}$ & $0.420^{\scalebox{0.8}{$+$0.020}}_{\scalebox{0.8}{$-$0.020}}$ & $0.2348$ & - & - \\[1ex] 
$0.5853$--$0.5933$ & $0.14790^{\scalebox{0.8}{$+$0.00090}}_{\scalebox{0.8}{$-$0.00094}}$ & $0.14694^{\scalebox{0.8}{$+$0.00089}}_{\scalebox{0.8}{$-$0.00092}}$ & $0.14646^{\scalebox{0.8}{$+$0.00089}}_{\scalebox{0.8}{$-$0.00093}}$ & $0.393^{\scalebox{0.8}{$+$0.024}}_{\scalebox{0.8}{$-$0.025}}$ & $0.2582$ & - & - \\[1ex] 
$0.5933$--$0.6013$ & $0.14586^{\scalebox{0.8}{$+$0.00067}}_{\scalebox{0.8}{$-$0.00066}}$ & $0.14491^{\scalebox{0.8}{$+$0.00067}}_{\scalebox{0.8}{$-$0.00069}}$ & $0.14442^{\scalebox{0.8}{$+$0.00066}}_{\scalebox{0.8}{$-$0.00067}}$ & $0.408^{\scalebox{0.8}{$+$0.018}}_{\scalebox{0.8}{$-$0.018}}$ & $0.2476$ & - & - \\[1ex] 
$0.6013$--$0.6173$ & $0.14638^{\scalebox{0.8}{$+$0.00055}}_{\scalebox{0.8}{$-$0.00052}}$ & $0.14545^{\scalebox{0.8}{$+$0.00055}}_{\scalebox{0.8}{$-$0.00052}}$ & $0.14497^{\scalebox{0.8}{$+$0.00054}}_{\scalebox{0.8}{$-$0.00053}}$ & $0.380^{\scalebox{0.8}{$+$0.014}}_{\scalebox{0.8}{$-$0.015}}$ & $0.2367$ & - & - \\[1ex] 
$0.6173$--$0.6253$ & $0.14542^{\scalebox{0.8}{$+$0.00062}}_{\scalebox{0.8}{$-$0.00061}}$ & $0.14450^{\scalebox{0.8}{$+$0.00062}}_{\scalebox{0.8}{$-$0.00061}}$ & $0.14400^{\scalebox{0.8}{$+$0.00061}}_{\scalebox{0.8}{$-$0.00060}}$ & $0.376^{\scalebox{0.8}{$+$0.017}}_{\scalebox{0.8}{$-$0.017}}$ & $0.2431$ & - & - \\[1ex] 
$0.6253$--$0.6333$ & $0.14688^{\scalebox{0.8}{$+$0.00100}}_{\scalebox{0.8}{$-$0.00093}}$ & $0.14591^{\scalebox{0.8}{$+$0.00100}}_{\scalebox{0.8}{$-$0.00092}}$ & $0.14546^{\scalebox{0.8}{$+$0.00102}}_{\scalebox{0.8}{$-$0.00094}}$ & $0.334^{\scalebox{0.8}{$+$0.026}}_{\scalebox{0.8}{$-$0.030}}$ & $0.2495$ & - & - \\[1ex] 
$0.6333$--$0.6413$ & $0.14598^{\scalebox{0.8}{$+$0.00072}}_{\scalebox{0.8}{$-$0.00068}}$ & $0.14508^{\scalebox{0.8}{$+$0.00074}}_{\scalebox{0.8}{$-$0.00068}}$ & $0.14462^{\scalebox{0.8}{$+$0.00071}}_{\scalebox{0.8}{$-$0.00070}}$ & $0.359^{\scalebox{0.8}{$+$0.019}}_{\scalebox{0.8}{$-$0.021}}$ & $0.2513$ & - & - \\[1ex] 
$0.6413$--$0.6493$ & $0.14583^{\scalebox{0.8}{$+$0.00056}}_{\scalebox{0.8}{$-$0.00058}}$ & $0.14488^{\scalebox{0.8}{$+$0.00057}}_{\scalebox{0.8}{$-$0.00057}}$ & $0.14442^{\scalebox{0.8}{$+$0.00056}}_{\scalebox{0.8}{$-$0.00057}}$ & $0.319^{\scalebox{0.8}{$+$0.016}}_{\scalebox{0.8}{$-$0.016}}$ & $0.2539$ & - & - \\[1ex] 
$0.6493$--$0.6573$ & $0.14529^{\scalebox{0.8}{$+$0.00083}}_{\scalebox{0.8}{$-$0.00077}}$ & $0.14439^{\scalebox{0.8}{$+$0.00086}}_{\scalebox{0.8}{$-$0.00080}}$ & $0.14390^{\scalebox{0.8}{$+$0.00083}}_{\scalebox{0.8}{$-$0.00077}}$ & $0.279^{\scalebox{0.8}{$+$0.022}}_{\scalebox{0.8}{$-$0.027}}$ & $0.3395$ & - & - \\[1ex] 
$0.6573$--$0.6653$ & $0.14600^{\scalebox{0.8}{$+$0.00054}}_{\scalebox{0.8}{$-$0.00051}}$ & $0.14508^{\scalebox{0.8}{$+$0.00056}}_{\scalebox{0.8}{$-$0.00052}}$ & $0.14461^{\scalebox{0.8}{$+$0.00056}}_{\scalebox{0.8}{$-$0.00053}}$ & $0.314^{\scalebox{0.8}{$+$0.014}}_{\scalebox{0.8}{$-$0.015}}$ & $0.2562$ & - & - \\[1ex] 
$0.6653$--$0.6733$ & $0.14586^{\scalebox{0.8}{$+$0.00074}}_{\scalebox{0.8}{$-$0.00071}}$ & $0.14493^{\scalebox{0.8}{$+$0.00075}}_{\scalebox{0.8}{$-$0.00070}}$ & $0.14449^{\scalebox{0.8}{$+$0.00073}}_{\scalebox{0.8}{$-$0.00071}}$ & $0.321^{\scalebox{0.8}{$+$0.020}}_{\scalebox{0.8}{$-$0.022}}$ & $0.2524$ & - & - \\[1ex] 
$0.6733$--$0.6813$ & $0.14563^{\scalebox{0.8}{$+$0.00059}}_{\scalebox{0.8}{$-$0.00057}}$ & $0.14475^{\scalebox{0.8}{$+$0.00059}}_{\scalebox{0.8}{$-$0.00057}}$ & $0.14427^{\scalebox{0.8}{$+$0.00059}}_{\scalebox{0.8}{$-$0.00058}}$ & $0.314^{\scalebox{0.8}{$+$0.016}}_{\scalebox{0.8}{$-$0.018}}$ & $0.2523$ & - & - \\[1ex] 
$0.6813$--$0.6893$ & $0.14497^{\scalebox{0.8}{$+$0.00068}}_{\scalebox{0.8}{$-$0.00062}}$ & $0.14408^{\scalebox{0.8}{$+$0.00067}}_{\scalebox{0.8}{$-$0.00063}}$ & $0.14362^{\scalebox{0.8}{$+$0.00067}}_{\scalebox{0.8}{$-$0.00062}}$ & $0.321^{\scalebox{0.8}{$+$0.018}}_{\scalebox{0.8}{$-$0.020}}$ & $0.2549$ & - & - \\[1ex] 
$0.6893$--$0.6973$ & $0.14378^{\scalebox{0.8}{$+$0.00055}}_{\scalebox{0.8}{$-$0.00056}}$ & $0.14287^{\scalebox{0.8}{$+$0.00058}}_{\scalebox{0.8}{$-$0.00056}}$ & $0.14242^{\scalebox{0.8}{$+$0.00056}}_{\scalebox{0.8}{$-$0.00056}}$ & $0.334^{\scalebox{0.8}{$+$0.016}}_{\scalebox{0.8}{$-$0.016}}$ & $0.2505$ & - & - \\[1ex] 
$0.6973$--$0.7053$ & $0.14509^{\scalebox{0.8}{$+$0.00071}}_{\scalebox{0.8}{$-$0.00067}}$ & $0.14420^{\scalebox{0.8}{$+$0.00071}}_{\scalebox{0.8}{$-$0.00067}}$ & $0.14374^{\scalebox{0.8}{$+$0.00071}}_{\scalebox{0.8}{$-$0.00066}}$ & $0.304^{\scalebox{0.8}{$+$0.020}}_{\scalebox{0.8}{$-$0.024}}$ & $0.2527$ & - & - \\[1ex] 
$0.7053$--$0.7133$ & $0.14491^{\scalebox{0.8}{$+$0.00056}}_{\scalebox{0.8}{$-$0.00055}}$ & $0.14402^{\scalebox{0.8}{$+$0.00056}}_{\scalebox{0.8}{$-$0.00056}}$ & $0.14358^{\scalebox{0.8}{$+$0.00057}}_{\scalebox{0.8}{$-$0.00055}}$ & $0.308^{\scalebox{0.8}{$+$0.015}}_{\scalebox{0.8}{$-$0.017}}$ & $0.2499$ & - & - \\[1ex] 
$0.7133$--$0.7213$ & $0.14489^{\scalebox{0.8}{$+$0.00053}}_{\scalebox{0.8}{$-$0.00054}}$ & $0.14399^{\scalebox{0.8}{$+$0.00056}}_{\scalebox{0.8}{$-$0.00056}}$ & $0.14352^{\scalebox{0.8}{$+$0.00054}}_{\scalebox{0.8}{$-$0.00055}}$ & $0.309^{\scalebox{0.8}{$+$0.015}}_{\scalebox{0.8}{$-$0.016}}$ & $0.2463$ & - & - \\[1ex] 
$0.7213$--$0.7293$ & $0.14557^{\scalebox{0.8}{$+$0.00061}}_{\scalebox{0.8}{$-$0.00059}}$ & $0.14470^{\scalebox{0.8}{$+$0.00064}}_{\scalebox{0.8}{$-$0.00060}}$ & $0.14424^{\scalebox{0.8}{$+$0.00061}}_{\scalebox{0.8}{$-$0.00062}}$ & $0.269^{\scalebox{0.8}{$+$0.017}}_{\scalebox{0.8}{$-$0.020}}$ & $0.2477$ & - & - \\[1ex] 
$0.7293$--$0.7373$ & $0.14453^{\scalebox{0.8}{$+$0.00054}}_{\scalebox{0.8}{$-$0.00055}}$ & $0.14365^{\scalebox{0.8}{$+$0.00053}}_{\scalebox{0.8}{$-$0.00055}}$ & $0.14319^{\scalebox{0.8}{$+$0.00054}}_{\scalebox{0.8}{$-$0.00055}}$ & $0.284^{\scalebox{0.8}{$+$0.016}}_{\scalebox{0.8}{$-$0.016}}$ & $0.2501$ & - & - \\[1ex] 
$0.7373$--$0.7453$ & $0.14614^{\scalebox{0.8}{$+$0.00058}}_{\scalebox{0.8}{$-$0.00059}}$ & $0.14524^{\scalebox{0.8}{$+$0.00058}}_{\scalebox{0.8}{$-$0.00059}}$ & $0.14479^{\scalebox{0.8}{$+$0.00058}}_{\scalebox{0.8}{$-$0.00058}}$ & $0.264^{\scalebox{0.8}{$+$0.017}}_{\scalebox{0.8}{$-$0.017}}$ & $0.2475$ & - & - \\[1ex] 
$0.7453$--$0.7533$ & $0.14487^{\scalebox{0.8}{$+$0.00058}}_{\scalebox{0.8}{$-$0.00061}}$ & $0.14400^{\scalebox{0.8}{$+$0.00059}}_{\scalebox{0.8}{$-$0.00061}}$ & $0.14353^{\scalebox{0.8}{$+$0.00059}}_{\scalebox{0.8}{$-$0.00060}}$ & $0.282^{\scalebox{0.8}{$+$0.017}}_{\scalebox{0.8}{$-$0.017}}$ & $0.2488$ & - & - \\[1ex] 
$0.7533$--$0.7613$ & $0.14429^{\scalebox{0.8}{$+$0.00071}}_{\scalebox{0.8}{$-$0.00072}}$ & $0.14344^{\scalebox{0.8}{$+$0.00073}}_{\scalebox{0.8}{$-$0.00072}}$ & $0.14300^{\scalebox{0.8}{$+$0.00070}}_{\scalebox{0.8}{$-$0.00071}}$ & $0.286^{\scalebox{0.8}{$+$0.020}}_{\scalebox{0.8}{$-$0.022}}$ & $0.2457$ & - & - \\[1ex] 
$0.7613$--$0.7693$ & $0.14744^{\scalebox{0.8}{$+$0.00108}}_{\scalebox{0.8}{$-$0.00108}}$ & $0.14656^{\scalebox{0.8}{$+$0.00110}}_{\scalebox{0.8}{$-$0.00105}}$ & $0.14617^{\scalebox{0.8}{$+$0.00107}}_{\scalebox{0.8}{$-$0.00106}}$ & $0.274^{\scalebox{0.8}{$+$0.030}}_{\scalebox{0.8}{$-$0.033}}$ & $0.2515$ & - & - \\[1ex] 
$0.7693$--$0.7773$ & $0.14577^{\scalebox{0.8}{$+$0.00061}}_{\scalebox{0.8}{$-$0.00060}}$ & $0.14488^{\scalebox{0.8}{$+$0.00061}}_{\scalebox{0.8}{$-$0.00061}}$ & $0.14445^{\scalebox{0.8}{$+$0.00060}}_{\scalebox{0.8}{$-$0.00061}}$ & $0.240^{\scalebox{0.8}{$+$0.018}}_{\scalebox{0.8}{$-$0.019}}$ & $0.2549$ & - & - \\[1ex] 
$0.7773$--$0.7853$ & $0.14428^{\scalebox{0.8}{$+$0.00063}}_{\scalebox{0.8}{$-$0.00060}}$ & $0.14345^{\scalebox{0.8}{$+$0.00063}}_{\scalebox{0.8}{$-$0.00062}}$ & $0.14299^{\scalebox{0.8}{$+$0.00063}}_{\scalebox{0.8}{$-$0.00061}}$ & $0.245^{\scalebox{0.8}{$+$0.018}}_{\scalebox{0.8}{$-$0.020}}$ & $0.2489$ & - & - \\[1ex] 
$0.7853$--$0.7933$ & $0.14495^{\scalebox{0.8}{$+$0.00092}}_{\scalebox{0.8}{$-$0.00083}}$ & $0.14408^{\scalebox{0.8}{$+$0.00095}}_{\scalebox{0.8}{$-$0.00083}}$ & $0.14363^{\scalebox{0.8}{$+$0.00093}}_{\scalebox{0.8}{$-$0.00083}}$ & $0.244^{\scalebox{0.8}{$+$0.024}}_{\scalebox{0.8}{$-$0.026}}$ & $0.2456$ & - & - \\[1ex] 
$0.7933$--$0.8013$ & $0.14368^{\scalebox{0.8}{$+$0.00071}}_{\scalebox{0.8}{$-$0.00070}}$ & $0.14287^{\scalebox{0.8}{$+$0.00074}}_{\scalebox{0.8}{$-$0.00070}}$ & $0.14243^{\scalebox{0.8}{$+$0.00070}}_{\scalebox{0.8}{$-$0.00069}}$ & $0.274^{\scalebox{0.8}{$+$0.020}}_{\scalebox{0.8}{$-$0.022}}$ & $0.2474$ & - & - \\[1ex] 
$0.8013$--$0.8093$ & $0.14417^{\scalebox{0.8}{$+$0.00093}}_{\scalebox{0.8}{$-$0.00097}}$ & $0.14334^{\scalebox{0.8}{$+$0.00093}}_{\scalebox{0.8}{$-$0.00097}}$ & $0.14291^{\scalebox{0.8}{$+$0.00096}}_{\scalebox{0.8}{$-$0.00095}}$ & $0.250^{\scalebox{0.8}{$+$0.027}}_{\scalebox{0.8}{$-$0.032}}$ & $0.2472$ & - & - \\[1ex] 
$0.8093$--$0.8173$ & $0.14464^{\scalebox{0.8}{$+$0.00085}}_{\scalebox{0.8}{$-$0.00076}}$ & $0.14385^{\scalebox{0.8}{$+$0.00084}}_{\scalebox{0.8}{$-$0.00077}}$ & $0.14341^{\scalebox{0.8}{$+$0.00083}}_{\scalebox{0.8}{$-$0.00077}}$ & $0.254^{\scalebox{0.8}{$+$0.022}}_{\scalebox{0.8}{$-$0.024}}$ & $0.2475$ & - & - \\[1ex] 
$0.8173$--$0.8253$ & $0.14501^{\scalebox{0.8}{$+$0.00087}}_{\scalebox{0.8}{$-$0.00083}}$ & $0.14420^{\scalebox{0.8}{$+$0.00087}}_{\scalebox{0.8}{$-$0.00085}}$ & $0.14377^{\scalebox{0.8}{$+$0.00085}}_{\scalebox{0.8}{$-$0.00085}}$ & $0.241^{\scalebox{0.8}{$+$0.024}}_{\scalebox{0.8}{$-$0.028}}$ & $0.2442$ & - & - \\[1ex] 
$0.8253$--$0.8333$ & $0.14477^{\scalebox{0.8}{$+$0.00063}}_{\scalebox{0.8}{$-$0.00064}}$ & $0.14394^{\scalebox{0.8}{$+$0.00061}}_{\scalebox{0.8}{$-$0.00062}}$ & $0.14352^{\scalebox{0.8}{$+$0.00063}}_{\scalebox{0.8}{$-$0.00064}}$ & $0.266^{\scalebox{0.8}{$+$0.018}}_{\scalebox{0.8}{$-$0.019}}$ & $0.2507$ & - & - \\[1ex] 
\textbf{\textit{TESS}} & & & & & & \\[1ex] 
$0.6000$--$1.0000$ & $0.14405^{\scalebox{0.8}{$+$0.00074}}_{\scalebox{0.8}{$-$0.00061}}$ & $0.14322^{\scalebox{0.8}{$+$0.00073}}_{\scalebox{0.8}{$-$0.00061}}$ & $0.14276^{\scalebox{0.8}{$+$0.00072}}_{\scalebox{0.8}{$-$0.00061}}$ & $0.6590$ & $-0.4538$ & $0.9531$ & $-0.4668$ \\ 
\vspace{-4pt} \\ 
\hline 
\end{tabular} 
\vspace{-4pt} \\ 
\caption{Measured spectrophotometric transit depths of WASP-6b for the G600B and G600RI datasets in addition to the weighted average transit depth of the \textit{TESS} photometry. Transit depths calculated following an activity correction based on the \textit{TESS} and \textit{AIT} photometry are also independently shown.} 
\label{slcparams} 
\end{table*} 

\begin{table*} 
\centering 
\fontsize{7}{9}\selectfont 
\begin{tabular}{l c c c c c c c} 
\hline 
\hline 
\vspace{-8pt} \\ 
Wavelength ($\mu$m) & $R_\textrm{p}/R_*$ & $R_\textrm{p}/R_{*,\textrm{\textit{TESS}}}$ & $R_\textrm{p}/R_{*,\textrm{\textit{AIT}}}$ & $c_1$ & $c_2$ & $c_3$ & $c_4$  \\ 
\hline 
\vspace{-5pt} \\ 
\textbf{STIS 430} & & & & & & & \\[1ex] 
$0.3250$--$0.4000$ & $0.14758^{\scalebox{0.8}{$+$0.00128}}_{\scalebox{0.8}{$-$0.00136}}$ & $0.14644^{\scalebox{0.8}{$+$0.00136}}_{\scalebox{0.8}{$-$0.00147}}$ & $0.14583^{\scalebox{0.8}{$+$0.00164}}_{\scalebox{0.8}{$-$0.00145}}$ & $0.4458$ & $-0.4545$ & $1.4520$ & $-0.5362$ \\[1ex] 
$0.4000$--$0.4400$ & $0.14714^{\scalebox{0.8}{$+$0.00073}}_{\scalebox{0.8}{$-$0.00075}}$ & $0.14610^{\scalebox{0.8}{$+$0.00085}}_{\scalebox{0.8}{$-$0.00103}}$ & $0.14554^{\scalebox{0.8}{$+$0.00072}}_{\scalebox{0.8}{$-$0.00068}}$ & $0.3943$ & $-0.1849$ & $1.1642$ & $-0.4810$ \\[1ex] 
$0.4400$--$0.4750$ & $0.14586^{\scalebox{0.8}{$+$0.00069}}_{\scalebox{0.8}{$-$0.00073}}$ & $0.14486^{\scalebox{0.8}{$+$0.00067}}_{\scalebox{0.8}{$-$0.00069}}$ & $0.14430^{\scalebox{0.8}{$+$0.00070}}_{\scalebox{0.8}{$-$0.00071}}$ & $0.4227$ & $-0.0653$ & $0.9769$ & $-0.4525$ \\[1ex] 
$0.4750$--$0.5000$ & $0.14581^{\scalebox{0.8}{$+$0.00104}}_{\scalebox{0.8}{$-$0.00093}}$ & $0.14469^{\scalebox{0.8}{$+$0.00105}}_{\scalebox{0.8}{$-$0.00097}}$ & $0.14470^{\scalebox{0.8}{$+$0.00106}}_{\scalebox{0.8}{$-$0.00108}}$ & $0.4290$ & $0.1635$ & $0.5462$ & $-0.2798$ \\[1ex] 
$0.5000$--$0.5250$ & $0.14607^{\scalebox{0.8}{$+$0.00084}}_{\scalebox{0.8}{$-$0.00089}}$ & $0.14509^{\scalebox{0.8}{$+$0.00089}}_{\scalebox{0.8}{$-$0.00090}}$ & $0.14460^{\scalebox{0.8}{$+$0.00084}}_{\scalebox{0.8}{$-$0.00084}}$ & $0.5095$ & $-0.1362$ & $0.8108$ & $-0.3523$ \\[1ex] 
$0.5250$--$0.5450$ & $0.14647^{\scalebox{0.8}{$+$0.00078}}_{\scalebox{0.8}{$-$0.00081}}$ & $0.14547^{\scalebox{0.8}{$+$0.00078}}_{\scalebox{0.8}{$-$0.00083}}$ & $0.14493^{\scalebox{0.8}{$+$0.00111}}_{\scalebox{0.8}{$-$0.00128}}$ & $0.4855$ & $0.0726$ & $0.5461$ & $-0.2855$ \\[1ex] 
$0.5450$--$0.5700$ & $0.14520^{\scalebox{0.8}{$+$0.00069}}_{\scalebox{0.8}{$-$0.00073}}$ & $0.14419^{\scalebox{0.8}{$+$0.00073}}_{\scalebox{0.8}{$-$0.00078}}$ & $0.14378^{\scalebox{0.8}{$+$0.00071}}_{\scalebox{0.8}{$-$0.00071}}$ & $0.5293$ & $-0.0370$ & $0.6028$ & $-0.2840$ \\[1ex] 
\textbf{STIS 750} & & & & & & & \\[1ex] 
$0.5500$--$0.5868$ & $0.14549^{\scalebox{0.8}{$+$0.00085}}_{\scalebox{0.8}{$-$0.00087}}$ & $0.14461^{\scalebox{0.8}{$+$0.00082}}_{\scalebox{0.8}{$-$0.00086}}$ & $0.14409^{\scalebox{0.8}{$+$0.00084}}_{\scalebox{0.8}{$-$0.00085}}$ & $0.5334$ & $0.0001$ & $0.5262$ & $-0.2553$ \\[1ex] 
$0.5868$--$0.5918$ & $0.14724^{\scalebox{0.8}{$+$0.00181}}_{\scalebox{0.8}{$-$0.00185}}$ & $0.14629^{\scalebox{0.8}{$+$0.00181}}_{\scalebox{0.8}{$-$0.00179}}$ & $0.14581^{\scalebox{0.8}{$+$0.00184}}_{\scalebox{0.8}{$-$0.00189}}$ & $0.5755$ & $-0.1509$ & $0.7106$ & $-0.3503$ \\[1ex] 
$0.5918$--$0.6200$ & $0.14451^{\scalebox{0.8}{$+$0.00087}}_{\scalebox{0.8}{$-$0.00087}}$ & $0.14355^{\scalebox{0.8}{$+$0.00084}}_{\scalebox{0.8}{$-$0.00081}}$ & $0.14311^{\scalebox{0.8}{$+$0.00085}}_{\scalebox{0.8}{$-$0.00083}}$ & $0.5635$ & $-0.0259$ & $0.4825$ & $-0.2387$ \\[1ex] 
$0.6200$--$0.6600$ & $0.14506^{\scalebox{0.8}{$+$0.00110}}_{\scalebox{0.8}{$-$0.00133}}$ & $0.14414^{\scalebox{0.8}{$+$0.00111}}_{\scalebox{0.8}{$-$0.00138}}$ & $0.14367^{\scalebox{0.8}{$+$0.00110}}_{\scalebox{0.8}{$-$0.00138}}$ & $0.5951$ & $-0.0778$ & $0.4802$ & $-0.2411$ \\[1ex] 
$0.6600$--$0.7000$ & $0.14494^{\scalebox{0.8}{$+$0.00094}}_{\scalebox{0.8}{$-$0.00096}}$ & $0.14413^{\scalebox{0.8}{$+$0.00095}}_{\scalebox{0.8}{$-$0.00099}}$ & $0.14365^{\scalebox{0.8}{$+$0.00092}}_{\scalebox{0.8}{$-$0.00092}}$ & $0.6087$ & $-0.1196$ & $0.4761$ & $-0.2288$ \\[1ex] 
$0.7000$--$0.7599$ & $0.14501^{\scalebox{0.8}{$+$0.00086}}_{\scalebox{0.8}{$-$0.00094}}$ & $0.14411^{\scalebox{0.8}{$+$0.00089}}_{\scalebox{0.8}{$-$0.00093}}$ & $0.14372^{\scalebox{0.8}{$+$0.00086}}_{\scalebox{0.8}{$-$0.00090}}$ & $0.6251$ & $-0.1738$ & $0.4843$ & $-0.2277$ \\[1ex] 
$0.7599$--$0.7769$ & $0.14743^{\scalebox{0.8}{$+$0.00125}}_{\scalebox{0.8}{$-$0.00127}}$ & $0.14660^{\scalebox{0.8}{$+$0.00125}}_{\scalebox{0.8}{$-$0.00126}}$ & $0.14625^{\scalebox{0.8}{$+$0.00123}}_{\scalebox{0.8}{$-$0.00120}}$ & $0.6373$ & $-0.2154$ & $0.4876$ & $-0.2239$ \\[1ex] 
$0.7769$--$0.8400$ & $0.14472^{\scalebox{0.8}{$+$0.00104}}_{\scalebox{0.8}{$-$0.00106}}$ & $0.14390^{\scalebox{0.8}{$+$0.00108}}_{\scalebox{0.8}{$-$0.00108}}$ & $0.14349^{\scalebox{0.8}{$+$0.00103}}_{\scalebox{0.8}{$-$0.00105}}$ & $0.6354$ & $-0.2178$ & $0.4696$ & $-0.2167$ \\[1ex] 
$0.8400$--$0.9200$ & $0.14443^{\scalebox{0.8}{$+$0.00085}}_{\scalebox{0.8}{$-$0.00085}}$ & $0.14367^{\scalebox{0.8}{$+$0.00086}}_{\scalebox{0.8}{$-$0.00086}}$ & $0.14328^{\scalebox{0.8}{$+$0.00085}}_{\scalebox{0.8}{$-$0.00086}}$ & $0.6486$ & $-0.2801$ & $0.4821$ & $-0.2159$ \\[1ex] 
$0.9200$--$1.0300$ & $0.14365^{\scalebox{0.8}{$+$0.00124}}_{\scalebox{0.8}{$-$0.00119}}$ & $0.14294^{\scalebox{0.8}{$+$0.00122}}_{\scalebox{0.8}{$-$0.00118}}$ & $0.14257^{\scalebox{0.8}{$+$0.00121}}_{\scalebox{0.8}{$-$0.00120}}$ & $0.6332$ & $-0.2595$ & $0.4444$ & $-0.2039$ \\[1ex] 
\textbf{WFC3 G141} & & & & & & & \\[1ex] 
$1.1308$--$1.1493$ & $0.14344^{\scalebox{0.8}{$+$0.00065}}_{\scalebox{0.8}{$-$0.00064}}$ & $0.14262^{\scalebox{0.8}{$+$0.00064}}_{\scalebox{0.8}{$-$0.00061}}$ & $0.14230^{\scalebox{0.8}{$+$0.00064}}_{\scalebox{0.8}{$-$0.00062}}$ & $0.5850$ & $-0.2405$ & $0.4859$ & $-0.2567$ \\[1ex] 
$1.1493$--$1.1678$ & $0.14282^{\scalebox{0.8}{$+$0.00060}}_{\scalebox{0.8}{$-$0.00062}}$ & $0.14198^{\scalebox{0.8}{$+$0.00066}}_{\scalebox{0.8}{$-$0.00074}}$ & $0.14165^{\scalebox{0.8}{$+$0.00066}}_{\scalebox{0.8}{$-$0.00069}}$ & $0.5739$ & $-0.1975$ & $0.4283$ & $-0.2336$ \\[1ex] 
$1.1678$--$1.1863$ & $0.14332^{\scalebox{0.8}{$+$0.00066}}_{\scalebox{0.8}{$-$0.00066}}$ & $0.14242^{\scalebox{0.8}{$+$0.00073}}_{\scalebox{0.8}{$-$0.00075}}$ & $0.14212^{\scalebox{0.8}{$+$0.00077}}_{\scalebox{0.8}{$-$0.00077}}$ & $0.5691$ & $-0.1734$ & $0.3892$ & $-0.2181$ \\[1ex] 
$1.1863$--$1.2048$ & $0.14400^{\scalebox{0.8}{$+$0.00085}}_{\scalebox{0.8}{$-$0.00083}}$ & $0.14322^{\scalebox{0.8}{$+$0.00080}}_{\scalebox{0.8}{$-$0.00082}}$ & $0.14296^{\scalebox{0.8}{$+$0.00080}}_{\scalebox{0.8}{$-$0.00081}}$ & $0.5643$ & $-0.1534$ & $0.3583$ & $-0.2049$ \\[1ex] 
$1.2048$--$1.2233$ & $0.14212^{\scalebox{0.8}{$+$0.00055}}_{\scalebox{0.8}{$-$0.00053}}$ & $0.14136^{\scalebox{0.8}{$+$0.00054}}_{\scalebox{0.8}{$-$0.00051}}$ & $0.14109^{\scalebox{0.8}{$+$0.00052}}_{\scalebox{0.8}{$-$0.00052}}$ & $0.5392$ & $-0.0570$ & $0.2359$ & $-0.1550$ \\[1ex] 
$1.2233$--$1.2418$ & $0.14379^{\scalebox{0.8}{$+$0.00058}}_{\scalebox{0.8}{$-$0.00051}}$ & $0.14301^{\scalebox{0.8}{$+$0.00056}}_{\scalebox{0.8}{$-$0.00051}}$ & $0.14274^{\scalebox{0.8}{$+$0.00054}}_{\scalebox{0.8}{$-$0.00052}}$ & $0.5287$ & $-0.0068$ & $0.1678$ & $-0.1276$ \\[1ex] 
$1.2418$--$1.2603$ & $0.14388^{\scalebox{0.8}{$+$0.00052}}_{\scalebox{0.8}{$-$0.00052}}$ & $0.14308^{\scalebox{0.8}{$+$0.00051}}_{\scalebox{0.8}{$-$0.00052}}$ & $0.14278^{\scalebox{0.8}{$+$0.00050}}_{\scalebox{0.8}{$-$0.00051}}$ & $0.5186$ & $0.0422$ & $0.0995$ & $-0.0998$ \\[1ex] 
$1.2603$--$1.2788$ & $0.14316^{\scalebox{0.8}{$+$0.00061}}_{\scalebox{0.8}{$-$0.00063}}$ & $0.14234^{\scalebox{0.8}{$+$0.00059}}_{\scalebox{0.8}{$-$0.00060}}$ & $0.14204^{\scalebox{0.8}{$+$0.00061}}_{\scalebox{0.8}{$-$0.00061}}$ & $0.5153$ & $0.0714$ & $0.0602$ & $-0.0882$ \\[1ex] 
$1.2788$--$1.2973$ & $0.14292^{\scalebox{0.8}{$+$0.00061}}_{\scalebox{0.8}{$-$0.00062}}$ & $0.14212^{\scalebox{0.8}{$+$0.00072}}_{\scalebox{0.8}{$-$0.00069}}$ & $0.14185^{\scalebox{0.8}{$+$0.00070}}_{\scalebox{0.8}{$-$0.00066}}$ & $0.5137$ & $0.1156$ & $-0.0193$ & $-0.0577$ \\[1ex] 
$1.2973$--$1.3158$ & $0.14356^{\scalebox{0.8}{$+$0.00056}}_{\scalebox{0.8}{$-$0.00054}}$ & $0.14279^{\scalebox{0.8}{$+$0.00055}}_{\scalebox{0.8}{$-$0.00053}}$ & $0.14248^{\scalebox{0.8}{$+$0.00056}}_{\scalebox{0.8}{$-$0.00052}}$ & $0.4957$ & $0.1714$ & $-0.0897$ & $-0.0212$ \\[1ex] 
$1.3158$--$1.3343$ & $0.14402^{\scalebox{0.8}{$+$0.00049}}_{\scalebox{0.8}{$-$0.00049}}$ & $0.14333^{\scalebox{0.8}{$+$0.00056}}_{\scalebox{0.8}{$-$0.00053}}$ & $0.14304^{\scalebox{0.8}{$+$0.00055}}_{\scalebox{0.8}{$-$0.00052}}$ & $0.4904$ & $0.2140$ & $-0.1587$ & $0.0094$ \\[1ex] 
$1.3343$--$1.3528$ & $0.14494^{\scalebox{0.8}{$+$0.00053}}_{\scalebox{0.8}{$-$0.00052}}$ & $0.14418^{\scalebox{0.8}{$+$0.00052}}_{\scalebox{0.8}{$-$0.00052}}$ & $0.14387^{\scalebox{0.8}{$+$0.00053}}_{\scalebox{0.8}{$-$0.00053}}$ & $0.4814$ & $0.2785$ & $-0.2558$ & $0.0510$ \\[1ex] 
$1.3528$--$1.3713$ & $0.14481^{\scalebox{0.8}{$+$0.00050}}_{\scalebox{0.8}{$-$0.00051}}$ & $0.14400^{\scalebox{0.8}{$+$0.00050}}_{\scalebox{0.8}{$-$0.00053}}$ & $0.14374^{\scalebox{0.8}{$+$0.00049}}_{\scalebox{0.8}{$-$0.00051}}$ & $0.4826$ & $0.3004$ & $-0.3031$ & $0.0739$ \\[1ex] 
$1.3713$--$1.3898$ & $0.14443^{\scalebox{0.8}{$+$0.00061}}_{\scalebox{0.8}{$-$0.00061}}$ & $0.14362^{\scalebox{0.8}{$+$0.00061}}_{\scalebox{0.8}{$-$0.00062}}$ & $0.14333^{\scalebox{0.8}{$+$0.00063}}_{\scalebox{0.8}{$-$0.00062}}$ & $0.4781$ & $0.3553$ & $-0.3914$ & $0.1116$ \\[1ex] 
$1.3898$--$1.4083$ & $0.14450^{\scalebox{0.8}{$+$0.00059}}_{\scalebox{0.8}{$-$0.00058}}$ & $0.14376^{\scalebox{0.8}{$+$0.00062}}_{\scalebox{0.8}{$-$0.00060}}$ & $0.14352^{\scalebox{0.8}{$+$0.00060}}_{\scalebox{0.8}{$-$0.00060}}$ & $0.4754$ & $0.4040$ & $-0.4798$ & $0.1539$ \\[1ex] 
$1.4083$--$1.4268$ & $0.14436^{\scalebox{0.8}{$+$0.00068}}_{\scalebox{0.8}{$-$0.00072}}$ & $0.14355^{\scalebox{0.8}{$+$0.00071}}_{\scalebox{0.8}{$-$0.00075}}$ & $0.14328^{\scalebox{0.8}{$+$0.00069}}_{\scalebox{0.8}{$-$0.00070}}$ & $0.4814$ & $0.4162$ & $-0.5192$ & $0.1739$ \\[1ex] 
$1.4268$--$1.4453$ & $0.14510^{\scalebox{0.8}{$+$0.00056}}_{\scalebox{0.8}{$-$0.00059}}$ & $0.14432^{\scalebox{0.8}{$+$0.00056}}_{\scalebox{0.8}{$-$0.00058}}$ & $0.14402^{\scalebox{0.8}{$+$0.00055}}_{\scalebox{0.8}{$-$0.00058}}$ & $0.4909$ & $0.4304$ & $-0.5695$ & $0.2001$ \\[1ex] 
$1.4453$--$1.4638$ & $0.14472^{\scalebox{0.8}{$+$0.00051}}_{\scalebox{0.8}{$-$0.00050}}$ & $0.14394^{\scalebox{0.8}{$+$0.00058}}_{\scalebox{0.8}{$-$0.00057}}$ & $0.14368^{\scalebox{0.8}{$+$0.00059}}_{\scalebox{0.8}{$-$0.00055}}$ & $0.5020$ & $0.4428$ & $-0.6133$ & $0.2215$ \\[1ex] 
$1.4638$--$1.4823$ & $0.14352^{\scalebox{0.8}{$+$0.00051}}_{\scalebox{0.8}{$-$0.00050}}$ & $0.14272^{\scalebox{0.8}{$+$0.00058}}_{\scalebox{0.8}{$-$0.00055}}$ & $0.14247^{\scalebox{0.8}{$+$0.00057}}_{\scalebox{0.8}{$-$0.00056}}$ & $0.5176$ & $0.4336$ & $-0.6300$ & $0.2323$ \\[1ex] 
$1.4823$--$1.5008$ & $0.14408^{\scalebox{0.8}{$+$0.00060}}_{\scalebox{0.8}{$-$0.00061}}$ & $0.14344^{\scalebox{0.8}{$+$0.00068}}_{\scalebox{0.8}{$-$0.00070}}$ & $0.14316^{\scalebox{0.8}{$+$0.00070}}_{\scalebox{0.8}{$-$0.00067}}$ & $0.5378$ & $0.4095$ & $-0.6377$ & $0.2437$ \\[1ex] 
$1.5008$--$1.5193$ & $0.14422^{\scalebox{0.8}{$+$0.00063}}_{\scalebox{0.8}{$-$0.00062}}$ & $0.14352^{\scalebox{0.8}{$+$0.00060}}_{\scalebox{0.8}{$-$0.00060}}$ & $0.14327^{\scalebox{0.8}{$+$0.00060}}_{\scalebox{0.8}{$-$0.00061}}$ & $0.5610$ & $0.3679$ & $-0.6216$ & $0.2451$ \\[1ex] 
$1.5193$--$1.5378$ & $0.14381^{\scalebox{0.8}{$+$0.00060}}_{\scalebox{0.8}{$-$0.00062}}$ & $0.14304^{\scalebox{0.8}{$+$0.00061}}_{\scalebox{0.8}{$-$0.00064}}$ & $0.14276^{\scalebox{0.8}{$+$0.00062}}_{\scalebox{0.8}{$-$0.00062}}$ & $0.5891$ & $0.3711$ & $-0.6882$ & $0.2845$ \\[1ex] 
$1.5378$--$1.5563$ & $0.14388^{\scalebox{0.8}{$+$0.00088}}_{\scalebox{0.8}{$-$0.00065}}$ & $0.14308^{\scalebox{0.8}{$+$0.00098}}_{\scalebox{0.8}{$-$0.00072}}$ & $0.14278^{\scalebox{0.8}{$+$0.00084}}_{\scalebox{0.8}{$-$0.00070}}$ & $0.6200$ & $0.3216$ & $-0.6681$ & $0.2839$ \\[1ex] 
$1.5563$--$1.5748$ & $0.14267^{\scalebox{0.8}{$+$0.00077}}_{\scalebox{0.8}{$-$0.00071}}$ & $0.14214^{\scalebox{0.8}{$+$0.00107}}_{\scalebox{0.8}{$-$0.00093}}$ & $0.14187^{\scalebox{0.8}{$+$0.00102}}_{\scalebox{0.8}{$-$0.00094}}$ & $0.6541$ & $0.2446$ & $-0.6119$ & $0.2700$ \\[1ex] 
$1.5748$--$1.5933$ & $0.14323^{\scalebox{0.8}{$+$0.00090}}_{\scalebox{0.8}{$-$0.00087}}$ & $0.14253^{\scalebox{0.8}{$+$0.00082}}_{\scalebox{0.8}{$-$0.00075}}$ & $0.14228^{\scalebox{0.8}{$+$0.00085}}_{\scalebox{0.8}{$-$0.00080}}$ & $0.6734$ & $0.1558$ & $-0.5064$ & $0.2283$ \\[1ex] 
$1.5933$--$1.6118$ & $0.14314^{\scalebox{0.8}{$+$0.00064}}_{\scalebox{0.8}{$-$0.00064}}$ & $0.14246^{\scalebox{0.8}{$+$0.00068}}_{\scalebox{0.8}{$-$0.00067}}$ & $0.14222^{\scalebox{0.8}{$+$0.00067}}_{\scalebox{0.8}{$-$0.00067}}$ & $0.7158$ & $0.1056$ & $-0.5040$ & $0.2401$ \\[1ex] 
$1.6118$--$1.6303$ & $0.14361^{\scalebox{0.8}{$+$0.00072}}_{\scalebox{0.8}{$-$0.00073}}$ & $0.14293^{\scalebox{0.8}{$+$0.00062}}_{\scalebox{0.8}{$-$0.00062}}$ & $0.14270^{\scalebox{0.8}{$+$0.00062}}_{\scalebox{0.8}{$-$0.00062}}$ & $0.7518$ & $0.0128$ & $-0.4181$ & $0.2107$ \\[1ex] 
$1.6303$--$1.6488$ & $0.14303^{\scalebox{0.8}{$+$0.00065}}_{\scalebox{0.8}{$-$0.00067}}$ & $0.14243^{\scalebox{0.8}{$+$0.00073}}_{\scalebox{0.8}{$-$0.00076}}$ & $0.14219^{\scalebox{0.8}{$+$0.00073}}_{\scalebox{0.8}{$-$0.00076}}$ & $0.7736$ & $-0.0330$ & $-0.3973$ & $0.2099$ \\[1ex] 
\vspace{-4pt} \\ 
\hline 
\end{tabular} 
\vspace{-4pt} \\ 
\caption{As in Table \ref{slcparams}, except for the STIS 430, STIS 750 and WFC3 G141 datasets.} 
\label{slcparams2} 
\end{table*}

\end{document}